\documentclass[11pt,letterpaper]{article}
\usepackage[letterpaper,margin=1in]{geometry}
\usepackage{amsmath,amssymb,amsthm,mathtools}
\usepackage{array,booktabs,longtable}
\usepackage{microtype}
\usepackage{enumitem}
\usepackage{hyperref}
\usepackage{xcolor}
\hypersetup{colorlinks=true,linkcolor=blue!55!black,citecolor=blue!55!black,urlcolor=blue!55!black,pdftitle={Lower Bounds for Products of Bounded-Degree Polynomials over Fp},pdfauthor={Bin Fu and Yunhui Fu},pdfsubject={Cubic and higher-degree product lower bounds over prime fields; self-contained characteristic-three core}}
\newtheorem{theorem}{Theorem}
\newtheorem{lemma}[theorem]{Lemma}
\newtheorem{corollary}[theorem]{Corollary}
\newtheorem{proposition}[theorem]{Proposition}

\theoremstyle{definition}\newtheorem{definition}[theorem]{Definition}
\theoremstyle{remark}\newtheorem{remark}[theorem]{Remark}
\newcommand{\F}{\mathbb F}
\newcommand{\E}{\mathbb E}

\newcommand{\MOD}{\operatorname{MOD}}

\newcommand{\xor}{\mathbin{\oplus}}
\title{Exponential Lower Bounds for Products of Cubic Polynomials over Prime Fields
}
\author{
Bin Fu\\
University of Texas Rio Grande Valley
\and
Yunhui Fu\\
Southwestern University
}
\date{November 2026}
\begin{document}
\begin{titlepage}
\thispagestyle{empty}
\maketitle
\begin{abstract}
We study exact representations of explicit signed modular predicates on
$\{-1,1\}^n$ as products of bounded-degree polynomials over prime fields,
measured by total variable-occurrence cost. Our main result is an exponential
lower bound for products of cubic polynomials. Over $\mathbb F_3$ the proof is
self-contained: every cubic factor in a nonvanishing product is a unit on the
sign cube, the unit identity forces bounded affine-observation complexity, and
a simultaneous clustering argument converts this structural compression into
an exponential occurrence lower bound. We also give an alternative
reconstruction route that isolates the quadratic structure behind the cubic
classification.

We then extend the cubic theorem to every fixed prime field $\mathbb F_p$ of
characteristic greater than three for odd target modulus $q$ coprime to $p$.
The only non-elementary structural input in this extension is Karam's
restricted-range rank theorem; the conversion from multiplication to parity and
the $p$-tuple clustering argument are proved explicitly. Finally, we establish
a coefficient-field-independent degree-lifting principle. If a lower bound is
known at one fixed degree, restriction to a large independent set yields a
quantitative lower bound at the next degree. In particular, the exponential
cubic theorem implies
$
L^{(p,q)}_4(n)=\Omega_{p,q}\!\left(\frac{n^4}{(\log(2+n))^3}\right).
$
For the characteristic-three core this already gives a lower bound for a
restricted mixed-modulus circuit class.  If a
$\MOD_2\circ\MOD_3\circ\MOD_2[2]$ circuit computing the signed
$\MOD_q$ predicate has middle-gate fan-ins $t_1,\ldots,t_m$, then
$
\sum_{j=1}^m (t_j+1)^2
 =\Omega_q\!\left(\frac{n^4}{(\log(2+n))^3}\right).
$
Consequently its number of bottom-to-middle wires is
$\Omega_q(n^2/(\log(2+n))^{3/2})$, and its gate count is
$\Omega_q(n^{4/3}/\log(2+n))$.  Finally, for every fixed $d\ge4$,
$
L^{(p,q)}_d(n)=\Omega_{p,q,d}\!\left(
\frac{n^{d/(d-3)}}{(\log(2+n))^{3/(d-3)}}\right).
$
The proofs apply to arbitrary nonsymmetric factor lists and require exact
agreement only on the sign cube.
\end{abstract}
\vfill
\end{titlepage}

\pagenumbering{roman}
\begingroup
\setcounter{tocdepth}{2}
\tableofcontents
\endgroup
\clearpage
\pagenumbering{arabic}
\setcounter{page}{1}

\section{Introduction}\label{sec:introduction}
How complicated must a modular predicate be when it is represented as a
product of low-degree polynomials over a prime field? This question isolates a
restricted but concrete mixed-modulus phenomenon: each individual factor is
algebraically simple, yet an unrestricted number of factors may interact
through multiplication. The polynomial method of Razborov and Smolensky gives
powerful lower bounds for bounded-depth circuits when the algebra and the gate
modulus are compatible~\cite{Razborov1987,Smolensky1987}; mixed moduli are much
more delicate, as reflected by the ACC literature~\cite{BeigelTarui1994,Yao1990}.
The Constant Degree Hypothesis of Barrington, Straubing, and Th\'erien provides
one important motivation for understanding such bounded-degree modular
representations~\cite{BarringtonStraubingTherien1990}.

A further motivation comes from three-layer mixed-modulus circuits.  In the
sign encoding, a bounded-fan-in $\MOD_2$ gate at the bottom produces a signed
monomial; a middle $\MOD_3$ gate can then be represented by a bounded-degree
polynomial predicate over $\mathbb F_3$; and a top $\MOD_2$ gate combines the
resulting signs multiplicatively.  Consequently, lower bounds for products of
bounded-degree polynomials over $\mathbb F_3$ provide an algebraic route toward
lower bounds for restricted circuits of the form
$\MOD_2\circ\MOD_3\circ\MOD_2[d]$, and more broadly illuminate the
mixed-modulus depth-three problem.  Strong lower bounds for such mixed-modulus
circuits have been a longstanding challenge in circuit complexity: the
Razborov--Smolensky polynomial method gives powerful bounds when the relevant
moduli are compatible prime powers, while circuits mixing distinct moduli lie
outside the reach of that theory in general~\cite{Razborov1987,Smolensky1987,Yao1990,BarringtonStraubingTherien1990}.
The product model studied here isolates a concrete algebraic core of this
barrier.  In the bounded-bottom-fan-in case, the algebraic lower bound does
translate back to an actual circuit lower bound.  More precisely, consider a
$\MOD_2\circ\MOD_3\circ\MOD_2[2]$ circuit, where every bottom
$\MOD_2$ gate has fan-in at most two.  If the $j$-th middle
$\MOD_3$ gate receives $t_j$ bottom-gate outputs, then its signed output can
be represented over $\mathbb F_3$ by a polynomial of degree at most four and
occurrence cost $O((t_j+1)^2)$.  The top parity gate multiplies these signed
outputs.  Hence the degree-four lower bound proved in this paper yields
\[
  \sum_j (t_j+1)^2
  =\Omega_q\!\left(\frac{n^4}{(\log(2+n))^3}\right).
\]
In particular, the number of bottom-to-middle wires is
$\Omega_q(n^2/(\log(2+n))^{3/2})$.  If circuit size counts gates, then with
$B$ bottom gates and $M$ middle gates one has
$\sum_j(t_j+1)^2=O(M(B+1)^2)$, so total gate size
$S=B+M+1$ satisfies
\[
  S=\Omega_q\!\left(\frac{n^{4/3}}{\log(2+n)}\right).
\]
Thus the product lower bound gives a superlinear gate lower bound for this
bounded-bottom-fan-in mixed-modulus class, without making any claim for the
unrestricted $\MOD_2\circ\MOD_3\circ\MOD_2$ model.

This paper focuses on the degree-three frontier. The central phenomenon is
structural: nonvanishing cubic factors on the sign cube cannot behave like
arbitrary cubic polynomials. In characteristic three, the identity $P^2=1$ on
$\{-1,1\}^n$ forces enough algebraic rigidity that every cubic unit depends on
only boundedly many affine observations. Once all factors have this form, a
random clustering argument aligns the observations while preserving a modular
obstruction in the target. The result is exponential occurrence complexity.

We deliberately separate this cubic theory from the direct quartic analysis.
The present paper therefore has one main technical narrative: classify or
compress cubic units, derive the exponential lower bound, extend the cubic
argument to general prime fields, and then use a field-independent restriction
principle to propagate the result to every higher fixed degree. This makes the
paper logically complete without mixing the qualitatively different quartic
support-hypergraph machinery into the cubic proof.

\subsection{Representation model}
Fix the characteristic-three core first. Let $p>3$ be prime and put
$X_n=\{-1,1\}^n\subseteq\mathbb F_3^n$. If $w(x)$ denotes the number of
negative coordinates, define
\[
F_{p,n}(x)=\begin{cases}-1,&p\mid w(x),\\1,&p\nmid w(x).\end{cases}
\]
We study exact functional identities
\[
F_{p,n}(x)=\prod_{j=1}^s P_j(x)\qquad(x\in X_n),
\]
where each $P_j$ has functional degree at most three. Equality is required on
$X_n$, not as a formal polynomial identity. The complexity measure is total
variable-occurrence cost in the original monomial listings. This functional
viewpoint is closely related to the framework of Forbes, Kumar, and
Saptharishi~\cite{ForbesKumarSaptharishi2016}, but the present model has a
single outer product and a specific syntactic occurrence measure.

Because the target never vanishes, every factor in a correct product is
nonzero on the sign cube. Over $\mathbb F_3$ this means
$P_j(x)\in\{-1,1\}$ and hence $P_j(x)^2=1$ for every $x\in X_n$. We call such
factors \emph{units on the sign cube}. The whole cubic argument begins from
this functional identity.

\subsection{Main contributions}
Our first theorem is an exponential lower bound in characteristic three. There
is a constant $\gamma>1$ such that every degree-at-most-three representation
of occurrence cost $C$ satisfies $\gamma^n\le C^2$ for all sufficiently large
$n$. Thus $L_3(n)=2^{\Omega_p(n)}$. The theorem quantifies over all
representations; no algorithmic-generation, symmetry, or disjointness
assumption is used.

The second contribution is a general-prime extension. For every fixed prime
$r>3$ and odd $q>1$ with $\gcd(r,q)=1$, exact products of degree-at-most-three
polynomials over $\mathbb F_r$ computing the signed modulo-$q$ predicate also
require exponential occurrence cost. Here Karam's restricted-range rank
result replaces the special characteristic-three unit identity as the main
structural input~\cite{Karam2023}.

The third contribution is a reusable degree-lifting theorem. A
$(d+1)$-degree representation of cost $S$ contains, after restriction to a
large independent set of its top-degree support hypergraph, a degree-$d$
representation of the same signed modular predicate and cost at most $S$.
This gives an implicit lifting inequality and closed-form consequences. In
particular, the exponential cubic lower bound over $\mathbb F_r$ yields
\[
L^{(r,q)}_d(n)=\Omega_{r,q,d}\!\left(
\frac{n^{d/(d-3)}}{(\log(2+n))^{3/(d-3)}}\right),\qquad d\ge4\text{ fixed}.
\]
The exponent remains strictly larger than one at every fixed degree.

\subsection{Method overview}
The characteristic-three proof has two complementary routes. Both begin by
turning nonvanishing into a unit identity. Quadratic bias and disjoint-minor
arguments then constrain the degree-two pieces. A classification of quadratic
units feeds a cubic elimination argument showing that a cubic unit is a
function of only constantly many affine forms. The direct route clusters the
affine observations of all factors and converts multiplication of their signs
to parity; the full coefficient of the target modulo predicate then supplies
the contradiction. The reconstruction route instead rewrites each cubic unit
through a controlled quadratic representation and reaches the same
exponential conclusion with a small polynomial overhead.

For a general coefficient field $\mathbb F_r$, bounded affine dependence is
obtained from Karam's range-to-rank theorem. The multiplicative group is then
projected to its $2$-primary quotient, the relevant binary digit is represented
as a bounded-degree parity polynomial, and a random $r$-tuple clustering
argument supplies the target obstruction.

\subsection{Relation to prior work and organization}
Clustering and degree reduction have important precedents in work of Grolmusz
and Tardos~\cite{Grolmusz2001,GT2000}; our contribution is to preserve the
sign-cube functional model and the explicit occurrence measure while tracking
what the unit condition forces. Structural results on biased low-degree
polynomials~\cite{GreenTao2009,HaramatyShpilka2010,KaufmanLovett2008} and on
restricted domains~\cite{GK2022,Karam2023} provide the closest algebraic
context. Correlation bounds for modular and low-degree polynomial tests are
also relevant~\cite{Bourgain2005,CaiGreenThierauf1996,Green2004}, but the
outer product of arbitrarily many factors requires a simultaneous argument.

Section~\ref{sec:model} fixes conventions. Section~\ref{sec:main-results}
states the main theorems. Section~\ref{sec:method-summary} gives a detailed
proof roadmap. The subsequent sections prove the characteristic-three cubic
structure and exponential lower bound, followed by the degree-lifting theorem
and the extension to general prime fields.  A near-final section then gives
the exact transfer from product lower bounds to restricted
$\MOD_2\circ\MOD_3\circ\MOD_2$ circuit lower bounds.  The final section
summarizes the scope and open problems.

\section{Problem, conventions, and complexity measure}\label{sec:model}
Let
\begin{equation}
  X_n=\{-1,1\}^n\subseteq \F_3^n,
\end{equation}
and let $w(x)=|\{j:x_j=-1\}|$ be the number of negative coordinates.  Define
\begin{equation}
F_{p,n}(x)=
\begin{cases}
-1,&p\mid w(x),\\
 1,&p\nmid w(x).
\end{cases}
\end{equation}
Thus $F_{p,n}$ is everywhere nonzero.

\begin{definition}[Cubic product representation]
A cubic representation of $F_{p,n}$ is a finite product $\prod_i P_i$ which agrees with $F_{p,n}$ at every point of $X_n$, where each $P_i\in\F_3[x_1,\dots,x_n]$ has multilinear degree at most three on $X_n$.
\end{definition}

On $X_n$ we have $x_j^2=1$.  Hence every polynomial has a unique multilinear representative modulo the ideal $(x_1^2-1,\dots,x_n^2-1)$.  Replacing a raw expression by this representative never increases functional degree.  We use the multilinear degree in the structural part of the proof, but we do \emph{not} reduce the syntactic cost that is being lower-bounded.

\begin{definition}[Occurrence cost]
For a listed monomial $a x_1^{e_1}\cdots x_n^{e_n}$ with $a\ne0$, its occurrence cost is $e_1+\cdots+e_n$.  The cost of a factor is the sum of these quantities over its listed monomial occurrences, and the cost $C$ of a product is the sum over all factors.  Constants have cost zero.
\end{definition}
Repeated terms, nonsymmetric factors, arbitrary powers, and either nonzero coefficient in $\F_3$ are allowed.  Thus the model is deliberately permissive.  A nonconstant listed factor has positive integer cost, so a representation of total cost $C$ contains at most $C$ positive-cost factors.

Because $F_{p,n}(x)\ne0$ for every $x$, no factor can vanish at any point of $X_n$.  Since the only nonzero elements of $\F_3$ are $1$ and $-1$, every factor satisfies
\begin{equation}
 P_i(x)\in\{1,-1\},\qquad P_i(x)^2=1\quad(x\in X_n).
\end{equation}
We call such a function a \emph{unit on the sign cube}.  The identity $P_i^2=1$ is the source of the coefficient relations used later.

\begin{definition}[Affine observation]
An affine observation is a form $A(x)=a_0+\sum_j a_jx_j$ over $\F_3$.  A function $f:X_n\to\F_3$ depends on $r$ affine observations if
\begin{equation}
 f(x)=H(A_1(x),\dots,A_r(x))
\end{equation}
for some function $H:\F_3^r\to\F_3$.  Only the values of $H$ on the attained image are relevant.
\end{definition}
This notion measures information rather than coordinate support.  A single affine observation may involve all $n$ variables.

\section{Main results}\label{sec:main-results}
Fix a prime target modulus $p>3$ in the characteristic-three model.

\begin{theorem}[Reconstruction form]\label{thm:reconstruction}
There exist constants $c>1$, $K\ge1$, and $n_0$ such that every cubic
representation of $F_{p,n}$ of occurrence cost $C$, for $n\ge n_0$, satisfies
\[
c^n\le (Kn^2C)^2.
\]
Consequently, for every real $a>0$ there is $N(a)$ such that $C\ge n^a$
whenever $n\ge N(a)$.
\end{theorem}

\begin{theorem}[Direct exponential form]\label{thm:direct}\label{prop:cubic}
There exist constants $\gamma>1$ and $n_1$ such that every cubic
representation of $F_{p,n}$ of original occurrence cost $C$, for $n\ge n_1$,
satisfies
\[
\gamma^n\le C^2.
\]
In particular $L_3(n)=2^{\Omega_p(n)}$.
\end{theorem}

\begin{corollary}[Generated families]\label{cor:generator}
Every procedure that outputs a correct cubic representation of $F_{p,n}$ for
each $n$ must output exponential occurrence cost at all sufficiently large
lengths. The conclusion is nonuniform: the running time of the procedure is
irrelevant.
\end{corollary}

The direct theorem is the main quantitative result. The reconstruction theorem
is retained because it exposes a different structural route and provides
useful checks on the cubic classification.

The general-prime theorem is stated later after the required rank notation is
introduced. In the notation of Section~\ref{sec:general-fp-cubic}, it gives an
exponential lower bound over every fixed $\mathbb F_r$, $r>3$, for odd target
modulus $q$ coprime to $r$. Section~\ref{sec:bounded-degree-lifting} then
propagates that cubic theorem to all higher fixed degrees.

\begin{theorem}[Circuit consequence of the degree-four bound]\label{thm:mod2-mod3-mod2-2}
Fix an odd prime $q>3$.  Let $C$ be a
$\MOD_2\circ\MOD_3\circ\MOD_2[2]$ circuit on $n$ inputs computing the
Boolean predicate $[q\mid |x|]$, equivalently its signed version $F_{q,n}$.
Assume every bottom $\MOD_2$ gate has fan-in at most two, and let $t_j$ be the
fan-in of the $j$-th middle $\MOD_3$ gate after constant middle gates are
removed.  Then
\begin{equation}\label{eq:circuit-quadratic-cost-lb}
  \sum_j (t_j+1)^2
  =\Omega_q\!\left(\frac{n^4}{(\log(2+n))^3}\right).
\end{equation}
Consequently, if $W$ is the number of bottom-to-middle wires, then
\begin{equation}\label{eq:circuit-wire-lb}
  W=\Omega_q\!\left(\frac{n^2}{(\log(2+n))^{3/2}}\right).
\end{equation}
If $S$ is the total number of gates, including bottom and middle gates, then
\begin{equation}\label{eq:circuit-gate-lb}
  S=\Omega_q\!\left(\frac{n^{4/3}}{\log(2+n)}\right).
\end{equation}
\end{theorem}

\section{Summary of the method}\label{sec:method-summary}
This section gives a compact map of the proof before the technical details.  The lower bound is obtained by passing through three different languages: products over $\F_3$, bounded-dimensional affine information, and finally multilinear polynomials over $\F_2$.  The transitions are exact; approximation is used only in the quadratic bias argument that supplies structure.

\subsection{Step 1: nonvanishing converts factors into units}
Suppose
\begin{equation}
        F_{q,n}(x)=\prod_{q=1}^{S}P_q(x)\qquad(x\in X_n),
\end{equation}
where every $P_q$ has multilinear degree at most three.  Since $F_{q,n}(x)\ne0$ everywhere, each $P_q$ is nonzero everywhere.  Over $\F_3$ this forces
\begin{equation}
        P_q(x)\in\{1,-1\},\qquad P_q(x)^2=1.
\end{equation}
Thus the lower-bound problem reduces to understanding low-degree polynomial functions satisfying the strong functional identity $P^2=1$ on the sign cube.

\subsection{Step 2: quadratic structure from bias and disjoint minors}
The cubic argument repeatedly produces quadratic slices.  We therefore first analyze a quadratic polynomial $Q$.  Write its off-diagonal quadratic coefficients as a symmetric matrix $C$.  If $C$ contains a large invertible minor with disjoint row and column sets, then after fixing all other variables the phase $\chi(Q)$ contains a nondegenerate bilinear term.  Two applications of Cauchy--Schwarz give exponential decay of its bias.  Consequently a quadratic with nonnegligible bias cannot contain a large disjoint invertible minor.

Maximality of such a minor then yields explicit algebraic reconstruction formulas for every remaining mixed coefficient.  The quadratic function can therefore be written, on the sign cube, as a bounded sum of products of affine forms.  This is the only bias-rank principle required by the direct proof.  It is a concrete degree-two counterpart of the broader finite-field structure theory in~\cite{GreenTao2009,KaufmanLovett2008,BhowmickLovett2022,GK2022}.

\subsection{Step 3: classify quadratic units}
For a quadratic unit $P$, slice in one coordinate $s$ and write
\begin{equation}
        P(s,y)=A(y)+sB(y),
\end{equation}
where $A$ is quadratic and $B$ is affine.  Evaluating at $s=\pm1$ and using $P^2=1$ gives
\begin{equation}
        AB=0,\qquad A^2+B^2=1
\end{equation}
on the remaining sign cube.  Coefficient comparisons force a sharp dichotomy: either
\begin{equation}
        P(x)=\varepsilon(1+L(x)^2)\qquad\text{for all }x\in X_n
\end{equation}
for an affine form $L$ and $\varepsilon\in\{\pm1\}$, or $P$ depends on at most eight original coordinates.  In either case a quadratic unit is controlled by bounded affine information.

\subsection{Step 4: compress every cubic unit to bounded affine dimension}
A cubic polynomial is sliced in three selected sign coordinates.  The eight slices are encoded by a small collection of coefficient functions on the remaining variables.  After affine information is separated, at most three genuinely quadratic coordinates remain.  If at least one quadratic coordinate remains, a suitable signed slice produces a nonzero linear combination $D$ of these coordinates that is nonvanishing on each fiber of the current affine observation map.

On a fixed fiber, one of the three translates $D-t$ has character bias at least a constant depending only on the number of current observations.  The quadratic bias-rank lemma then shows that $D$ is determined by boundedly many additional affine forms.  This permits one quadratic coordinate to be eliminated.  Since there are initially at most three, the process terminates after at most three rounds.  Hence there is an absolute constant $R$ such that every cubic unit has the form
\begin{equation}
        P(x)=H(A_1(x),\ldots,A_R(x))
\end{equation}
for affine $A_i:X_n\to\F_3$.

This is the structural heart of the paper.  The number $R$ may be very large, but it is independent of $n$ and of the particular factor.

\subsection{Step 5: convert multiplication to parity}
For a unit $P_q$, define its Boolean sign bit $f_q$ by
\begin{equation}
        P_q(x)=(-1)^{f_q(x)}.
\end{equation}
Then
\begin{equation}
        \prod_{q=1}^{S}P_q(x)=(-1)^{\sum_q f_q(x)},
\end{equation}
so equality with $F_{q,n}$ is equivalent to a modulo-two sum of the bits $f_q$.  The change of variables
\begin{equation}
        x_i=1+z_i\qquad(z_i\in\{0,1\}\subset\F_3)
\end{equation}
turns each affine observation in the sign variables into an affine observation in the Boolean variables.  Therefore every $f_q$ has an $R$-observation representation over $\F_3$.

\subsection{Step 6: cluster variables into triples}
The finite-counting clustering lemma is proved for $N=3\cdot2^d$.  Divide the $N$ Boolean variables into fixed-size chunks.  For one function of $R$ affine observations, a pigeonhole argument inside each chunk finds many coordinates with the same coefficient vector in $\F_3^R$.  A triple of such coordinates has coefficient sum zero in characteristic three.  A uniformly selected local triple partition therefore has a fixed positive probability of making that function omit at least one clustered variable.

If there are fewer than $\gamma_R^N$ summands, independent choices across chunks and a union bound show that one global triple partition works simultaneously for every summand.  After identifying each triple, the entire modulo-two sum becomes \emph{2-simple}: every summand is independent of at least one of the new variables.  This finite-counting argument is a specialization of the clustering philosophy developed for modular circuits by Grolmusz and Tardos~\cite{GT2000}.

\subsection{Step 7: the full coefficient of MOD~$q$ forbids 2-simplicity}
Let $N=3M$ and identify the three coordinates in each triple.  A Hamming weight $t$ in the $M$ new variables corresponds to weight $3t$ in the original variables.  Since $q\ne3$, the element $3$ is invertible modulo $q$,
\begin{equation}
        q\mid 3t\quad\Longleftrightarrow\quad q\mid t.
\end{equation}
Hence the clustered target is again the MOD~$q$ bit on $M$ variables.

When $M=2^d$, the coefficient of the full monomial $z_1\cdots z_M$ in the unique multilinear $\F_2$ representation of this target is nonzero.  By contrast, any function that omits one variable has zero full coefficient, and therefore every 2-simple sum has zero full coefficient.  This contradiction proves that the number of nonconstant factors is at least exponential in $N$.

\subsection{Step 8: charge occurrence cost and pass to arbitrary lengths}
Every nonconstant listed factor has positive integer occurrence cost.  Thus
\begin{equation}
        S\le C.
\end{equation}
The special-length lower bound therefore gives $C\ge\gamma_R^N$.  For an arbitrary sufficiently large $n$, choose a special length $N=3\cdot2^d$ with $N\le n<2N$ and set the surplus sign coordinates to $+1$.  This restriction preserves the MOD~$q$ target on the remaining variables and cannot increase degree or occurrence cost.  The exponential bound at length $N$ then implies
\begin{equation}
        \gamma^n\le C^2
\end{equation}
for a constant $\gamma>1$.

\subsection{The alternative reconstruction route}
The paper also gives a second route.  Once a unit is known to depend on $R$ affine observations, its outer function on $\F_3^R$ can be reconstructed by a bounded product of quadratic hyperplane tests.  Substitution produces a quadratic-unit representation with an $O(n^2)$ syntactic overhead.  One may then invoke the Grolmusz--Tardos clustering theorem~\cite{GT2000}.  This yields Theorem~\ref{thm:reconstruction}.  The direct route avoids reconstruction and is therefore quantitatively stronger.

\section{Quadratic bias and affine-product rank}
Put $\omega=e^{2\pi i/3}$ and $\chi(t)=\omega^t$.  The bias of $Q:X_n\to\F_3$ is $|\E_x\chi(Q(x))|$.  We will prove that a quadratic with nonnegligible bias is a function of boundedly many affine observations.

\subsection{A restricted bilinear estimate}
\begin{lemma}\label{lem:bilinear}
Let $A$ be an invertible $r\times r$ matrix over $\F_3$, and let $U,V:\{-1,1\}^r\to\F_3$ be arbitrary.  For independent uniform signs $x,y$,
\begin{equation}
\left|\E_{x,y}\chi\!\left(U(x)+V(y)+\sum_{i=1}^r x_i(Ay)_i\right)\right|^4\le(3/4)^r.
\end{equation}
\end{lemma}
\begin{proof}
Set
\begin{equation}
 \beta:=\E_{x,y}\chi\!\left(U(x)+V(y)+x^{\mathsf T}Ay\right).
\end{equation}
We spell out the two Cauchy--Schwarz steps because this estimate is the analytic input used later in the proof of Lemma~\ref{lem:cubic}.

First average in $y$ and regard the result as a function of $x$:
\begin{equation}
 \beta=\E_x\left[\chi(U(x))\,
 \E_y\chi\!\left(V(y)+x^{\mathsf T}Ay\right)\right].
\end{equation}
To make the first use of Cauchy--Schwarz completely explicit, define
\begin{equation}
 G(x):=\E_y\chi\!\left(V(y)+x^{\mathsf T}Ay\right).
\end{equation}
Then
\begin{equation}
 \beta=\E_x[\chi(U(x))G(x)].
\end{equation}
We apply Cauchy--Schwarz on the finite probability space $X_r=\{-1,1\}^r$ equipped with the uniform measure.  In the standard form
\begin{equation}
 |\E_x[a(x)b(x)]|^2\le \E_x|a(x)|^2\,\E_x|b(x)|^2,
\end{equation}
choose
\begin{equation}
 a(x)=\chi(U(x)),\qquad b(x)=G(x).
\end{equation}
Because every additive character has modulus one,
\begin{equation}
 |a(x)|^2=|\chi(U(x))|^2=1
\end{equation}
for every $x$.  Therefore the first factor on the right side of Cauchy--Schwarz is exactly one:
\begin{equation}
 \E_x|a(x)|^2=1.
\end{equation}
Substitution into Cauchy--Schwarz yields
\begin{equation}
 |\beta|^2\le \E_x|G(x)|^2.
\end{equation}
We next expand the right side rather than hiding this step in notation.  For fixed $x$, put
\begin{equation}
 g_x(y):=\chi\!\left(V(y)+x^{\mathsf T}Ay\right),
\end{equation}
so that
\begin{equation}
 G(x)=\E_y g_x(y).
\end{equation}
Let $z$ be an independent copy of $y$.  The elementary identity for the squared modulus of an average is
\begin{equation}
 |\E_y g_x(y)|^2
 =\E_{y,z}g_x(y)\overline{g_x(z)}.
\end{equation}
Since $\overline{\chi(q)}=\chi(-q)$ for $q\in\F_3$, we have
\begin{equation}
 g_x(y)\overline{g_x(z)}
 =\chi\!\left(V(y)-V(z)+x^{\mathsf T}A(y-z)\right).
\end{equation}
Consequently,
\begin{align}
 |\beta|^2
 &\le \E_x|G(x)|^2\\
 &=\E_{x,y,z}\chi\!\left(V(y)-V(z)+x^{\mathsf T}A(y-z)\right).
\end{align}
Thus the first Cauchy--Schwarz application has only one inequality: it replaces the average of $\chi(U(x))G(x)$ by the $L^2$ norm of $G$.  All subsequent steps in the preceding display are exact identities obtained by introducing an independent copy $z$ and expanding the squared modulus.

Now regard the last expression as an average in the pair $(y,z)$.  Applying Cauchy--Schwarz a second time eliminates the remaining phase $\chi(V(y)-V(z))$, whose modulus is one.  We obtain
\begin{align}
 |\beta|^4
 &\le \E_{y,z}\left|\E_x\chi\!\left(x^{\mathsf T}A(y-z)\right)\right|^2. \label{eq:cs-second}
\end{align}
This is the precise form of the ``two applications of Cauchy--Schwarz'' used in the argument.

For a fixed vector $u\in\F_3^r$, independence of the sign coordinates gives
\begin{equation}
 \E_x\chi(x^{\mathsf T}u)=\prod_{i=1}^r \E_{x_i\in\{-1,1\}}\chi(x_i u_i).
\end{equation}
For $a\in\F_3$ and a uniform sign $s$,
\begin{equation}
 \E_s\chi(as)=\begin{cases}1,&a=0,\\-1/2,&a\ne0.\end{cases}
\end{equation}
Indeed, if $a\ne0$, then $\chi(a)+\chi(-a)=\omega+\omega^2=-1$.  Hence
\begin{equation}
 \left|\E_x\chi(x^{\mathsf T}u)\right|^2
 =\prod_{i=1}^r h(u_i),
 \qquad h(0)=1,\quad h(1)=h(-1)=1/4.
\end{equation}
Define
\begin{equation}
 H(u):=\prod_{i=1}^r h(u_i).
\end{equation}
Substituting $u=A(y-z)$ into equation~\eqref{eq:cs-second} gives
\begin{equation}
 |\beta|^4\le \E_{y,z}H(A(y-z)).
\end{equation}

It remains to bound this expectation.  For one coordinate, $y_i-z_i$ equals $0$ with probability $1/2$ and equals each of $1,-1$ with probability $1/4$.  Therefore every vector $v\in\F_3^r$ has
\begin{equation}
 \Pr[y-z=v]\le 2^{-r}.
\end{equation}
Because $A$ is invertible, $v\mapsto Av$ is a bijection of $\F_3^r$, so every value of $A(y-z)$ also has probability at most $2^{-r}$.  Since $H\ge0$,
\begin{align}
 \E H(A(y-z))
 &\le 2^{-r}\sum_{u\in\F_3^r}H(u)\\
 &=2^{-r}\prod_{i=1}^r\bigl(h(0)+h(1)+h(-1)\bigr)\\
 &=2^{-r}(1+1/4+1/4)^r\\
 &=(3/4)^r.
\end{align}
Thus $|\beta|^4\le(3/4)^r$, as claimed.  The case $r=0$ follows from the empty-product convention.
\end{proof}

\subsection{Disjoint minors}
For a quadratic polynomial $Q$, let $C_{ij}$ be the coefficient of $x_ix_j$ for $i\ne j$ and put $C_{ii}=0$.  Thus $C$ is symmetric.  Before stating the lemma, we make precise the phrase ``agrees on $X_n$.''  If $P$ and $Q$ are polynomials, then $P$ and $Q$ agree on $X_n$ means pointwise equality of the functions that they induce on the sign cube:
\begin{equation}
 P(x)=Q(x)\qquad\text{for every }x\in X_n.
\end{equation}
This does not require $P$ and $Q$ to be identical as formal polynomials in $\F_3[x_1,\ldots,x_n]$.  For example, $x_i^2$ and $1$ are different formal polynomials, but they agree on $X_n$ because
\begin{equation}
 x_i^2=1\qquad\text{for every }x\in X_n.
\end{equation}
This distinction is exactly what permits us to absorb diagonal quadratic terms into the affine remainder at the end of the proof.

\begin{lemma}\label{lem:minor}
There are disjoint index sets $I,J$, each of size $r$, for which $D=C(I,J)$ is invertible, and there are affine forms $A_1,B_1,\ldots,A_m,B_m$, with $m\le 3r+1$, such that
\begin{equation}
 Q(x)=P(x):=\sum_{\ell=1}^{m}A_\ell(x)B_\ell(x)\qquad\text{for every }x\in X_n.
\end{equation}
Thus the second polynomial in the assertion is the explicitly constructed polynomial $P=\sum_\ell A_\ell B_\ell$; the conclusion is equality of $Q$ and $P$ as functions on $X_n$, not necessarily equality as formal polynomials.
\end{lemma}
\begin{proof}
Choose disjoint index sets
\begin{equation}
 I=\{u_1,\ldots,u_r\},\qquad J=\{v_1,\ldots,v_r\},\qquad I\cap J=\varnothing,
\end{equation}
so that
\begin{equation}
 D=C(I,J)=\bigl(C_{u_bv_a}\bigr)_{b,a=1}^{r}
\end{equation}
is invertible and $r$ is as large as possible.  The empty minor is allowed, so such a maximal choice always exists.  Put
\begin{equation}
 S=I\cup J,
\end{equation}
and, because $I$ and $J$ are disjoint,
\begin{equation}
 |S|=2r.
\end{equation}

\paragraph{Step 1: maximality determines all coefficients outside $S$.}
Fix distinct $i,j\notin S$.  Adjoin row $i$ and column $j$ to $D$.  The resulting $(r+1)\times(r+1)$ matrix is
\begin{equation}
 M_{ij}=\begin{pmatrix}D&C(I,j)\\ C(i,J)&C_{ij}\end{pmatrix},
\end{equation}
where
\begin{equation}
 C(I,j)=\begin{pmatrix}C_{u_1j}\\ \vdots\\ C_{u_rj}\end{pmatrix}
\end{equation}
and
\begin{equation}
 C(i,J)=\begin{pmatrix}C_{iv_1}&\cdots&C_{iv_r}\end{pmatrix}.
\end{equation}
Because $i,j\notin S$, the enlarged row set $I\cup\{i\}$ and enlarged column set $J\cup\{j\}$ are still disjoint.  If $M_{ij}$ were invertible, it would be a larger disjoint invertible minor, contradicting maximality.  Hence
\begin{equation}
 \det M_{ij}=0.
\end{equation}

We now spell out the Schur-complement determinant identity.  For any invertible square matrix $D$, column vector $u$, row vector $v^{\mathsf T}$, and scalar $a$,
\begin{equation}
 \det\begin{pmatrix}D&u\\ v^{\mathsf T}&a\end{pmatrix}
 =\det(D)\bigl(a-v^{\mathsf T}D^{-1}u\bigr).
\end{equation}
To see this directly, multiply on the left by the block lower-triangular matrix
\begin{equation}
 L=\begin{pmatrix}I&0\\ -v^{\mathsf T}D^{-1}&1\end{pmatrix},
\end{equation}
which has determinant $1$.  Then
\begin{equation}
 L\begin{pmatrix}D&u\\ v^{\mathsf T}&a\end{pmatrix}
 =\begin{pmatrix}D&u\\ 0&a-v^{\mathsf T}D^{-1}u\end{pmatrix}.
\end{equation}
Taking determinants gives the displayed Schur-complement identity.

Apply it to $M_{ij}$ with
\begin{equation}
 u=C(I,j),\qquad v^{\mathsf T}=C(i,J),\qquad a=C_{ij}.
\end{equation}
We obtain
\begin{equation}
 0=\det(D)\left(C_{ij}-C(i,J)D^{-1}C(I,j)\right).
\end{equation}
Since $D$ is invertible,
\begin{equation}
 \det(D)\ne0,
\end{equation}
and therefore
\begin{equation}
 C_{ij}=C(i,J)D^{-1}C(I,j).
\end{equation}
Writing this scalar identity with indices gives
\begin{equation}
 C_{ij}=\sum_{a=1}^{r}\sum_{b=1}^{r}C_{iv_a}(D^{-1})_{ab}C_{u_bj}.
\end{equation}
This is the key consequence of maximality: every mixed coefficient whose two indices lie outside $S$ factors through the fixed minor $D$.

\paragraph{Step 2: $r$ products reproduce all quadratic terms completely outside $S$.}
For $b=1,\ldots,r$, define linear forms
\begin{equation}
 A_b(x)=\sum_i\sum_{a=1}^{r}C_{iv_a}(D^{-1})_{ab}x_i
\end{equation}
and
\begin{equation}
 B_b(x)=\sum_j C_{u_bj}x_j.
\end{equation}
Set
\begin{equation}
 T(x)=2\sum_{b=1}^{r}A_b(x)B_b(x).
\end{equation}
Thus $T$ is a sum of exactly $r$ products of affine forms.

Fix distinct $i,j\notin S$.  The contribution to the coefficient of $x_ix_j$ obtained by taking $x_i$ from $A_b$ and $x_j$ from $B_b$ is
\begin{equation}
 2\sum_{a,b}C_{iv_a}(D^{-1})_{ab}C_{u_bj}=2C_{ij}.
\end{equation}
The reversed choice, taking $x_j$ from $A_b$ and $x_i$ from $B_b$, contributes
\begin{equation}
 2\sum_{a,b}C_{jv_a}(D^{-1})_{ab}C_{u_bi}=2C_{ji}.
\end{equation}
Because $C$ is symmetric,
\begin{equation}
 C_{ji}=C_{ij}.
\end{equation}
Hence the total coefficient of $x_ix_j$ in $T$ is
\begin{equation}
 2C_{ij}+2C_{ji}=4C_{ij}=C_{ij}\qquad\text{in }\F_3,
\end{equation}
since $4\equiv1\pmod3$.  Thus $T$ has exactly the same off-diagonal quadratic coefficient as $Q$ whenever both indices lie outside $S$.

Define
\begin{equation}
 R(x)=Q(x)-T(x).
\end{equation}
If $R_{ij}$ denotes the coefficient of $x_ix_j$ in $R$, then
\begin{equation}
 R_{ij}=0\qquad\text{whenever }i\ne j\text{ and }i,j\notin S.
\end{equation}
Therefore every remaining off-diagonal quadratic monomial in $R$ has at least one endpoint in $S$.

\paragraph{Step 3: $2r$ further products reproduce every remaining off-diagonal term.}
Define
\begin{equation}
 U(x)=\sum_{i\in S}x_i\left(\sum_{j\notin S}R_{ij}x_j+2\sum_{j\in S}R_{ij}x_j\right).
\end{equation}
There is one product for each $i\in S$, so $U$ is a sum of
\begin{equation}
 |S|=2r
\end{equation}
products of affine forms.

If $i\in S$ and $j\notin S$, the monomial $x_ix_j$ occurs exactly once in $U$, with coefficient
\begin{equation}
 R_{ij}.
\end{equation}
If instead $i,j\in S$ are distinct, the summand indexed by $i$ contributes $2R_{ij}$ and the summand indexed by $j$ contributes $2R_{ji}$.  Since $R$ is symmetric, the total is
\begin{equation}
 2R_{ij}+2R_{ji}=4R_{ij}=R_{ij}\qquad\text{in }\F_3.
\end{equation}
Thus $U$ has exactly the same off-diagonal quadratic coefficients as $R$.

\paragraph{Step 4: the remainder is affine as a function on $X_n$.}
The polynomial
\begin{equation}
 W(x)=R(x)-U(x)
\end{equation}
has no off-diagonal quadratic terms.  Consequently it has the form
\begin{equation}
 W(x)=c'+\sum_i d_i'x_i+\sum_i e_i x_i^2.
\end{equation}
For every $x\in X_n$, $x_i^2=1$, so
\begin{equation}
 W(x)=\left(c'+\sum_i e_i\right)+\sum_i d_i'x_i\qquad(x\in X_n).
\end{equation}
Define the affine form
\begin{equation}
 L(x)=\left(c'+\sum_i e_i\right)+\sum_i d_i'x_i.
\end{equation}
Then
\begin{equation}
 W(x)=L(x)\qquad\text{for every }x\in X_n.
\end{equation}
Equivalently,
\begin{equation}
 Q(x)=T(x)+U(x)+L(x)\qquad\text{for every }x\in X_n.
\end{equation}
Finally write the affine remainder itself as one product,
\begin{equation}
 L(x)=1\cdot L(x).
\end{equation}
The right-hand side therefore contains at most
\begin{equation}
 r+2r+1=3r+1
\end{equation}
products of affine forms.  Taking this right-hand side to be the polynomial $P(x)$ from the statement gives
\begin{equation}
 Q(x)=P(x)\qquad\text{for every }x\in X_n,
\end{equation}
which is the asserted pointwise agreement on the sign cube.
\end{proof}

\begin{proposition}[Restricted quadratic bias--rank]\label{prop:biasrank}
For every $\epsilon>0$ there is an integer $B(\epsilon)$ such that every degree-at-most-two polynomial $Q$ satisfying
\begin{equation}
 \left|\E_{x\in X_n}\omega^{Q(x)}\right|\ge\epsilon
\end{equation}
agrees on $X_n$ with a sum of at most $\max(1,B(\epsilon))$ products of affine forms.  Consequently $Q$ depends on at most $2\max(1,B(\epsilon))$ affine observations.
\end{proposition}
\begin{proof}
Let $C$ be the symmetric off-diagonal coefficient matrix of $Q$, and let $C(I,J)$ be any invertible disjoint minor of size $r$.  Write the corresponding variables as $x_I$ and $x_J$, and let $w$ denote all remaining signs.  For each fixed value of $w$, the restricted quadratic has the form
\begin{equation}
 Q_w(x_I,x_J)=U_w(x_I)+V_w(x_J)+x_I^{\mathsf T}A x_J,
\end{equation}
where $A=C(I,J)$ is still invertible.  Terms involving only $x_I$, only $x_J$, or the fixed variables are absorbed into the arbitrary functions $U_w,V_w$.

Lemma~\ref{lem:bilinear} therefore gives, for every $w$,
\begin{equation}
 \left|\E_{x_I,x_J}\omega^{Q_w(x_I,x_J)}\right|^4\le(3/4)^r.
\end{equation}
Put
\begin{equation}
 b(w):=\E_{x_I,x_J}\omega^{Q_w(x_I,x_J)}.
\end{equation}
The global bias is $|\E_w b(w)|$.  By Jensen's inequality for the convex function $z\mapsto |z|^4$ (or by repeated Cauchy--Schwarz),
\begin{equation}
 \left|\E_w b(w)\right|^4\le \E_w|b(w)|^4\le(3/4)^r.
\end{equation}
Thus a global bias at least $\epsilon$ forces
\begin{equation}
 \epsilon^4\le(3/4)^r.
\end{equation}
Equivalently,
\begin{equation}
 r\le \frac{4\log(1/\epsilon)}{\log(4/3)}.
\end{equation}
We do not need the sharp integer rounding.  Choose any integer $R(\epsilon)$ satisfying
\begin{equation}
 (3/4)^{R(\epsilon)}<\epsilon^4.
\end{equation}
Then an invertible disjoint minor can have size only $r<R(\epsilon)$.  In particular a maximal such minor has bounded size depending only on $\epsilon$, not on $n$.

Lemma~\ref{lem:minor} now converts this matrix bound into an algebraic one: $Q$ agrees on the sign cube with a sum of at most $3r+1<3R(\epsilon)+1$ products of affine forms.  We may therefore take
\begin{equation}
 B(\epsilon)=3R(\epsilon)+1.
\end{equation}
Each product uses two affine factors, so the resulting function is determined
by at most
\[
 2\max\{1,B(\epsilon)\}
\]
affine observations. The outer maximum covers the affine and constant cases.
\end{proof}

\section{Quadratic units}
A \emph{junta} is a function determined by a specified bounded set of coordinates.

\begin{lemma}[Quadratic-unit dichotomy]\label{lem:quadunit}
Let $P$ have a quadratic representation and satisfy $P(x)\ne0$ on $X_n$.  Then at least one of the following holds:
\begin{enumerate}[label=(\arabic*),leftmargin=2em]
\item $P(x)=\varepsilon(1+L(x)^2)$ for every $x\in X_n$, for an affine form $L$ and $\varepsilon\in\{1,-1\}$;
\item $P$ is an eight-junta.
\end{enumerate}
The statement also applies to a signed raw factor each of whose monomials has odd exponent on at most two coordinates.
\end{lemma}
\begin{proof}
Write the unique square-free (multilinear) representative of $P$ on the sign cube as
\begin{equation}
 P=c+\sum_i d_ix_i+\sum_{i<j}a_{ij}x_ix_j,
\end{equation}
where, as usual, we extend $a_{ij}$ symmetrically by $a_{ji}=a_{ij}$ and put $a_{ii}=0$.

\paragraph{Step 1: slicing in one coordinate and deriving the unit identities.}
Choose one of the original coordinates of the input vector.  After relabeling coordinates if necessary, take this distinguished coordinate to be
\begin{equation}
 s:=x_1.
\end{equation}
Thus $s$ is \emph{not} a new variable: it is the first coordinate of the original input.  Denote the remaining coordinates by
\begin{equation}
 y:=(x_2,\ldots,x_n),
\end{equation}
so that the original input vector is
\begin{equation}
 x=(s,y).
\end{equation}
Because the multilinear representative of $P$ has degree at most two, collecting the terms that contain $s$ gives
\begin{equation}
 P(s,y)=A(y)+sB(y),
\end{equation}
where $A$ has degree at most two and
\begin{equation}
 B(y)=b_0+\sum_{j=2}^n b_jx_j
\end{equation}
is affine.  Fix temporarily a point $y$ of the remaining sign cube and abbreviate
\begin{equation}
 a:=A(y),\qquad b:=B(y).
\end{equation}
Because $P$ is nowhere zero and the only nonzero elements of $\F_3$ are $1$ and $-1$, both slices are units:
\begin{equation}
 a+b=P(1,y)\in\{1,-1\},\qquad a-b=P(-1,y)\in\{1,-1\}.
\end{equation}
Every nonzero element of $\F_3$ has square $1$.  Hence
\begin{equation}
 (a+b)^2=1,\qquad (a-b)^2=1.
\end{equation}
Expanding gives
\begin{align}
 a^2+2ab+b^2&=1,\\
 a^2-2ab+b^2&=1.
\end{align}
Subtracting the second equation from the first gives $4ab=0$.  Since $4=1$ in $\F_3$, we obtain $ab=0$.  Adding the two equations gives
$2(a^2+b^2)=2$; since $2$ is invertible in $\F_3$, this gives $a^2+b^2=1$.  The point $y$ was arbitrary, so as functions on the remaining sign cube,
\begin{equation}\label{eq:unit-slice-identities}
 AB=0,\qquad A^2+B^2=1.
\end{equation}
This is the detailed derivation of the two identities used below.  Notice that they are, at this stage, \emph{functional identities on the sign cube}.  Whenever we compare coefficients, we first pass to the unique multilinear representative modulo $x_i^2-1$.

\paragraph{Step 2: the cubic coefficients of $AB$.}
Let $j,k,l$ be three distinct remaining coordinates.  In the multilinear reduction of $AB$, a term $x_jx_kx_l$ can only be produced by multiplying one of the three quadratic terms
$a_{jk}x_jx_k$, $a_{jl}x_jx_l$, $a_{kl}x_kx_l$ in $A$ by the complementary linear term in $B$.  Therefore the coefficient of $x_jx_kx_l$ in $AB$ is
\begin{equation}
 a_{jk}b_l+a_{jl}b_k+a_{kl}b_j.
\end{equation}
Since $AB$ is the zero function and the multilinear representative of the zero function has all coefficients zero, we obtain
\begin{equation}\label{eq:triangle}
 a_{jk}b_l+a_{jl}b_k+a_{kl}b_j=0
\end{equation}
for every triple of distinct indices.  This coefficient comparison is legitimate because multilinear functions on $\{-1,1\}^m$ have unique multilinear representatives over $\F_3$: equivalently, their coefficients can be recovered from alternating sums over coordinate slices, and $2$ is invertible in $\F_3$.

\paragraph{Step 3: a coordinate with at least five nonzero $b_j$'s forces rank one.}
Assume that the distinguished coordinate $s$ has
\begin{equation}
 S:=\{j:b_j\ne0\},\qquad |S|\ge5.
\end{equation}
For distinct $j,k\in S$ define
\begin{equation}
 g_{jk}:=\frac{a_{jk}}{b_jb_k}.
\end{equation}
Dividing~\eqref{eq:triangle} by $b_jb_kb_l$ for distinct $j,k,l\in S$ gives
\begin{equation}
 g_{jk}+g_{jl}+g_{kl}=0.
\end{equation}
We now justify carefully why all the edge weights $g_{jk}$ are equal.  Take four distinct support indices $i,j,k,l$.  Applying above to the four triples $ijk,ijl,ikl,jkl$ and eliminating the common terms gives
\begin{equation}
 g_{ij}=g_{kl},\qquad g_{ik}=g_{jl},\qquad g_{il}=g_{jk}.
\end{equation}
Thus disjoint edges have equal weight.  If two edges share a vertex, say $ij$ and $ik$, choose two further support vertices $l,m$; this is possible because $|S|\ge5$.  The edge $lm$ is disjoint from both $ij$ and $ik$, and hence
\begin{equation}
 g_{ij}=g_{lm}=g_{ik}.
\end{equation}
Consequently every support edge has one common value, say $t$:
\begin{equation}
 a_{jk}=t b_jb_k\qquad(j\ne k,\ j,k\in S).
\end{equation}

We next show that no quadratic coefficient can leave the support.  Let $l\notin S$ and choose distinct $j,k\in S$.  Since $b_l=0$, equation~\eqref{eq:triangle} gives
\begin{equation}
 a_{jl}b_k+a_{kl}b_j=0.
\end{equation}
Put $r_j=a_{jl}/b_j$ for $j\in S$.  Then $r_j+r_k=0$ for every distinct $j,k\in S$.  Using three support indices gives $r_j=-r_k=r_m$ and also $r_j=-r_m$, whence $2r_j=0$ and therefore $r_j=0$.  Thus
\begin{equation}
 a_{jl}=0\qquad(j\in S,\ l\notin S).
\end{equation}
Finally, if $k,l\notin S$, choose $j\in S$ in~\eqref{eq:triangle}.  The first two terms vanish and we get $a_{kl}b_j=0$, hence $a_{kl}=0$.  We have proved the global relation
\begin{equation}
 a_{jk}=t b_jb_k\qquad(j\ne k)
\end{equation}
with the convention that the right side is zero when one of the $b$'s is zero.

\paragraph{Step 4: removing the quadratic part of $A$.}
Consider
\begin{equation}
 D:=A+tB^2.
\end{equation}
For $j\ne k$, the coefficient of $x_jx_k$ in $B^2$ is $2b_jb_k$.  By the preceding relation, the corresponding coefficient in $D$ is
\begin{equation}
 tb_jb_k+2tb_jb_k=3tb_jb_k=0.
\end{equation}
The terms $b_j^2x_j^2$ become constants on the sign cube.  Hence $D$ is affine as a function on the cube.

Multiply the definition of $D$ by $B$.  From $AB=0$ in~\eqref{eq:unit-slice-identities} and from the pointwise identity $B^3=B$ over $\F_3$, we obtain
\begin{equation}
 DB=AB+tB^3=tB,
\end{equation}
so
\begin{equation}
 (D-t)B=0.
\end{equation}
We use the following elementary fact in this particular situation: if $E=e_0+\sum_j e_jx_j$ is affine, $B=b_0+\sum_jb_jx_j$ has at least three nonzero linear coefficients, and $EB=0$ on the sign cube, then $E=0$.  Indeed, for distinct $j,k\in S$, the coefficient of $x_jx_k$ in the multilinear product is
\begin{equation}
 e_jb_k+e_kb_j=0.
\end{equation}
Writing $q_j=e_j/b_j$ gives $q_j+q_k=0$ for all distinct $j,k\in S$; three support indices force every $q_j=0$.  Hence $e_j=0$ on $S$.  If $k\notin S$, comparison of the coefficient of $x_jx_k$ with any $j\in S$ gives $e_kb_j=0$, so $e_k=0$ as well.  Thus $E=e_0$ is constant, and $EB=0$ together with the fact that $B$ is not the zero function forces $e_0=0$.

Apply this fact to $E=D-t$ in the preceding identity.  We get
\begin{equation}
 D=t,
\end{equation}
and therefore
\begin{equation}
 A=t(1-B^2).
\end{equation}
Because at least two linear coefficients of $B$ are nonzero, $B$ assumes all three values of $\F_3$ as the corresponding two signs vary (with the other signs fixed appropriately); in particular there is an input with $B=0$.  At such an input the preceding identity gives $A=t$.  If $t=0$, then $A=0$ everywhere and $P=sB$ would vanish at an input with $B=0$, contradicting the hypothesis.  Hence
\begin{equation}
 t\ne0,
\end{equation}
so $t^2=1$ in $\F_3$.

Substituting the preceding identity into $P=A+sB$ and expanding in characteristic three gives
\begin{equation}
 P=t(1-B^2)+sB=-t\bigl(1+(B+ts)^2\bigr).
\end{equation}
Indeed, $(B+ts)^2=B^2+2tsB+t^2s^2=B^2-t sB+1$, and multiplication by $-t$ gives $t-tB^2+sB$.  Thus alternative~(1) holds with
\begin{equation}
 \varepsilon=-t,\qquad L=B+ts.
\end{equation}

\paragraph{Step 5: the complementary case and the four-index identity.}
The preceding argument applies whenever, for some choice of distinguished coordinate $s=x_i$, the associated affine coefficient $B$ has at least five nonzero linear coefficients.  In the original quadratic expansion, those coefficients are exactly $a_{ij}$ with $j\ne i$.  Therefore, if alternative~(1) has not already occurred, every vertex in the quadratic coefficient graph has degree at most four.

We also record the identity used to control that graph.  Take four distinct indices $i,j,k,l$.  Distinguish $s=x_i$.  Then the coefficient of $x_j$ in the corresponding $B$ is $b_j=a_{ij}$.  Applying~\eqref{eq:triangle} to $j,k,l$ gives
\begin{equation}
 a_{jk}a_{il}+a_{jl}a_{ik}+a_{kl}a_{ij}=0,
\end{equation}
which, after reordering the three summands, is
\begin{equation}
 a_{ij}a_{kl}+a_{ik}a_{jl}+a_{il}a_{jk}=0.
\end{equation}

Let $G$ be the graph on the coordinates whose edges are the pairs $ij$ with $a_{ij}\ne0$.  Fix an edge $uv$.  We claim that every nonisolated vertex belongs to $N(u)\cup N(v)$.  Suppose instead that a nonisolated vertex $w$ is adjacent to neither $u$ nor $v$.  Choose a neighbor $z$ of $w$.  Since $w$ is adjacent to neither endpoint of $uv$, we have $z\ne u,v,w$.  Apply the four-index identity to $u,v,w,z$.  The two cross terms contain $a_{uw}$ or $a_{vw}$ and vanish, leaving
\begin{equation}
 a_{uv}a_{wz}=0,
\end{equation}
contrary to the choice of the two edges.  Thus the claim holds.  If the maximum degree of $G$ is at most four, the set of nonisolated vertices is contained in $N(u)\cup N(v)$ and consequently has size at most
\begin{equation}
 |N(u)|+|N(v)|\le8.
\end{equation}
So a quadratic unit with no linear terms depends on at most eight coordinates unless alternative~(1) holds.

\paragraph{Step 6: absorbing the linear terms.}
For a general quadratic unit $P$, introduce one fresh sign $z$ and define
\begin{equation}
 \widetilde P(z,x):=P(zx_1,\ldots,zx_n).
\end{equation}
On the sign cube, a linear term $d_ix_i$ becomes $d_i z x_i$, while a quadratic term satisfies
\begin{equation}
 a_{ij}(zx_i)(zx_j)=a_{ij}z^2x_ix_j=a_{ij}x_ix_j.
\end{equation}
Thus $\widetilde P$ is a nonvanishing quadratic unit with no linear terms in the enlarged set of variables $(z,x_1,\ldots,x_n)$; the old linear coefficients appear as quadratic edges incident with $z$.

Apply the preceding graph argument to $\widetilde P$.  If some vertex has degree at least five, Step~3 gives alternative~(1) for $\widetilde P$; setting $z=1$ leaves an affine-square representation of the same form for $P$.  Otherwise the coefficient graph of $\widetilde P$ has at most eight nonisolated vertices.  Every variable on which $\widetilde P$ actually depends is among those vertices (apart from the constant term), so $\widetilde P$ is an eight-junta.  Restricting $z=1$ cannot introduce dependence on a new coordinate, and hence $P$ is also an eight-junta.  This proves alternative~(2).

Finally, for the stated raw-factor variant, reduce every exponent modulo two on the sign cube.  By hypothesis each resulting monomial is supported on at most two coordinates, so the resulting function has a quadratic multilinear representation.  The dichotomy just proved therefore applies.
\end{proof}

\section{Cubic units have bounded affine dependence}
This is the structural heart of the argument.  We give the proof in a form that uses only explicitly defined finite sets and finite averages.
\begin{lemma}\label{lem:cubic}
There is an absolute constant $R$ such that every cubic unit on every $X_n$ depends on at most $R$ affine observations.
\end{lemma}
\begin{proof}
We first treat the case in which the multilinear representative of the unit has a nonzero cubic coefficient.

\paragraph{Step 1: isolate one nonzero cubic coefficient.}
After permuting variables, assume that the coefficient of $s_1s_2s_3$ is a nonzero element $c\in\F_3^\times$.  Write the remaining coordinates as
\begin{equation}
 x=(x_1,\ldots,x_m)\in X_m,\qquad m=n-3,
\end{equation}
so the full input is $(s_1,s_2,s_3,x)$.  Collecting terms according to their degree in $s_1,s_2,s_3$ gives
\begin{equation}
 f(s,x)=T(x)+\sum_{i=1}^3s_iQ_i(x)
 +\sum_{i=1}^3\left(\prod_{j\ne i}s_j\right)L_i(x)+cs_1s_2s_3,
\end{equation}
where
\begin{equation}
 \deg T\le3,\qquad \deg Q_i\le2,\qquad \deg L_i\le1.
\end{equation}
In fact the identity $f^2=1$ will force $T$ to be a combination of the $Q_i$ with affine coefficients, and hence no independent cubic information remains in $T$.

Because $f$ is a unit, for every point of the sign cube
\begin{equation}
 f(s,x)^2=1.
\end{equation}
After multilinear reduction modulo $s_i^2-1$ and $x_j^2-1$, the coefficient of every nonconstant monomial in $f^2$ is therefore zero.  Consider the coefficient of $s_1s_2s_3$.  The terms contributing to it are exactly
\begin{equation}
 2cT+2\sum_{i=1}^3L_iQ_i.
\end{equation}
Since $2\ne0$ in $\F_3$, this coefficient vanishes if and only if
\begin{equation}\label{eq:cubic-T-relation}
 cT+\sum_{i=1}^3L_iQ_i=0.
\end{equation}
As $c\ne0$, we may solve for $T$:
\begin{equation}
 T=-c^{-1}\sum_{i=1}^3L_iQ_i.
\end{equation}
Substituting into the expansion of $f$ gives
\begin{equation}\label{eq:cubicidentity}
 f(s,x)=\sum_{i=1}^3\bigl(s_i-c^{-1}L_i(x)\bigr)Q_i(x)+B_0(s,L(x)),
\end{equation}
where, writing $L(x)=(L_1(x),L_2(x),L_3(x))$,
\begin{equation}
 B_0(s,l):=\sum_{i=1}^3\left(\prod_{j\ne i}s_j\right)l_i+cs_1s_2s_3.
\end{equation}
Equation~\eqref{eq:cubicidentity} holds for every $s\in\{-1,1\}^3$ and every $x\in X_m$.

\paragraph{Step 2: induction invariant for the remaining quadratic information.}
Put
\begin{equation}
 Q(x):=(Q_1(x),Q_2(x),Q_3(x))^{\mathsf T}\in\F_3^3.
\end{equation}
We maintain an identity
\begin{equation}\label{eq:Q-invariant}
 Q(x)=M U(x)+H(A(x)).
\end{equation}
Here $A=(A_1,\ldots,A_r)$ is a list of affine forms on $X_m$ which contains $L_1,L_2,L_3$ among its entries,
\begin{equation}
 A:X_m\longrightarrow\F_3^r,
\end{equation}
$U=(U_1,\ldots,U_k)^{\mathsf T}$ is a vector of quadratic polynomials,
\begin{equation}
 M:\F_3^k\longrightarrow\F_3^3
\end{equation}
is an injective linear map, and $H:\F_3^r\to\F_3^3$ is an arbitrary function on the values attained by $A$.  Initially we take
\begin{equation}
 r=3,\qquad A=L,\qquad k=3,\qquad U=Q,\qquad M=I_3,\qquad H=0.
\end{equation}
Thus the invariant holds before any elimination.

Assume now that $k>0$.  We will enlarge the list $A$ by a bounded number of affine forms and replace~\eqref{eq:Q-invariant} by another identity of the same type with $k$ replaced by $k-1$.

\paragraph{Step 3: choose a large preimage set of the observation map.}
For each $a\in\F_3^r$, define explicitly
\begin{equation}\label{eq:Sa-def}
 S_a:=\{x\in X_m:A(x)=a\}.
\end{equation}
The sets $S_a$ are pairwise disjoint and their union is $X_m$.  At most $3^r$ of them are nonempty.  Since
\begin{equation}
 |X_m|=2^m=\sum_{a\in\F_3^r}|S_a|,
\end{equation}
there is at least one value $a\in\F_3^r$ satisfying
\begin{equation}\label{eq:Sa-size}
 |S_a|\ge \frac{2^m}{3^r}.
\end{equation}
Fix such an $a$.  Define its relative size in the whole sign cube by
\begin{equation}\label{eq:rho-def}
 \rho_a:=\frac{|S_a|}{2^m}.
\end{equation}
Then~\eqref{eq:Sa-size} is exactly
\begin{equation}\label{eq:rho-lower}
 \rho_a\ge3^{-r}.
\end{equation}
Because the three forms $L_i$ occur among the entries of $A$, their values are constant on $S_a$.  Write these constants as
\begin{equation}
 l_i:=L_i(x)\quad(x\in S_a),\qquad l=(l_1,l_2,l_3).
\end{equation}

\paragraph{Step 4: select a nonzero quadratic linear combination.}
For each $s=(s_1,s_2,s_3)\in\{-1,1\}^3$, define the row vector
\begin{equation}\label{eq:v-s-def}
 v_s:=\bigl(s_1-c^{-1}l_1,\ s_2-c^{-1}l_2,\ s_3-c^{-1}l_3\bigr)\in\F_3^3.
\end{equation}
We claim that the eight vectors $v_s$ span $\F_3^3$.  Indeed, if $s$ and $s'$ differ only in coordinate $i$, then
\begin{equation}
 v_s-v_{s'}=(s_i-s_i')e_i.
\end{equation}
Here $s_i-s_i'\in\{1,-1\}\subset\F_3^\times$, because changing a sign from $1$ to $-1$ changes it by $2=-1$ in $\F_3$.  Hence each standard basis vector $e_i$ lies in the span of differences of the vectors $v_s$, and therefore
\begin{equation}
 \operatorname{span}\{v_s:s\in\{-1,1\}^3\}=\F_3^3.
\end{equation}
We now make the linear-algebra notation completely explicit.  Regard $M$ as its $3\times k$ matrix and write its columns as
\begin{equation}\label{eq:M-columns}
 M=\begin{pmatrix} |&&|\\ M_1&\cdots&M_k\\ |&&|\end{pmatrix},
 \qquad M_j\in\F_3^3.
\end{equation}
Thus, for the column vector $U(x)=(U_1(x),\ldots,U_k(x))^{\mathsf T}$,
\begin{equation}\label{eq:MU-columns}
 MU(x)=\sum_{j=1}^k M_jU_j(x)\in\F_3^3.
\end{equation}
Because $M:\F_3^k\to\F_3^3$ is injective and $k>0$, at least one column $M_j$ is nonzero.  Since the vectors $v_s$ span $\F_3^3$, it cannot happen that $v_s\cdot M_j=0$ for every $s$ and every $j$; otherwise every vector of $\F_3^3$ would be orthogonal to every column of $M$, forcing every $M_j=0$.  Therefore there exists $s^*\in\{-1,1\}^3$ such that at least one of the scalars
\begin{equation}\label{eq:alpha-components}
 \alpha_j:=v_{s^*}\cdot M_j\in\F_3,
 \qquad 1\le j\le k,
\end{equation}
is nonzero.  Define the row vector
\begin{equation}\label{eq:alpha-nonzero}
 \alpha:=(\alpha_1,\ldots,\alpha_k)\in\F_3^k.
\end{equation}
Then $\alpha\ne0$.  In conventional matrix notation this row vector is exactly $v_{s^*}M$, but we will use the componentwise definition~\eqref{eq:alpha-components} to avoid confusing matrix multiplication with composition of maps.

For $x\in S_a$, the invariant~\eqref{eq:Q-invariant} gives
\begin{equation}
 Q(x)=MU(x)+H(a).
\end{equation}
Substitute this into~\eqref{eq:cubicidentity} and fix $s=s^*$.  Since $L(x)=l$ on $S_a$, we obtain
\begin{align}\label{eq:f-on-Sa}
 f(s^*,x)
 &=v_{s^*}\cdot(MU(x))+v_{s^*}\cdot H(a)+B_0(s^*,l)\\
 &=\sum_{j=1}^k (v_{s^*}\cdot M_j)U_j(x)
   +v_{s^*}\cdot H(a)+B_0(s^*,l)\\
 &=\sum_{j=1}^k\alpha_jU_j(x)
   +v_{s^*}\cdot H(a)+B_0(s^*,l),
 \qquad x\in S_a.
\end{align}
Define the scalar quadratic polynomial
\begin{equation}\label{eq:D-def}
 D(x):=\sum_{j=1}^k\alpha_jU_j(x)=\alpha\cdot U(x).
\end{equation}
Thus $D$ is one nonzero linear combination, at the coefficient-vector level, of the remaining quadratic coordinates $U_1,\ldots,U_k$.  The equality above is the precise meaning of the abbreviated expression $v_{s^*}MU(x)$ used in the preceding version.  Define also the constant
\begin{equation}\label{eq:beta-def}
 \beta:=v_{s^*}H(a)+B_0(s^*,l)\in\F_3.
\end{equation}
Then~\eqref{eq:f-on-Sa} becomes the particularly simple identity
\begin{equation}\label{eq:f-D-beta}
 f(s^*,x)=D(x)+\beta\qquad(x\in S_a).
\end{equation}
Since $f$ is a unit, its value is never $0$.  Therefore~\eqref{eq:f-D-beta} implies
\begin{equation}\label{eq:D-omits}
 D(x)\ne-\beta\qquad\text{for every }x\in S_a.
\end{equation}
Thus the set of values attained by $D$ on $S_a$ has cardinality at most two:
\begin{equation}
 |D(S_a)|\le2.
\end{equation}
This is the precise consequence of the unit condition that we need.

\paragraph{Step 5: a two-valued function has a large nontrivial character average.}
For $u\in\F_3$ define
\begin{equation}\label{eq:mu-def}
 \mu_u:=\frac{|\{x\in S_a:D(x)=u\}|}{|S_a|}.
\end{equation}
Then
\begin{equation}
 \mu_0+\mu_1+\mu_{-1}=1,
\end{equation}
and by~\eqref{eq:D-omits}, at least one of the three numbers $\mu_u$ equals $0$.  For $t\in\F_3$, define the finite character average
\begin{equation}\label{eq:Ct-def}
 C_t:=\frac1{|S_a|}\sum_{x\in S_a}\omega^{tD(x)}
      =\sum_{u\in\F_3}\mu_u\omega^{tu}.
\end{equation}
We claim that for at least one $t\in\F_3^\times=\{1,-1\}$,
\begin{equation}\label{eq:Ct-half}
 |C_t|\ge\frac12.
\end{equation}
To verify this directly, suppose for example that $\mu_{-1}=0$; the other cases are identical after relabeling the field elements.  Put $p:=\mu_0$, so $\mu_1=1-p$.  Then
\begin{equation}
 C_1=p+(1-p)\omega.
\end{equation}
Using $\omega+\overline\omega=-1$ and $|\omega|=1$,
\begin{align}
 |C_1|^2
 &=p^2+(1-p)^2-p(1-p)\\
 &=3\left(p-\frac12\right)^2+\frac14\\
 &\ge\frac14.
\end{align}
Hence $|C_1|\ge1/2$.  This proves~\eqref{eq:Ct-half}.  Fix a nonzero $t$ for which~\eqref{eq:Ct-half} holds.

The average over $S_a$ can be written as an average over the whole sign cube by inserting the indicator of the set $S_a$:
\begin{equation}\label{eq:indicator-average}
 \frac1{2^m}\sum_{x\in X_m}1_{S_a}(x)\omega^{tD(x)}
 =\rho_a C_t.
\end{equation}
Taking absolute values and using~\eqref{eq:Ct-half} and~\eqref{eq:rho-lower} gives
\begin{equation}\label{eq:restricted-character-lower}
 \left|\frac1{2^m}\sum_{x\in X_m}1_{S_a}(x)\omega^{tD(x)}\right|
 \ge\frac{\rho_a}{2}
 \ge\frac1{2\cdot3^r}.
\end{equation}

\paragraph{Step 6: remove the condition $A(x)=a$ by character orthogonality.}
For $y\in\F_3$, the elementary identity
\begin{equation}
 1_{\{y=0\}}=\frac13\sum_{\lambda\in\F_3}\omega^{\lambda y}
\end{equation}
holds.  Applying it independently to the $r$ coordinates of $A(x)-a$ yields, for every $x\in X_m$,
\begin{equation}\label{eq:indicator-orthogonality}
 1_{S_a}(x)=1_{\{A(x)=a\}}
 =3^{-r}\sum_{\lambda\in\F_3^r}\omega^{\lambda\cdot(A(x)-a)}.
\end{equation}
Substitute~\eqref{eq:indicator-orthogonality} into the left side of~\eqref{eq:restricted-character-lower}.  We obtain
\begin{align}
 &\frac1{2^m}\sum_{x\in X_m}1_{S_a}(x)\omega^{tD(x)}\\
 &\qquad=3^{-r}\sum_{\lambda\in\F_3^r}
 \omega^{-\lambda\cdot a}
 \left(\frac1{2^m}\sum_{x\in X_m}
 \omega^{tD(x)+\lambda\cdot A(x)}\right).
\end{align}
By the triangle inequality,
\begin{align}
 &\left|\frac1{2^m}\sum_{x\in X_m}1_{S_a}(x)\omega^{tD(x)}\right|\\
 &\qquad\le
 \max_{\lambda\in\F_3^r}
 \left|\frac1{2^m}\sum_{x\in X_m}
 \omega^{tD(x)+\lambda\cdot A(x)}\right|.
\end{align}
Combining this inequality with~\eqref{eq:restricted-character-lower}, there exists some $\lambda\in\F_3^r$ such that
\begin{equation}\label{eq:global-character-lower}
 \left|\frac1{2^m}\sum_{x\in X_m}
 \omega^{tD(x)+\lambda\cdot A(x)}\right|
 \ge\frac1{2\cdot3^r}.
\end{equation}
Notice that every average in~\eqref{eq:global-character-lower} is now over the entire sign cube $X_m$; no restriction to $S_a$ remains.

\paragraph{Step 7: apply the quadratic character-sum proposition globally.}
Define
\begin{equation}\label{eq:P-def-L11}
 P(x):=tD(x)+\lambda\cdot A(x).
\end{equation}
Because $D$ is quadratic and $A_1,\ldots,A_r$ are affine, $P$ has degree at most two.  Equation~\eqref{eq:global-character-lower} states exactly that
\begin{equation}\label{eq:P-char-lower}
 \left|\frac1{2^m}\sum_{x\in X_m}\omega^{P(x)}\right|
 \ge\epsilon_r,
 \qquad
 \epsilon_r:=\frac1{2\cdot3^r}.
\end{equation}
Proposition~\ref{prop:biasrank} therefore implies that $P$ is determined by at most
\begin{equation}\label{eq:new-observation-count}
 q_r:=2\max\{1,B(\epsilon_r)\}
\end{equation}
affine observations.  Thus there exist affine forms $C_1,\ldots,C_{q_r}$ and a function $G$ such that
\begin{equation}
 P(x)=G(C_1(x),\ldots,C_{q_r}(x))
 \qquad(x\in X_m).
\end{equation}
Since $\lambda\cdot A(x)$ is already determined by the old observations $A_1(x),\ldots,A_r(x)$ and $t\ne0$, equation~\eqref{eq:P-def-L11} gives
\begin{equation}\label{eq:D-observed}
 D(x)=t^{-1}\bigl(P(x)-\lambda\cdot A(x)\bigr).
\end{equation}
Consequently $D$ is determined by the enlarged list of affine observations
\begin{equation}
 A_1,\ldots,A_r,C_1,\ldots,C_{q_r}.
\end{equation}
The important point is that the number $q_r$ depends only on $r$, not on $m$ or $n$.

\paragraph{Step 8: eliminate one quadratic coordinate from $U$.}
Write
\begin{equation}
 \alpha=(\alpha_1,\ldots,\alpha_k).
\end{equation}
Because $\alpha\ne0$, choose $j$ with $\alpha_j\ne0$.  From $D=\sum_i\alpha_iU_i$,
\begin{equation}\label{eq:Uj-eliminate}
 U_j=\alpha_j^{-1}\left(D-\sum_{i\ne j}\alpha_iU_i\right).
\end{equation}
Let $M_i\in\F_3^3$ denote the $i$th column of $M$.  Substituting~\eqref{eq:Uj-eliminate} into $MU=\sum_iM_iU_i$ yields
\begin{equation}\label{eq:MU-eliminate}
 MU=M_j\alpha_j^{-1}D+
 \sum_{i\ne j}\left(M_i-\frac{\alpha_i}{\alpha_j}M_j\right)U_i.
\end{equation}
The first term on the right of~\eqref{eq:MU-eliminate} is determined by the enlarged affine-observation list and can therefore be absorbed into the function $H$.  For $i\ne j$, define
\begin{equation}\label{eq:Mprime-def}
 M_i':=M_i-\frac{\alpha_i}{\alpha_j}M_j.
\end{equation}
We verify that the $k-1$ columns $M_i'$ are linearly independent.  Suppose
\begin{equation}
 \sum_{i\ne j}c_iM_i'=0.
\end{equation}
Using~\eqref{eq:Mprime-def}, this becomes
\begin{equation}
 \sum_{i\ne j}c_iM_i-
 \left(\sum_{i\ne j}c_i\frac{\alpha_i}{\alpha_j}\right)M_j=0.
\end{equation}
The original columns $M_1,\ldots,M_k$ are linearly independent because $M$ is injective.  Therefore every coefficient in the last relation is zero, and in particular
\begin{equation}
 c_i=0\qquad(i\ne j).
\end{equation}
Hence the new linear map with columns $M_i'$ is injective.  We have recovered an identity of the form~\eqref{eq:Q-invariant}, but with $k$ replaced by $k-1$.

\paragraph{Step 9: iterate at most three times.}
Initially $k=3$ and the list $A$ has $r_0=3$ entries.  If the observation list has $r_j$ entries at the $j$th elimination, Step~7 adds at most
\begin{equation}
 2\max\left\{1,B\left((2\cdot3^{r_j})^{-1}\right)\right\}
\end{equation}
new affine forms.  Define recursively
\begin{equation}\label{eq:r-recursion}
 r_{j+1}=r_j+2\max\left\{1,B\left((2\cdot3^{r_j})^{-1}\right)\right\},
 \qquad 0\le j<3.
\end{equation}
After at most three applications of the elimination step, $k=0$.  Equation~\eqref{eq:Q-invariant} then says that the entire vector $Q(x)$ is determined by at most $r_3$ affine observations of $x$.  Equation~\eqref{eq:cubicidentity} shows that $f(s,x)$ is consequently determined by those observations together with the three coordinates $s_1,s_2,s_3$.  Since each coordinate function $s_i$ is itself an affine observation, $f$ depends on at most
\begin{equation}
 3+r_3
\end{equation}
affine observations.

\paragraph{Step 10: the case with no cubic term.}
If the cubic part of the multilinear representative of $f$ is zero, then $f$ is a quadratic unit.  Lemma~\ref{lem:quadunit} says that either $f$ is a function of one affine form or $f$ depends on at most eight original coordinates.  Each original coordinate is an affine observation.  Therefore this case requires at most eight affine observations.

Combining the two cases, we may take
\begin{equation}\label{eq:R-final}
 R:=\max\{8,3+r_3\}.
\end{equation}
The recursion~\eqref{eq:r-recursion} involves only absolute constants and is applied only three times.  Thus $R$ is an absolute constant independent of $n$.
\end{proof}

\begin{remark}
The numerical value of $R$ obtained from this proof is not intended to be useful.  Only its independence of $n$ matters.  This distinction is important: the later clustering argument loses a constant in its exponential base as $R$ grows, but any fixed $R$ still yields a base strictly larger than one.
\end{remark}

\section{Quadratic reconstruction: an alternative route}
The next lemma is not needed for the strongest direct theorem, but it gives a useful second proof and clarifies why bounded affine dependence can be converted back into low-degree algebra.

\paragraph{Definition ($R$-observed function).}\label{def:R-observed}
Let $R\ge 0$.  A function $f:X_n\to\F_3$ is called \emph{$R$-observed} if there exist affine forms
\begin{equation}\label{eq:Robs-A}
 A_1,\ldots,A_R:X_n\to\F_3
\end{equation}
and a function
\begin{equation}\label{eq:Robs-h}
 h:\F_3^R\to\F_3
\end{equation}
such that
\begin{equation}\label{eq:Robs-representation}
 f(x)=h\bigl(A_1(x),\ldots,A_R(x)\bigr)
 \qquad\text{for every }x\in X_n.
\end{equation}
Equivalently, if
\begin{equation}\label{eq:Robs-map}
 A(x):=\bigl(A_1(x),\ldots,A_R(x)\bigr)\in\F_3^R,
\end{equation}
then $f=h\circ A$ on $X_n$.  If $f$ is a unit, then $f(x)\in\{1,-1\}$ for every $x\in X_n$.  In that case we call $f$ an \emph{$R$-observed unit}.  A function depending on fewer than $R$ affine observations is also regarded as $R$-observed by adjoining zero observations.

Thus ``$R$-observed'' means that all information about $x$ used by $f$ is compressed into the vector $A(x)\in\F_3^R$.  The affine forms $A_i$ may individually involve all $n$ original variables.

\begin{lemma}[Quadratic reconstruction of an $R$-observed unit]\label{lem:reconstruct}
Fix $R\ge1$.  Let $f:X_n\to\{1,-1\}$ be an $R$-observed unit.  Then $f$ can be represented on $X_n$ as a product of quadratic units with total occurrence cost at most
\begin{equation}\label{eq:K-def}
 Kn^2,
 \qquad
 K:=\frac{3(3^R-1)}2.
\end{equation}
Consequently, if a product of cubic units has original occurrence cost $C$, and each cubic unit is $R$-observed, then the same product function has a representation by quadratic units of total occurrence cost at most
\begin{equation}\label{eq:reconstruct-product-cost}
 Kn^2C.
\end{equation}
\end{lemma}

\begin{proof}
We divide the proof into five steps.

\paragraph{Step 1: represent the unit as a function on $\F_3^R$.}
By Definition~\ref{def:R-observed}, there are affine forms $A_1,\ldots,A_R$ and a function $h$ such that
\begin{equation}\label{eq:recon-f-hA}
 f(x)=h(A(x)),
 \qquad
 A(x)=\bigl(A_1(x),\ldots,A_R(x)\bigr).
\end{equation}
Let
\begin{equation}\label{eq:attained-image}
 Y:=A(X_n)\subseteq\F_3^R
\end{equation}
be the set of values of the observation vector that actually occur on the sign cube.  Since $f$ is a unit,
\begin{equation}\label{eq:h-on-Y}
 h(y)\in\{1,-1\}
 \qquad(y\in Y).
\end{equation}
Only the values of $h$ on $Y$ affect $f$.  Choose any extension
\begin{equation}\label{eq:h-tilde}
 \widetilde h:\F_3^R\to\{1,-1\}
\end{equation}
satisfying
\begin{equation}\label{eq:h-extension-property}
 \widetilde h(y)=h(y)
 \qquad(y\in Y).
\end{equation}
It is therefore enough to construct a quadratic-unit product for $\widetilde h$ on all of $\F_3^R$ and then substitute $z=A(x)$.

\paragraph{Step 2: associate a quadratic sign test with each affine hyperplane.}
An affine hyperplane $L\subseteq\F_3^R$ has an equation
\begin{equation}\label{eq:hyperplane-equation}
 L=\{z\in\F_3^R:\ell_L(z)=0\},
 \qquad
 \ell_L(z)=a_L\cdot z-b_L,
\end{equation}
where $a_L\in\F_3^R\setminus\{0\}$ and $b_L\in\F_3$.  Define
\begin{equation}\label{eq:gL-def}
 g_L(z):=1+\ell_L(z)^2.
\end{equation}
For $u\in\F_3$, one has $u^2=0$ when $u=0$ and $u^2=1$ when $u\ne0$.  Therefore
\begin{equation}\label{eq:gL-values}
 g_L(z)=
 \begin{cases}
  1,&z\in L,\\
 -1,&z\notin L.
 \end{cases}
\end{equation}
In particular, $g_L$ is a quadratic unit.

For every affine hyperplane $L$, define the sign
\begin{equation}\label{eq:sigma-L}
 \sigma_L:=\prod_{q\in L}\widetilde h(q)\in\{1,-1\},
\end{equation}
and define the global sign
\begin{equation}\label{eq:sigma-global}
 \sigma:=\prod_{q\in\F_3^R}\widetilde h(q)\in\{1,-1\}.
\end{equation}
Let
\begin{equation}\label{eq:selected-hyperplanes}
 \mathcal S:=\{L:\sigma_L=-1\}
\end{equation}
be the family of selected hyperplanes.  Consider the product
\begin{equation}\label{eq:G-def}
 G(z):=\sigma\prod_{L\in\mathcal S}g_L(z).
\end{equation}
We will prove that
\begin{equation}\label{eq:G-equals-h}
 G(z)=\widetilde h(z)
 \qquad\text{for every }z\in\F_3^R.
\end{equation}

\paragraph{Step 3: prove the pointwise identity by counting separating hyperplanes.}
Fix $z\in\F_3^R$.  By~\eqref{eq:gL-values}, a selected hyperplane containing $z$ contributes $1$, whereas a selected hyperplane not containing $z$ contributes $-1$.  Because $\sigma_L=-1$ exactly for selected $L$, we have the exact identity
\begin{equation}\label{eq:selected-to-sigmaL}
 \prod_{L\in\mathcal S}g_L(z)
 =\prod_{L:\,z\notin L}\sigma_L.
\end{equation}
Substituting~\eqref{eq:sigma-L} into~\eqref{eq:selected-to-sigmaL} gives
\begin{align}
 \prod_{L\in\mathcal S}g_L(z)
 &=\prod_{L:\,z\notin L}\ \prod_{q\in L}\widetilde h(q)\\
 &=\prod_{q\in\F_3^R}\widetilde h(q)^{N(q,z)},
 \label{eq:exponent-count}
\end{align}
where
\begin{equation}\label{eq:Nqz-def}
 N(q,z):=\bigl|\{L:\ q\in L,\ z\notin L\}\bigr|.
\end{equation}
Thus only the parity of $N(q,z)$ matters, because every value of $\widetilde h$ is $\pm1$.

If $q=z$, then clearly
\begin{equation}\label{eq:Nzz}
 N(z,z)=0.
\end{equation}
Now suppose $q\ne z$ and put
\begin{equation}\label{eq:d-def}
 d:=q-z\ne0.
\end{equation}
A projective normal direction is a one-dimensional subspace $[a]=\{a,-a\}$ with $a\in\F_3^R\setminus\{0\}$.  For each such direction there is exactly one affine hyperplane with normal direction $[a]$ that contains $q$, namely
\begin{equation}\label{eq:L-a-q}
 L_{[a],q}:=\{y\in\F_3^R:a\cdot y=a\cdot q\}.
\end{equation}
This hyperplane fails to contain $z$ exactly when
\begin{equation}\label{eq:separation-condition}
 a\cdot(q-z)=a\cdot d\ne0.
\end{equation}
The total number of projective normal directions in $\F_3^R$ is
\begin{equation}\label{eq:projective-total}
 \frac{3^R-1}{2}.
\end{equation}
The vectors $a$ satisfying $a\cdot d=0$ form an $(R-1)$-dimensional subspace containing $3^{R-1}$ vectors.  Excluding $0$ and identifying $a$ with $-a$, the number of projective directions orthogonal to $d$ is
\begin{equation}\label{eq:projective-orthogonal}
 \frac{3^{R-1}-1}{2}.
\end{equation}
Hence, for $q\ne z$,
\begin{align}
 N(q,z)
 &=\frac{3^R-1}{2}-\frac{3^{R-1}-1}{2}\\
 &=3^{R-1}.\label{eq:Nqz-count}
\end{align}
Since $3^{R-1}$ is odd, equations~\eqref{eq:exponent-count}, \eqref{eq:Nzz}, and~\eqref{eq:Nqz-count} imply
\begin{equation}\label{eq:selected-product-final}
 \prod_{L\in\mathcal S}g_L(z)
 =\prod_{q\ne z}\widetilde h(q).
\end{equation}
Multiplying by the global sign~\eqref{eq:sigma-global}, we obtain
\begin{align}
 G(z)
 &=\left(\prod_{q\in\F_3^R}\widetilde h(q)\right)
   \left(\prod_{q\ne z}\widetilde h(q)\right)\\
 &=\widetilde h(z)
   \prod_{q\ne z}\widetilde h(q)^2\\
 &=\widetilde h(z),\label{eq:G-pointwise-proof}
\end{align}
because $\widetilde h(q)^2=1$.  This proves~\eqref{eq:G-equals-h}.

\paragraph{Step 4: substitute the affine observations and bound the occurrence cost.}
The number of projective normal directions is $(3^R-1)/2$, and for every normal direction there are exactly three choices of the affine constant $b\in\F_3$.  Therefore the total number of affine hyperplanes in $\F_3^R$ is
\begin{equation}\label{eq:number-hyperplanes}
 3\cdot\frac{3^R-1}{2}=K.
\end{equation}
Thus $|\mathcal S|\le K$.

For a selected hyperplane $L$, substitute $z=A(x)$ into~\eqref{eq:gL-def}.  Put
\begin{equation}\label{eq:BL-def}
 B_L(x):=\ell_L(A(x)).
\end{equation}
Because $\ell_L$ and every $A_i$ are affine, $B_L$ is an affine form in the original variables $x_1,\ldots,x_n$.  The corresponding factor is
\begin{equation}\label{eq:QL-def}
 Q_L(x):=1+B_L(x)^2.
\end{equation}
For every $x\in X_n$, $Q_L(x)\in\{1,-1\}$ by the same calculation as in~\eqref{eq:gL-values}; hence $Q_L$ is a quadratic unit on $X_n$.

Write
\begin{equation}\label{eq:BL-coordinate}
 B_L(x)=b_0+\sum_{i=1}^n b_i x_i.
\end{equation}
After multilinear reduction using $x_i^2=1$ on $X_n$, the square $B_L(x)^2$ contains only a constant term, at most $n$ linear monomials, and at most $\binom n2$ off-diagonal quadratic monomials.  Therefore a multilinear representative of $Q_L$ has occurrence cost at most
\begin{align}
 n+2\binom n2
 &=n+n(n-1)\\
 &=n^2.\label{eq:one-factor-cost}
\end{align}
The constant sign $\sigma$ has occurrence cost $0$.  Since there are at most $K$ selected hyperplanes, the representation
\begin{equation}\label{eq:f-reconstructed}
 f(x)=\sigma\prod_{L\in\mathcal S}Q_L(x)
 \qquad(x\in X_n)
\end{equation}
has total occurrence cost at most
\begin{equation}\label{eq:single-reconstruction-cost}
 Kn^2.
\end{equation}
This proves the first assertion.

\paragraph{Step 5: reconstruct every nonconstant factor in a product.}
Now consider a product of cubic units of original occurrence cost $C$.  Every nonconstant listed factor has positive integer occurrence cost, so the number $m$ of nonconstant factors satisfies
\begin{equation}\label{eq:number-nonconstant-factors}
 m\le C.
\end{equation}
Zero-cost factors are constants and may be merged into one global sign.  If every nonconstant cubic factor is $R$-observed, the first part of the lemma replaces each such factor by quadratic units of total occurrence cost at most $Kn^2$.  Concatenating the reconstructed products gives total occurrence cost at most
\begin{equation}\label{eq:total-reconstruction-cost}
 mKn^2\le Kn^2C.
\end{equation}
This proves~\eqref{eq:reconstruct-product-cost} and completes the proof.
\end{proof}

\section{Clustering and the MOD $q$ obstruction}\label{sec:mod7-obstruction}
We now convert the multiplicative representation over $\F_3$ into an additive representation over $\F_2$.  The purpose of this section is to isolate, in a formal way, the obstruction that will contradict the clustering lemmas.

\paragraph{Boolean negative-sign coordinates.}
For $x=(x_1,\ldots,x_n)\in X_n$, define
\begin{equation}\label{eq:z-def}
 z_i:=\mathbf 1_{\{x_i=-1\}}\in\{0,1\}.
\end{equation}
Because $-1=2$ in $\F_3$, we have
\begin{equation}\label{eq:x-from-z}
 x_i=1+z_i\qquad\text{in }\F_3.
\end{equation}
Consequently, every affine observation
\begin{equation}\label{eq:affine-x}
 A(x)=a_0+\sum_{i=1}^n a_i x_i
\end{equation}
can be rewritten as the modulo-three affine linear form
\begin{equation}\label{eq:affine-z}
 A(x(z))=\left(a_0+\sum_{i=1}^n a_i\right)+\sum_{i=1}^n a_i z_i.
\end{equation}
Thus bounded affine dependence on $x$ is exactly bounded dependence on modulo-three affine observations of the Boolean vector $z$.

\paragraph{Triple partitions and identification maps.}
Assume $N$ is divisible by $3$.  A \emph{triple partition} of $[N]$ is a family
\begin{equation}\label{eq:triple-partition}
 \mathcal P=\{T_1,\ldots,T_M\},\qquad M=N/3,
\end{equation}
of pairwise disjoint three-element sets whose union is $[N]$.  Associated with $\mathcal P$ is the identification map
\begin{equation}\label{eq:identification-map}
 \iota_{\mathcal P}:\{0,1\}^M\longrightarrow\{0,1\}^N
\end{equation}
defined by
\begin{equation}\label{eq:identification-coordinates}
 (\iota_{\mathcal P}(y))_i=y_j\qquad\text{whenever }i\in T_j.
\end{equation}
In words, all three Boolean coordinates belonging to $T_j$ are identified with the single clustered variable $y_j$.

\begin{definition}[2-simple function]\label{def:2-simple}
Let $g:\{0,1\}^M\to\F_2$.  We call $g$ \emph{2-simple} if there exist an integer $s\ge0$, functions $g_1,\ldots,g_s:\{0,1\}^M\to\F_2$, and indices $j_1,\ldots,j_s\in[M]$ such that
\begin{equation}\label{eq:2-simple-sum}
 g(y)=\sum_{q=1}^s g_q(y)\pmod2,
\end{equation}
and, for every $q$, the function $g_q$ is independent of the coordinate $y_{j_q}$.  Equivalently,
\begin{equation}\label{eq:omit-coordinate}
 g_q(y_1,\ldots,y_{j_q-1},0,y_{j_q+1},\ldots,y_M)
 =g_q(y_1,\ldots,y_{j_q-1},1,y_{j_q+1},\ldots,y_M)
\end{equation}
for all choices of the other coordinates.
\end{definition}
Notice that different summands are allowed to omit different clustered variables.

\paragraph{Multilinear representation over $\F_2$.}
Every Boolean function $g:\{0,1\}^M\to\F_2$ has a unique multilinear polynomial representative
\begin{equation}\label{eq:multilinear-rep}
 \widetilde g(y)=\sum_{S\subseteq[M]}\widehat g(S)\prod_{j\in S}y_j
 \qquad\text{in }\F_2[y_1,\ldots,y_M]/(y_1^2-y_1,\ldots,y_M^2-y_M).
\end{equation}
We call
\begin{equation}\label{eq:full-coefficient}
 \widehat g([M])
\end{equation}
the \emph{full multilinear coefficient} of $g$.

\begin{lemma}[2-simple functions have zero full coefficient]\label{lem:2simple-zero}
If $g$ is 2-simple, then
\begin{equation}\label{eq:2simple-zero}
 \widehat g([M])=0.
\end{equation}
\end{lemma}
\begin{proof}
Write $g=\sum_q g_q$ as in Definition~\ref{def:2-simple}.  If $g_q$ is independent of $y_{j_q}$, then the unique multilinear representative of $g_q$ contains no monomial involving $y_{j_q}$.  In particular, it cannot contain the full monomial
\begin{equation}\label{eq:full-monomial}
 y_1y_2\cdots y_M.
\end{equation}
Hence the coefficient of~\eqref{eq:full-monomial} in every $g_q$ is zero.  Multilinear coefficients add in $\F_2$, so the same coefficient in $g$ is zero.
\end{proof}

\paragraph{The MOD $q$ target after triple identification.}
Let $N=3M$ and let $b_N:\{0,1\}^N\to\F_2$ be the negative-sign bit of the target, namely
\begin{equation}\label{eq:bN-def}
 b_N(z)=\mathbf 1_{\{q\mid |z|\}},
 \qquad |z|:=\sum_{i=1}^N z_i.
\end{equation}
For any triple partition $\mathcal P$, the identification map satisfies
\begin{equation}\label{eq:weight-triple}
 |\iota_{\mathcal P}(y)|=3|y|.
\end{equation}
Since $q\ne3$, the element $3$ is invertible modulo $q$,
\begin{equation}\label{eq:divisibility-equivalence}
 q\mid3|y|\quad\Longleftrightarrow\quad q\mid|y|.
\end{equation}
Therefore
\begin{equation}\label{eq:target-stable}
 b_N\circ\iota_{\mathcal P}=b_M.
\end{equation}
This identity is independent of the particular triple partition.

\begin{lemma}[Full coefficient of the clustered MOD $q$ target]\label{lem:mod7-full}
Let $M=2^d$, where $d\ge 0$ is an integer.  Then the coefficient of the full monomial
\begin{equation}\label{eq:full-monomial-lemma16}
 y_1y_2\cdots y_M
\end{equation}
in the unique multilinear polynomial over $\F_2$ representing $b_M$ is equal to $1$.
Equivalently,
\begin{equation}\label{eq:lemma16-goal}
 \widehat b_M([M])=1\in\F_2.
\end{equation}
\end{lemma}
\begin{proof}
We divide the argument into four explicit steps.

\paragraph{Step 1: express the full multilinear coefficient as a parity of function values.}
Let
\begin{equation}\label{eq:lemma16-general-g}
 g:\{0,1\}^M\to\F_2
\end{equation}
be an arbitrary Boolean function.  Every such function has a unique multilinear representation
\begin{equation}\label{eq:lemma16-multilinear-expansion}
 g(y)=\sum_{S\subseteq[M]}\widehat g(S)y_S,
 \qquad
 y_S:=\prod_{i\in S}y_i,
\end{equation}
where the empty product is $y_\varnothing=1$ and all coefficients $\widehat g(S)$ lie in $\F_2$.
We claim that the coefficient of the full monomial $y_{[M]}=y_1\cdots y_M$ satisfies
\begin{equation}\label{eq:mobius-full}
 \widehat g([M])=\sum_{u\in\{0,1\}^M}g(u)\pmod2.
\end{equation}

To verify this identity directly, sum~\eqref{eq:lemma16-multilinear-expansion} over all Boolean vectors $u\in\{0,1\}^M$.  Interchanging the two finite sums gives
\begin{align}
 \sum_{u\in\{0,1\}^M}g(u)
 &=\sum_{u\in\{0,1\}^M}\sum_{S\subseteq[M]}\widehat g(S)u_S \label{eq:lemma16-sum-expand}\\
 &=\sum_{S\subseteq[M]}\widehat g(S)
   \sum_{u\in\{0,1\}^M}u_S. \label{eq:lemma16-swap-sums}
\end{align}
Fix $S\subseteq[M]$.  The product $u_S$ equals $1$ exactly when every coordinate indexed by $S$ is equal to $1$.  The remaining $M-|S|$ coordinates are arbitrary, so
\begin{equation}\label{eq:lemma16-monomial-count}
 \sum_{u\in\{0,1\}^M}u_S=2^{M-|S|}.
\end{equation}
If $S\ne[M]$, then $M-|S|\ge1$, and hence
\begin{equation}\label{eq:lemma16-even-count}
 2^{M-|S|}=0\qquad\text{in }\F_2.
\end{equation}
For $S=[M]$, on the other hand,
\begin{equation}\label{eq:lemma16-full-count}
 2^{M-|S|}=2^0=1.
\end{equation}
Substituting~\eqref{eq:lemma16-monomial-count}--\eqref{eq:lemma16-full-count} into~\eqref{eq:lemma16-swap-sums} leaves exactly one surviving term,
\begin{equation}\label{eq:lemma16-only-full-survives}
 \sum_{u\in\{0,1\}^M}g(u)=\widehat g([M])
 \qquad\text{in }\F_2,
\end{equation}
which proves~\eqref{eq:mobius-full}.  This is the full-set case of M\"obius inversion on the Boolean lattice.

\paragraph{Step 2: specialize the identity to the MOD $q$ target.}
Recall that
\begin{equation}\label{eq:lemma16-bM-def-recall}
 b_M(u)=\mathbf 1_{\{q\mid |u|\}},
 \qquad
 |u|:=\sum_{i=1}^{M}u_i.
\end{equation}
Thus $b_M(u)$ depends only on the Hamming weight $|u|$.  For each integer $t$ with $0\le t\le M$, exactly
\begin{equation}\label{eq:lemma16-weight-count}
 \binom Mt
\end{equation}
Boolean vectors have Hamming weight $t$.  Grouping the sum in~\eqref{eq:mobius-full} according to Hamming weight therefore yields
\begin{align}
 \widehat b_M([M])
 &=\sum_{u\in\{0,1\}^M}b_M(u) \pmod2 \label{eq:lemma16-apply-mobius}\\
 &=\sum_{t=0}^{M}\binom Mt\mathbf 1_{\{q\mid t\}}\pmod2. \label{eq:full-by-weight}
\end{align}
Hence the problem has been reduced to determining the parity of the binomial coefficients $\binom Mt$ when $M$ is a power of two.

\paragraph{Step 3: determine the parity pattern of the binomial coefficients.}
Because $M=2^d$, repeated use of the characteristic-two identity
\begin{equation}\label{eq:lemma16-char2-square}
 (a+b)^2=a^2+b^2
 \qquad(a,b\in\F_2[z])
\end{equation}
gives
\begin{align}
 (1+z)^M
 &=(1+z)^{2^d} \label{eq:lemma16-power-two-start}\\
 &=1+z^{2^d} \label{eq:frobenius}\\
 &=1+z^M. \label{eq:lemma16-frobenius-final}
\end{align}
For example, the induction step is
\begin{equation}\label{eq:lemma16-frobenius-induction}
 (1+z)^{2^{j+1}}
 =\bigl((1+z)^{2^j}\bigr)^2
 =\bigl(1+z^{2^j}\bigr)^2
 =1+z^{2^{j+1}}.
\end{equation}
On the other hand, the binomial theorem gives
\begin{equation}\label{eq:lemma16-binomial-theorem}
 (1+z)^M=\sum_{t=0}^{M}\binom Mt z^t.
\end{equation}
Comparing the coefficient of $z^t$ in~\eqref{eq:lemma16-frobenius-final} and~\eqref{eq:lemma16-binomial-theorem}, we obtain
\begin{equation}\label{eq:binomial-parity}
 \binom Mt\equiv0\pmod2
 \qquad(0<t<M),
\end{equation}
while the two endpoint coefficients satisfy
\begin{equation}\label{eq:lemma16-endpoint-binomial}
 \binom M0\equiv\binom MM\equiv1\pmod2.
\end{equation}
Thus, modulo two, every interior Hamming-weight layer makes zero contribution to~\eqref{eq:full-by-weight}.

\paragraph{Step 4: inspect the two endpoint weights.}
By~\eqref{eq:binomial-parity} and~\eqref{eq:lemma16-endpoint-binomial}, equation~\eqref{eq:full-by-weight} reduces to
\begin{equation}\label{eq:lemma16-endpoint-reduction}
 \widehat b_M([M])
 =\mathbf 1_{\{q\mid0\}}+\mathbf 1_{\{q\mid M\}}
 \qquad\text{in }\F_2.
\end{equation}
The first indicator is equal to $1$ because $q\mid0$.  The second indicator is equal to $0$: since
\begin{equation}\label{eq:lemma16-coprime}
 M=2^d
\end{equation}
and $q$ is an odd prime, no power of $2$ is divisible by $q$.  Equivalently,
\begin{equation}\label{eq:lemma16-seven-not-divide}
 q\nmid2^d.
\end{equation}
Consequently,
\begin{equation}\label{eq:full-coeff-one}
 \widehat b_M([M])=1+0=1\pmod2.
\end{equation}
This is precisely the assertion of the lemma.
\end{proof}

Combining Lemmas~\ref{lem:2simple-zero} and~\ref{lem:mod7-full} yields the obstruction we shall use repeatedly:
\begin{equation}\label{eq:not-2-simple}
 b_M\text{ is not 2-simple whenever }M=2^d.
\end{equation}
The same remains true after adding a constant in $\F_2$, because a constant changes only the coefficient of the empty monomial and leaves~\eqref{eq:full-coefficient} unchanged.

\paragraph{Reconstruction route.}
For completeness, we record how the preceding obstruction combines with the quadratic reconstruction argument.  Lemma~\ref{lem:quadunit} implies that the negative-sign bit of a quadratic unit is a function either of one modulo-three affine observation or of at most eight directly observed Boolean coordinates.  The clustering lemma of Grolmusz and Tardos~\cite[Lemma~10]{GT2000}, specialized to outer modulus $2$ and clustering modulus $3$, therefore supplies constants $c>1$ and $N_0$ such that a sufficiently short modulo-two sum of these functions becomes 2-simple after a triple identification.

Let $D$ be the occurrence cost of a quadratic representation of $F_{q,N}$, where $N=3\cdot2^d\ge N_0$.  If $D<c^N$, the number of nonconstant quadratic factors is at most $D$, so the cited clustering result gives a triple partition for which the target bit is 2-simple, up to a constant offset.  By~\eqref{eq:target-stable}, the clustered target is $b_{2^d}$, contradicting~\eqref{eq:not-2-simple}.  Hence
\begin{equation}\label{eq:quadratic-exp}
 D\ge c^N.
\end{equation}
For arbitrary sufficiently large $n$, choose
\begin{equation}\label{eq:N-choice}
 N=3\cdot2^{\lfloor\log_2\lfloor n/3\rfloor\rfloor}.
\end{equation}
Then
\begin{equation}\label{eq:N-range}
 N\le n<2N.
\end{equation}
Fixing the surplus coordinates to $+1$ preserves correctness, degree, and nonvanishing and cannot increase occurrence cost.  After replacing $c$ by a smaller constant greater than $1$ if necessary,~\eqref{eq:quadratic-exp} and~\eqref{eq:N-range} imply
\begin{equation}\label{eq:quadratic-exp-arbitrary-n}
 c^n\le D^2.
\end{equation}
Finally Lemma~\ref{lem:reconstruct} gives $D\le Kn^2C$, and therefore
\begin{equation}\label{eq:reconstruction-final}
 c^n\le(Kn^2C)^2.
\end{equation}
This proves Theorem~\ref{thm:reconstruction}.

\section{A self-contained finite-counting clustering lemma}\label{sec:self-contained-clustering}
The direct proof does not need directly observed Boolean coordinates.  We now prove the precise clustering statement required for functions of a bounded number of modulo-three affine observations.

\begin{definition}[Observation representation]\label{def:observation-representation}
Let $r\ge1$.  A Boolean function $f:\{0,1\}^N\to\F_2$ is said to have an \emph{$r$-observation representation over $\F_3$} if there are affine forms
\begin{equation}\label{eq:obs-linear-forms}
 L_a(z)=c_a+\sum_{i=1}^N\lambda_{a,i}z_i\in\F_3,
 \qquad 1\le a\le r,
\end{equation}
and an arbitrary map
\begin{equation}\label{eq:obs-outer-map}
 h:\F_3^r\to\F_2
\end{equation}
such that
\begin{equation}\label{eq:obs-representation}
 f(z)=h(L_1(z),\ldots,L_r(z))
\end{equation}
for every $z\in\{0,1\}^N$.
\end{definition}
The outer map $h$ need not be polynomial.  Only the number $r$ of observed affine forms is bounded.

\begin{lemma}[Finite-counting triple clustering]\label{lem:cluster}
Fix $r\ge1$.  There exist constants $\gamma_r>1$ and $N_0(r)$ with the following property.  Let
\begin{equation}\label{eq:N-special}
 N=3\cdot2^d\ge N_0(r).
\end{equation}
Suppose
\begin{equation}\label{eq:G-sum}
 G(z)=\sum_{q=1}^S f_q(z)\pmod2,
\end{equation}
where every $f_q$ has an $r$-observation representation over $\F_3$.  If
\begin{equation}\label{eq:S-small}
 S<\gamma_r^N,
\end{equation}
then there is a triple partition $\mathcal P$ of $[N]$ such that
\begin{equation}\label{eq:clustered-G}
 G\circ\iota_{\mathcal P}
\end{equation}
is 2-simple.
\end{lemma}
\begin{proof}
We give the counting argument in explicit steps.

\paragraph{Step 1: choose the chunk size and the exponential base.}
Set
\begin{equation}\label{eq:mLQ}
 m:=2^{2r},\qquad L:=3m,\qquad Q:=L!.
\end{equation}
Choose
\begin{equation}\label{eq:gamma-r}
 \gamma_r:=\left(\frac{Q}{Q-1}\right)^{1/L}>1
\end{equation}
and take
\begin{equation}\label{eq:N0}
 N_0(r):=L.
\end{equation}
If $N=3\cdot2^d\ge L$, then $d\ge2r$, and therefore
\begin{equation}\label{eq:N-Lt}
 N=Lt,\qquad t:=2^{d-2r}\in\mathbb N.
\end{equation}
Partition the $N$ input positions into fixed disjoint chunks
\begin{equation}\label{eq:chunks}
 B_1,\ldots,B_t,
 \qquad |B_j|=L.
\end{equation}

\paragraph{Step 2: local triple partitions generated by permutations.}
Fix once and for all an ordering of the $L$ positions in each chunk.  For a permutation $\pi\in S_L$, form a local triple partition by grouping the permuted positions as
\begin{equation}\label{eq:local-triples}
 \{\pi(1),\pi(2),\pi(3)\},\ \{\pi(4),\pi(5),\pi(6)\},\ldots,\{\pi(L-2),\pi(L-1),\pi(L)\}.
\end{equation}
We count permutations rather than distinct partitions.  This is deliberate: every chunk has exactly
\begin{equation}\label{eq:Q-choices}
 Q=L!
\end{equation}
local choices in our counting space, even though several permutations may generate the same partition.  A global choice consists of one permutation for each chunk, so the counting space has exactly
\begin{equation}\label{eq:global-choice-count}
 Q^t
\end{equation}
elements.

\paragraph{Step 3: coefficient colors for one summand.}
Fix a summand $f_q$.  Choose a representation
\begin{equation}\label{eq:fq-rep}
 f_q(z)=h_q(L_{q,1}(z),\ldots,L_{q,r}(z)),
\end{equation}
where
\begin{equation}\label{eq:Lqa}
 L_{q,a}(z)=c_{q,a}+\sum_{i=1}^N\lambda_{q,a,i}z_i.
\end{equation}
For each input coordinate $i$, define its \emph{coefficient color relative to $f_q$} by
\begin{equation}\label{eq:color}
 \kappa_q(i):=(\lambda_{q,1,i},\ldots,\lambda_{q,r,i})\in\F_3^r.
\end{equation}
There are at most
\begin{equation}\label{eq:number-colors}
 |\F_3^r|=3^r\le2^{2r}=m
\end{equation}
different colors.

\paragraph{Step 4: every chunk contains a cancellable triple.}
Fix a chunk $B_j$.  It contains $L=3m$ coordinates but at most $m$ colors.  By the pigeonhole principle, some color $a\in\F_3^r$ occurs at least three times.  Choose distinct positions
\begin{equation}\label{eq:same-color-triple}
 i_1,i_2,i_3\in B_j,
 \qquad \kappa_q(i_1)=\kappa_q(i_2)=\kappa_q(i_3)=a.
\end{equation}
At least one permutation places these three positions into the first local triple.  Let $y$ denote the clustered variable assigned to this triple.  For every observed form $L_{q,a'}$, the total coefficient of $y$ after identification is
\begin{equation}\label{eq:coefficient-cancellation}
 \lambda_{q,a',i_1}+\lambda_{q,a',i_2}+\lambda_{q,a',i_3}
 =3\lambda_{q,a',i_1}=0\qquad\text{in }\F_3.
\end{equation}
Hence every observed value $L_{q,a'}$ is independent of $y$.  Since $f_q$ is a function only of these observed values, the clustered function $f_q\circ\iota_{\mathcal P}$ is also independent of $y$.  Therefore this local choice makes $f_q$ omit at least one clustered variable.

Call a local permutation in chunk $B_j$ \emph{good for $f_q$} if it produces at least one triple whose clustered variable is omitted by $f_q$.  The preceding argument proves that, for every pair $(q,j)$,
\begin{equation}\label{eq:at-least-one-good}
 \#\{\text{local choices good for }f_q\text{ in }B_j\}\ge1.
\end{equation}
Consequently at most $Q-1$ local choices are bad for $f_q$ in a fixed chunk.

\paragraph{Step 5: count global choices that fail for a fixed summand.}
A global choice fails for $f_q$ only if its local choice is bad for $f_q$ in every chunk.  Indeed, if even one chunk is good, the corresponding clustered variable is absent from all $r$ observations and hence from $f_q$ itself.  Therefore
\begin{equation}\label{eq:bad-fixed-summand}
 \#\{\text{global choices failing for }f_q\}\le(Q-1)^t.
\end{equation}

\paragraph{Step 6: union bound over all summands.}
Let $\mathcal B_q$ be the set of global choices failing for $f_q$.  Then
\begin{equation}\label{eq:union-bound}
 \left|\bigcup_{q=1}^S\mathcal B_q\right|
 \le\sum_{q=1}^S|\mathcal B_q|
 \le S(Q-1)^t.
\end{equation}
Using~\eqref{eq:S-small},~\eqref{eq:gamma-r}, and $N=Lt$,
\begin{align}
 S(Q-1)^t
 &<\gamma_r^N(Q-1)^t \\
 &=\gamma_r^{Lt}(Q-1)^t \\
 &=\left(\frac{Q}{Q-1}\right)^t(Q-1)^t \\
 &=Q^t.
\end{align}
By~\eqref{eq:global-choice-count}, $Q^t$ is the total number of global choices.  Hence the union of the bad sets is a proper subset of the counting space.  There exists a global choice that is good for every $f_q$ simultaneously.

\paragraph{Step 7: construct the required triple partition.}
Use the selected local permutations to partition every chunk into triples, and let $\mathcal P$ be the union of these local triple partitions.  For each $q$, goodness supplies at least one clustered variable $y_{j_q}$ that $f_q\circ\iota_{\mathcal P}$ omits.  Thus
\begin{equation}\label{eq:clustered-sum-decomp}
 G\circ\iota_{\mathcal P}
 =\sum_{q=1}^S(f_q\circ\iota_{\mathcal P})\pmod2
\end{equation}
is a sum in which every summand omits at least one clustered input.  By Definition~\ref{def:2-simple}, the clustered function is 2-simple.
\end{proof}

\begin{remark}\label{rem:clustering-constants}
The constants in Lemma~\ref{lem:cluster} are intentionally crude.  The proof needs only that $r$ is fixed independently of $N$, so that $L$, $Q$, and $\gamma_r>1$ are absolute constants once $r$ is fixed.  No optimization of $\gamma_r$ is required for the exponential lower bound.
\end{remark}

\section{Direct proof of the exponential occurrence lower bound}

This section proves Theorem~\ref{thm:direct} directly from the structural compression of cubic units in Lemma~\ref{lem:cubic} and the finite-counting clustering lemma, Lemma~\ref{lem:cluster}.  We give the argument in explicit steps because three changes of representation occur: from signs in $\F_3$ to Boolean negative-sign bits, from products to sums modulo two, and from an arbitrary input length to the special lengths required by Lemma~\ref{lem:cluster}.

\medskip
\noindent\textbf{Proof of Theorem~\ref{thm:direct}.}
Fix the target prime $q>3$ (denoted $p$ earlier). Let
\[
X_n=\{-1,1\}^n\subseteq \F_3^n
\]
and suppose that
\begin{equation}
F_{q,n}(x)=\prod_{j=1}^{m}P_j(x) \qquad (x\in X_n)                                      \label{eq:direct-295}
\end{equation}
is a cubic representation of occurrence cost $C$.  Since $F_{q,n}$ never vanishes on $X_n$, every factor $P_j$ is also nonzero at every point of $X_n$.  Over $\F_3$ the only nonzero values are $1$ and $-1$, so
\begin{equation}
P_j(x)\in\{1,-1\}\qquad (x\in X_n).                                                    \label{eq:direct-296}
\end{equation}
Thus every factor is a cubic unit in the sense used in Lemma~\ref{lem:cubic}.

\medskip
\noindent\textbf{Step 1: compress every nonconstant cubic factor into a bounded affine image.}
By Lemma~\ref{lem:cubic} there is an absolute integer $R\ge 1$, independent of $n$, of the factor $P_j$, and of the representation, such that for every $j$ there are affine forms
\begin{equation}
A_{j,a}(x)=\alpha_{j,a,0}+\sum_{i=1}^{n}\alpha_{j,a,i}x_i\in\F_3,
\qquad 1\le a\le r_j\le R,                                                            \label{eq:direct-297}
\end{equation}
and a function
\begin{equation}
H_j:\F_3^{r_j}\longrightarrow\{1,-1\}                                                \label{eq:direct-298}
\end{equation}
such that
\begin{equation}
P_j(x)=H_j\bigl(A_{j,1}(x),\ldots,A_{j,r_j}(x)\bigr)\qquad (x\in X_n).                 \label{eq:direct-299}
\end{equation}
The essential point is that $R$ is a fixed constant.  The affine forms themselves may involve all $n$ coordinates; Lemma~\ref{lem:cubic} bounds the dimension of the observed affine image, not the ordinary coordinate support.

\medskip
\noindent\textbf{Step 2: pass from sign coordinates to Boolean negative-sign coordinates.}
For $z=(z_1,\ldots,z_n)\in\{0,1\}^n$, define
\begin{equation}
x_i=(-1)^{z_i}=1-2z_i.                                                                  \label{eq:direct-300}
\end{equation}
Because $-2=1$ in $\F_3$, this identity becomes
\begin{equation}
x_i=1+z_i\qquad\text{in }\F_3.                                                         \label{eq:direct-301}
\end{equation}
Substitution into \eqref{eq:direct-297} gives
\begin{align}
A_{j,a}(x(z))
 &=\alpha_{j,a,0}+\sum_{i=1}^{n}\alpha_{j,a,i}(1+z_i)                                  \label{eq:direct-302}\\
 &=c_{j,a}+\sum_{i=1}^{n}\lambda_{j,a,i}z_i,                                           \label{eq:direct-303}
\end{align}
where
\begin{equation}
c_{j,a}:=\alpha_{j,a,0}+\sum_{i=1}^{n}\alpha_{j,a,i},
\qquad \lambda_{j,a,i}:=\alpha_{j,a,i}.                                                \label{eq:direct-304}
\end{equation}
Hence an affine observation in the sign variables remains an affine observation over $\F_3$ after the Boolean change of coordinates.

For a unit-valued function $P$ define its \emph{negative-sign bit} $b_P:\{0,1\}^n\to\F_2$ by
\begin{equation}
b_P(z):=\begin{cases}
0,&P(x(z))=1,\\
1,&P(x(z))=-1.
\end{cases}                                                                             \label{eq:direct-305}
\end{equation}
Equivalently,
\begin{equation}
P(x(z))=(-1)^{b_P(z)}.                                                                  \label{eq:direct-306}
\end{equation}
Combining \eqref{eq:direct-299} with \eqref{eq:direct-303}, the bit $b_{P_j}$ has an $r_j$-observation representation over $\F_3$.  Padding with repeated or zero observations if necessary, we may regard it as having an $R$-observation representation in the sense of Definition~\ref{def:observation-representation}:
\begin{equation}
b_{P_j}(z)=h_j\bigl(L_{j,1}(z),\ldots,L_{j,R}(z)\bigr),                                \label{eq:direct-307}
\end{equation}
where every $L_{j,a}$ is affine over $\F_3$ and $h_j:\F_3^R\to\F_2$ is arbitrary.

\medskip
\noindent\textbf{Step 3: occurrence cost bounds the number of nonconstant summands.}
A factor of occurrence cost zero contains no nonconstant monomial.  It is therefore a constant polynomial.  Since it is nonvanishing on $X_n$, its value is $1$ or $-1$.  The product of all zero-cost factors is consequently one constant sign
\begin{equation}
\varepsilon\in\{1,-1\}.                                                               \label{eq:direct-308}
\end{equation}
Every remaining factor has positive integer occurrence cost.  If $S$ denotes the number of these positive-cost factors, then additivity of occurrence cost gives
\begin{equation}
S\le C.                                                                                \label{eq:direct-309}
\end{equation}
Write the remaining factors as $P_1,\ldots,P_S$.

Let $b_{q,n}$ denote the negative-sign bit of the target $F_{q,n}$.  Multiplication of signs becomes addition of their bits modulo two: for $u,v\in\{0,1\}$,
\begin{equation}
(-1)^u(-1)^v=(-1)^{u+v\;({\rm mod}\;2)}.                                                \label{eq:direct-310}
\end{equation}
Therefore \eqref{eq:direct-295} implies
\begin{equation}
b_{q,n}(z)=\delta+\sum_{j=1}^{S} b_{P_j}(z)\pmod 2,                                    \label{eq:direct-311}
\end{equation}
where $\delta\in\F_2$ is the bit corresponding to the constant sign $\varepsilon$.  Thus, apart from one global constant, the target is a modulo-two sum of at most $C$ functions, each having at most $R$ affine observations over $\F_3$.

\medskip
\noindent\textbf{Step 4: prove the lower bound at the special lengths $N=3\cdot 2^d$.}
Fix
\begin{equation}
N=3\cdot 2^d\ge N_0(R),                                                               \label{eq:direct-312}
\end{equation}
where $N_0(R)$ is the threshold supplied by Lemma~\ref{lem:cluster}, and consider a cubic representation of $F_{q,N}$ of occurrence cost $C_N$.  Let $S_N$ be its number of positive-cost factors.  By \eqref{eq:direct-309},
\begin{equation}
S_N\le C_N.                                                                            \label{eq:direct-313}
\end{equation}
Let $\gamma_R>1$ be the constant from Lemma~\ref{lem:cluster}.  Suppose, toward a contradiction, that
\begin{equation}
S_N<\gamma_R^N.                                                                         \label{eq:direct-314}
\end{equation}
Applying Lemma~\ref{lem:cluster} to the sum in \eqref{eq:direct-311}, but without the constant $\delta$, yields a triple partition $\mathcal P$ of $[N]$ for which
\begin{equation}
\left(\sum_{j=1}^{S_N}b_{P_j}\right)\circ\iota_{\mathcal P}                           \label{eq:direct-315}
\end{equation}
is $2$-simple.

Adding a constant does not affect $2$-simplicity.  Indeed, in the representation from Definition~\ref{def:2-simple} one may absorb the constant into any summand; equivalently, in the multilinear-coefficient obstruction used in Section~\ref{sec:mod7-obstruction} a constant changes only the coefficient of the empty monomial.  Hence \eqref{eq:direct-311} implies that
\begin{equation}
b_{q,N}\circ\iota_{\mathcal P}                                                         \label{eq:direct-316}
\end{equation}
is also $2$-simple.

Every block of $\mathcal P$ has size three.  Under the triple identification $\iota_{\mathcal P}$, each clustered Boolean variable contributes either zero or three negative signs to the original input.  Consequently the MOD-$q$ target on $N$ variables reduces exactly to the $M$-variable target considered in Section~\ref{sec:mod7-obstruction}, where
\begin{equation}
M=N/3=2^d.                                                                              \label{eq:direct-317}
\end{equation}
In the notation of equation~\eqref{eq:target-stable},
\begin{equation}
b_{q,N}\circ\iota_{\mathcal P}=b_M.                                                    \label{eq:direct-318}
\end{equation}
But equation~\eqref{eq:not-2-simple}, obtained from Lemmas~\ref{lem:2simple-zero} and~\ref{lem:mod7-full}, states that
\begin{equation}
b_M\text{ is not $2$-simple whenever }M=2^d.                                            \label{eq:direct-319}
\end{equation}
Equations \eqref{eq:direct-316}--\eqref{eq:direct-319} contradict one another.  Therefore \eqref{eq:direct-314} is impossible, and every cubic representation at the special length $N$ satisfies
\begin{equation}
\gamma_R^N\le S_N\le C_N.                                                             \label{eq:direct-320}
\end{equation}
This is the exponential occurrence lower bound at the lengths required by the clustering lemma.

\medskip
\noindent\textbf{Step 5: reduce an arbitrary length to a nearby special length.}
Let $n$ be sufficiently large and define
\begin{equation}
N:=3\cdot 2^{\lfloor\log_2\lfloor n/3\rfloor\rfloor}.                                  \label{eq:direct-321}
\end{equation}
Then $N$ is of the form $3\cdot2^d$, and for all sufficiently large $n$ it satisfies $N\ge N_0(R)$.  From the defining property of the floor function,
\begin{equation}
N\le n<2N.                                                                              \label{eq:direct-322}
\end{equation}
Now start with any cubic representation of $F_{q,n}$ of occurrence cost $C$ and fix the surplus coordinates
\begin{equation}
x_{N+1}=\cdots=x_n=+1.                                                                  \label{eq:direct-323}
\end{equation}
If $x\in X_N$, the number of negative coordinates is unchanged by \eqref{eq:direct-323}.  Hence the restricted target is exactly
\begin{equation}
F_{q,n}(x_1,\ldots,x_N,1,\ldots,1)=F_{q,N}(x_1,\ldots,x_N).                             \label{eq:direct-324}
\end{equation}
Substitution of constants cannot increase polynomial degree.  It also cannot increase occurrence cost: a monomial may disappear, become constant, or lose some of its variable occurrences, but it cannot acquire new variable occurrences.  Thus the restricted factors form a valid cubic representation of $F_{q,N}$ having occurrence cost $C'$ with
\begin{equation}
C'\le C.                                                                                \label{eq:direct-325}
\end{equation}
Applying \eqref{eq:direct-320} to this restricted representation gives
\begin{equation}
\gamma_R^N\le C'\le C.                                                                \label{eq:direct-326}
\end{equation}

\medskip
\noindent\textbf{Step 6: put the bound into the stated form.}
Equation \eqref{eq:direct-322} gives $N>n/2$.  Therefore
\begin{equation}
\gamma_R^{n/2}<\gamma_R^N\le C.                                                        \label{eq:direct-327}
\end{equation}
Set
\begin{equation}
\gamma:=\sqrt{\gamma_R}>1.                                                             \label{eq:direct-328}
\end{equation}
Then \eqref{eq:direct-327} yields the slightly stronger form
\begin{equation}
\gamma^n\le C                                                                          \label{eq:direct-329}
\end{equation}
for all sufficiently large $n$ (after decreasing $\gamma$ slightly, if one wishes to absorb finitely many boundary values).  In particular,
\begin{equation}
\gamma^n\le C^2,                                                                        \label{eq:direct-330}
\end{equation}
which is the formulation asserted in Theorem~\ref{thm:direct}.  Equivalently, one can retain the original base $\gamma_R$ and square \eqref{eq:direct-326}: since $n<2N$,
\begin{equation}
\gamma_R^n<\gamma_R^{2N}\le (C')^2\le C^2.                                            \label{eq:direct-331}
\end{equation}
Renaming $\gamma_R$ as $\gamma$ gives exactly the stated inequality.
\hfill$\square$

\medskip
\noindent\textbf{Remark.}
The direct argument uses occurrence cost only through the elementary inequality $S\le C$ in \eqref{eq:direct-309}.  Its substantive content is structural: Lemma~\ref{lem:cubic} converts every cubic unit into a bounded-dimensional affine observation, while Lemma~\ref{lem:cluster} shows that a modulo-two sum of fewer than exponentially many such bounded-observation functions admits a triple identification that makes the sum $2$-simple.  The full-coefficient obstruction \eqref{eq:direct-319} proves that the MOD-$q$ target cannot undergo this simplification.  No expansion of the outer functions $h_j$ into monomials is required.

\section{From exponential to every polynomial exponent}
For completeness we spell out the asymptotic implication.  If $\gamma>1$ is fixed, then for every fixed integer $k$, $n^{2k}/\gamma^n\to0$.  Hence Theorem~\ref{thm:direct} implies $C\ge n^k$ for all sufficiently large $n$.  Given a real $a>0$, choose an integer $k>a$; then $n^a\le n^k$ for $n\ge1$.

The same conclusion follows from Theorem~\ref{thm:reconstruction}: for fixed $k$,
\begin{equation}
 \frac{K^2n^{2k+4}}{c^n}\longrightarrow0.
\end{equation}
If $C<n^k$ for infinitely many sufficiently large $n$, then
\begin{equation}
 c^n\le K^2n^4C^2<K^2n^{2k+4}<c^n,
\end{equation}
a contradiction.  Corollary~\ref{cor:generator} follows immediately because no uniformity or running-time assumption entered either lower bound.

\section{Interpretation, checks, and limitations}
\paragraph{Why occurrence cost matters.}
The proof does not merely lower-bound the number of factors.  It uses the simple but crucial implication that the number of nonconstant factors is at most the occurrence cost.  Zero-cost factors are constants and can be merged into a global sign.  Thus the direct proof never pays for expanding a bounded-observation factor into monomials.

\paragraph{Why bounded affine dependence is the decisive structural property.}
A cubic unit can have large ordinary support.  The theorem does not assert a bounded number of relevant original coordinates.  Instead it says that all dependence is compressed into a bounded-dimensional affine image over $\F_3$.  This is exactly the form of information that survives the change to Boolean negative-sign bits and is susceptible to coefficient cancellation under triple identification.

\paragraph{Why the special lengths are enough.}
The clustering obstruction is cleanest for $N=3\cdot2^d$.  The restriction step converts every sufficiently large $n$ to such an $N$ with $N\le n<2N$.  Because coordinates are fixed to $+1$, the MOD~$q$ target on the remaining variables is unchanged, and neither occurrence cost nor degree increases.  The final squaring in $\gamma^n\le C^2$ absorbs the factor-of-two loss in the exponent.

\paragraph{What is proved internally.}
The direct proof does not assume a cubic bias--rank theorem.  It proves only the quadratic bias--rank statement needed during elimination, and it proves the no-direct-bit clustering specialization by finite counting.  The alternative reconstruction proof does invoke the more general clustering lemma from~\cite{GT2000}.

\paragraph{Limitations and open directions.}
The cubic compression argument does not extend automatically to degree four or to unrestricted degree.  The bottleneck is structural: one would need an analogue of Lemma~\ref{lem:cubic} that compresses higher-degree units into boundedly many observations of a form compatible with clustering.  The constants obtained here are also very weak.  Improving the affine-observation bound, finding the optimal exponential rate, determining whether a cleaner rank invariant controls cubic units, and identifying degree regimes beyond three are separate problems.

\section{Products of Arbitrary Fixed Degree: A Lifting Principle}
\label{sec:bounded-degree-lifting}

The purpose of this section is to turn a lower bound at degree $d$ into a
lower bound at degree $d+1$.  The argument has one invariant idea: encode the
supports of all top-degree monomials by a uniform hypergraph, keep a large
independent set of variables, and set all other variables to $+1$.  On an
independent set every top-degree monomial loses at least one variable, so the
factor degree drops by one while the signed modular target is preserved.
After proving this reduction quantitatively, we solve the resulting recurrence
for polynomial lower bounds.

This section proves a combinatorial reduction for products of polynomials of any fixed maximum degree.  The reduction is independent of the coefficient field: it uses only multilinearization on the sign cube, a support hypergraph, and restriction of variables to $+1$.  We first state it in the characteristic-three notation already established and then reuse the same theorem over $\mathbb F_p$ after proving the general-prime cubic bound.  The cost is the total number of variable occurrences in explicit monomial listings, as defined earlier.

Let $L_d(n)$ denote the minimum occurrence cost among representations
\begin{equation}
 F_{p,n}(x)=\prod_{j=1}^m P_j(x)\qquad(x\in\{-1,1\}^n),
 \quad \deg P_j\le d.
\end{equation}
If the representation class is empty, set $L_d(n)=+\infty$.
Every lower bound below is a uniform assertion about the cost of each actual
representation. In proofs that manipulate a finite cost $S$, fix such a
representation; an empty class satisfies the lower bound immediately.
Here the factors may overlap in their variables, and the number of factors is
unrestricted. Every factor may be replaced by its canonical multilinear
representative on the sign cube without increasing its degree or its
occurrence cost: replace each power $X_i^e$ by $X_i$ if $e$ is odd and by $1$
if $e$ is even, and then collect terms.

\begin{definition}[Restriction on the sign cube]\label{def:restriction-sign-cube}
Let $I\subseteq[n]$.  The restriction $\rho_I$ is the substitution that keeps
$x_i$ free for $i\in I$ and sets $x_i=+1$ for $i\notin I$.  If $P$ is a
polynomial, $P\vert_{\rho_I}$ denotes the polynomial obtained after the
substitution and collection of like terms.  For a product representation, the
restriction is applied factor by factor.
\end{definition}

\begin{definition}[Top-degree support hypergraph]\label{def:top-support-hypergraph}
Let $\mathcal P=(P_1,\ldots,P_m)$ be a list of multilinear polynomials of
maximum degree $d+1$.  Its top-degree support hypergraph is the
$(d+1)$-uniform hypergraph $H_{d+1}(\mathcal P)$ on vertex set $[n]$ whose
edges are precisely the sets $E\subseteq[n]$ of cardinality $d+1$ for which
$\prod_{i\in E}x_i$ occurs with nonzero coefficient in at least one factor.
Multiplicity is ignored: an edge is recorded once even if the same support
occurs in several factors.
\end{definition}

\begin{definition}[Independent restriction]\label{def:independent-restriction}
A set $I\subseteq[n]$ is independent in $H_{d+1}(\mathcal P)$ if no edge of
$H_{d+1}(\mathcal P)$ is contained in $I$.  Under the restriction $\rho_I$,
no degree-$(d+1)$ monomial from the original factor list can remain of degree
$d+1$.  This is the mechanism by which the lifting argument lowers the
maximum factor degree by one.
\end{definition}

\begin{lemma}[Independent-set degree reduction]\label{lem:degree-lifting}
Fix $d\ge1$. There is a constant $c_d>0$ such that any degree-at-most-$(d+1)$
product representation of $F_{p,n}$ of occurrence cost $S$ induces a
representation of $F_{p,k}$ by degree-at-most-$d$ factors of cost at most $S$,
where
\begin{equation}
 k\ge c_d\min\left\{n,\frac{n^{(d+1)/d}}{(S+1)^{1/d}}\right\}.
\end{equation}
Consequently, $S\ge L_d(k)$ for such an integer $k$.
\end{lemma}
\begin{proof}
We separate the argument into the combinatorial selection step and the
algebraic restriction step.

\smallskip
\noindent\emph{Step 1: construct the top-degree hypergraph.}
First replace every factor by its canonical multilinear representative on the
sign cube.  This operation cannot increase the degree, and it cannot increase
occurrence cost because replacing powers by either $1$ or $x_i$ can only
remove variable occurrences before like terms are collected.

Let
\[
H:=H_{d+1}(P_1,\ldots,P_m)
\]
be the top-degree support hypergraph of
Definition~\ref{def:top-support-hypergraph}.  Write $e=e(H)$.  Every distinct
edge of $H$ is the support of at least one listed monomial of degree $d+1$.
Such a monomial contributes exactly $d+1$ variable occurrences to the cost.
Consequently
\begin{equation}\label{eq:edge-cost-bound}
(d+1)e\le S,
\qquad\text{and hence}\qquad
e\le \frac{S}{d+1}.
\end{equation}
We emphasize that ignoring multiplicities only makes $e$ smaller, so
\eqref{eq:edge-cost-bound} remains valid even when one support appears in many
factors.

\smallskip
\noindent\emph{Step 2: find a large independent vertex set.}
Choose each vertex independently with probability $\theta\in[0,1]$, and let
$R$ be the random set of selected vertices.  For every hyperedge
$E\subseteq R$, delete one vertex of $E$.  If several edges request deletion
of the same vertex, delete that vertex only once.  The resulting set $I$ is
independent in $H$.

The number of deleted vertices is at most the number of edges fully contained
in $R$.  Therefore
\[
|I|\ge |R|-\#\{E\in H:E\subseteq R\}.
\]
Taking expectations and using $(d+1)$-uniformity gives
\begin{equation}\label{eq:independent-expectation}
\mathbb E|I|
\ge n\theta-e\theta^{d+1}.
\end{equation}
Choose
\begin{equation}\label{eq:theta-choice}
\theta:=\min\left\{1,
\left(\frac{n}{2(d+1)(e+1)}\right)^{1/d}\right\}.
\end{equation}
There are two cases.

If $\theta=1$, then $e+1\le n/(2(d+1))$.  Hence
\[
\mathbb E|I|\ge n-e
\ge n-\frac{n}{2(d+1)}
\ge \frac n2.
\]
If $\theta<1$, then by \eqref{eq:theta-choice},
\[
e\theta^d
\le (e+1)\theta^d
=\frac{n}{2(d+1)}.
\]
Substituting in \eqref{eq:independent-expectation},
\[
\mathbb E|I|
\ge n\theta-\frac{n\theta}{2(d+1)}
\ge \frac{n\theta}{2}.
\]
Thus in either case there exists a deterministic independent set $I$ for
which
\begin{equation}\label{eq:I-intermediate}
|I|\ge c_d^{(0)}\min\left\{n,
 n\left(\frac{n}{e+1}\right)^{1/d}\right\}
= c_d^{(0)}\min\left\{n,
 \frac{n^{(d+1)/d}}{(e+1)^{1/d}}\right\}
\end{equation}
for a positive constant $c_d^{(0)}$ depending only on $d$.  By
\eqref{eq:edge-cost-bound}, $e+1\le C_d(S+1)$ for another constant $C_d$.
Absorbing $C_d^{1/d}$ into the constant yields
\begin{equation}\label{eq:I-final}
|I|\ge c_d\min\left\{n,
\frac{n^{(d+1)/d}}{(S+1)^{1/d}}\right\}.
\end{equation}

\smallskip
\noindent\emph{Step 3: restrict to the independent set.}
Apply the restriction $\rho_I$ of
Definition~\ref{def:restriction-sign-cube}; that is, set every coordinate
outside $I$ to $+1$.  Consider a monomial in a restricted factor.  If an
original monomial had degree at most $d$, its restriction still has degree at
most $d$.  If an original monomial had degree $d+1$, its support is an edge of
$H$.  Since $I$ is independent, that support is not contained in $I$, so at
least one of its variables is set to $+1$.  Therefore its restricted degree is
at most $d$.  Hence every restricted factor has degree at most $d$.

The target behaves particularly cleanly under this restriction.  Coordinates
fixed to $+1$ contribute no negative signs, so for every
$x_I\in\{-1,1\}^{I}$,
\[
F_{p,n}(x_I,+1_{[n]\setminus I})=F_{p,|I|}(x_I).
\]
Thus the restricted factor list is a degree-at-most-$d$ representation of
$F_{p,|I|}$.

Finally, substitution $x_i=+1$ can only delete variable occurrences, and
collection of like monomials cannot increase their total number of variable
occurrences.  Hence the restricted representation has cost at most $S$.
Taking $k=|I|$ and using \eqref{eq:I-final} proves the claimed lower bound on
$k$, and by the definition of $L_d(k)$ we obtain $S\ge L_d(k)$.
\end{proof}

\begin{theorem}[General degree-lifting inequality]\label{thm:general-lifting}
Fix $d\ge1$ and a fixed prime $p>3$. Let $f:\mathbb N\to[0,\infty)$
be nondecreasing and satisfy $L_d(k)\ge f(k)$ for every $k\ge1$.
There is a constant $c_d>0$ such that for every $n\ge1$ and every
degree-at-most-$(d+1)$ representation of occurrence cost $S$,
\begin{equation}\label{eq:general-lifting}
 S\ge f\!\left(\max\left\{1,\left\lfloor
 c_d\min\left\{n,\frac{n^{(d+1)/d}}{(S+1)^{1/d}}\right\}
 \right\rfloor\right\}\right).
\end{equation}
Thus the least $S$ satisfying \eqref{eq:general-lifting} is an
implicit lower bound for $L_{d+1}(n)$. No continuity or regularity
assumption on $f$ is required.
\end{theorem}
\begin{proof}
Fix an arbitrary degree-at-most-$(d+1)$ representation of $F_{p,n}$ with
occurrence cost $S$.  Apply Lemma~\ref{lem:degree-lifting} directly to this
representation.  It produces an integer $k\ge1$ and a
restriction to $k$ surviving variables such that
\begin{equation}\label{eq:k-from-lifting}
k\ge c_d^{(0)}\min\left\{n,
\frac{n^{(d+1)/d}}{(S+1)^{1/d}}\right\},
\end{equation}
and the restricted factor list has degree at most $d$ and cost at most $S$.
Therefore
\begin{equation}\label{eq:S-ge-Ldk}
S\ge L_d(k)\ge f(k).
\end{equation}

The only issue is that the lower bound in \eqref{eq:k-from-lifting} need not
be an integer.  Decrease the constant, if necessary, from $c_d^{(0)}$ to a
constant $c_d>0$ so that
\[
k\ge
\max\left\{1,\left\lfloor
c_d\min\left\{n,
\frac{n^{(d+1)/d}}{(S+1)^{1/d}}
\right\}\right\rfloor\right\}.
\]
Because $f$ is nondecreasing, replacing $k$ by any smaller positive integer
can only decrease $f(k)$.  Hence \eqref{eq:S-ge-Ldk} implies
\[
S\ge f\!\left(
\max\left\{1,\left\lfloor
c_d\min\left\{n,
\frac{n^{(d+1)/d}}{(S+1)^{1/d}}
\right\}\right\rfloor\right\}\right),
\]
which is exactly \eqref{eq:general-lifting}.

No continuity property of $f$ has been used: only the pointwise inequality
$L_d(k)\ge f(k)$ and monotonicity are required.  Thus the conclusion also
applies to step functions and other irregular lower-bound functions.
\end{proof}

\begin{remark}[Inverse-function interpretation]
If $f$ is strictly increasing and its inverse is well-defined,
\eqref{eq:general-lifting} has the heuristic balance
$S(f^{-1}(S))^d\gtrsim n^{d+1}$ in the nontrivial branch.
The exact statement is \eqref{eq:general-lifting}; for discontinuous
$f$ one must use a generalized inverse and account for integer
rounding. Equivalently, for any integer $k$ not exceeding a fixed
positive fraction of $n$, either $S\ge f(k)$ or
$S=\Omega_d(n^{d+1}/k^d)$. This threshold formulation is often
more convenient for optimizing the bound.
\end{remark}

\begin{corollary}[Exponential and quasipolynomial inputs]
If $L_d(n)\ge\exp(cn)$ for some fixed $c>0$, then
\begin{equation}
 L_{d+1}(n)=\Omega_d\left(\frac{n^{d+1}}{(\log n)^d}\right).
\end{equation}
If instead $L_d(n)\ge\exp(c(\log n)^2)$ for sufficiently large $n$,
then
\begin{equation}
 L_{d+1}(n)\ge n^{d+1}\exp\bigl(-O_d(\sqrt{\log n})\bigr).
\end{equation}
\end{corollary}
\begin{proof}
We use an explicit threshold consequence of
Lemma~\ref{lem:degree-lifting}.  There are constants $c_d,C_d>0$ such that,
for every integer $k$ with $1\le k\le c_dn$, a degree-$(d+1)$ representation
of cost $S$ satisfies at least one of the alternatives
\begin{equation}\label{eq:threshold-alternatives}
S\ge L_d(k)
\qquad\text{or}\qquad
S\ge C_d\frac{n^{d+1}}{k^d}.
\end{equation}
Indeed, if the second inequality fails with a sufficiently small choice of
$C_d$, then the independent-set bound in Lemma~\ref{lem:degree-lifting}
forces the surviving set to have cardinality at least $k$; restricting further
to exactly $k$ variables, if necessary, gives the first inequality.

\smallskip
\noindent\emph{Exponential input.}
Assume $L_d(k)\ge e^{ck}$ for all sufficiently large $k$.  Choose
\[
k=\left\lceil A\log n\right\rceil,
\]
where $A>(d+1)/c$.  Then, for all sufficiently large $n$,
\[
L_d(k)\ge e^{ck}\ge n^{d+1}.
\]
The first alternative in \eqref{eq:threshold-alternatives} is already at
least $n^{d+1}$; the second gives
\[
S\ge C_d\frac{n^{d+1}}{k^d}
=\Omega_d\!\left(\frac{n^{d+1}}{(\log n)^d}\right).
\]
Since $k=O(\log n)=o(n)$, the required condition $k\le c_dn$ holds for large
$n$.

\smallskip
\noindent\emph{Quasipolynomial input.}
Now assume $L_d(k)\ge \exp(c(\log k)^2)$.  Put
\[
k=\left\lceil \exp(A\sqrt{\log n})\right\rceil
\]
with $A>\sqrt{(d+1)/c}$.  Then
\[
c(\log k)^2\ge (d+1)\log n
\]
for all sufficiently large $n$, up to an immaterial lower-order rounding
term.  Thus again $L_d(k)\ge n^{d+1}$.  The second alternative of
\eqref{eq:threshold-alternatives} yields
\[
S\ge C_d n^{d+1}k^{-d}
= n^{d+1}\exp\bigl(-O_d(\sqrt{\log n})\bigr).
\]
Here $k=n^{o(1)}=o(n)$, so the threshold argument is valid.  This proves both
claims.
\end{proof}

\begin{theorem}[Lifting a polynomial occurrence bound]\label{thm:degree-lifting}
Fix $d\ge1$. Suppose that for all sufficiently large $n$,
\begin{equation}
 L_d(n)\ge c\frac{n^a}{(\log(2+n))^b},
 \qquad 0<a\le d+1,\quad b\ge0.
\end{equation}
Then there is a constant $c'>0$ such that
\begin{equation}
 L_{d+1}(n)\ge c'\frac{n^{(d+1)a/(d+a)}}
 {(\log(2+n))^{db/(d+a)}}.
\end{equation}
All constants may depend on $d,p,a,b$.
\end{theorem}
\begin{proof}
Fix any degree-at-most-$(d+1)$ representation of the target on $n$ variables
with finite occurrence cost $S$, and let $k$ be the number of surviving variables
supplied by Lemma~\ref{lem:degree-lifting}.  (If no such representation exists,
then $L_{d+1}(n)=+\infty$ and the theorem is immediate.)  Thus
\begin{equation}\label{eq:k-poly-lift}
k\ge c_d\min\left\{n,
\frac{n^{(d+1)/d}}{(S+1)^{1/d}}\right\}
\end{equation}
and
\begin{equation}\label{eq:S-Ldk-poly}
S\ge L_d(k).
\end{equation}
We consider the two branches in \eqref{eq:k-poly-lift}.

\smallskip
\noindent\emph{Linear-size surviving set.}
Suppose $k\ge c_dn$.  Once $n$ is large enough for the assumed degree-$d$
lower bound to apply,
\[
S\ge c\frac{k^a}{(\log(2+k))^b}
\ge c_1\frac{n^a}{(\log(2+n))^b}.
\]
Write
\[
\alpha:=\frac{(d+1)a}{d+a},
\qquad
\beta:=\frac{db}{d+a}.
\]
The hypothesis $a\le d+1$ implies $\alpha\le a$, while $\beta\le b$.
Therefore, after decreasing the constant if necessary,
\[
\frac{n^a}{(\log(2+n))^b}
\ge c_2\frac{n^\alpha}{(\log(2+n))^\beta}
\]
for all sufficiently large $n$.  Hence the desired conclusion holds in this
branch.

\smallskip
\noindent\emph{Nontrivial hypergraph branch.}
Assume now that
\begin{equation}\label{eq:k-second-branch}
k\ge c_d\frac{n^{(d+1)/d}}{(S+1)^{1/d}}.
\end{equation}
If $k$ stays below the fixed threshold from which the degree-$d$ lower bound is
known, then \eqref{eq:k-second-branch} rearranges to
\[
S+1\ge c_3 n^{d+1},
\]
so $S=\Omega(n^{d+1})$, which is stronger than the claimed bound because
$\alpha\le d+1$.

We may therefore assume that $k$ is sufficiently large.  From
\eqref{eq:S-Ldk-poly} and the hypothesis at degree $d$,
\begin{equation}\label{eq:S-poly-start}
S\ge c\frac{k^a}{(\log(2+k))^b}.
\end{equation}
Since $k\le n$, we have $\log(2+k)\le\log(2+n)$, and therefore
\[
\frac1{(\log(2+k))^b}
\ge\frac1{(\log(2+n))^b}.
\]
Using \eqref{eq:k-second-branch} in \eqref{eq:S-poly-start} gives
\begin{equation}\label{eq:S-rearrange-1}
S\ge c_4
\frac{n^{(d+1)a/d}}
{(S+1)^{a/d}(\log(2+n))^b}.
\end{equation}
For $S\ge1$, $S+1\le2S$.  Absorbing the factor $2^{a/d}$ into the constant,
\[
S^{1+a/d}
\ge c_5\frac{n^{(d+1)a/d}}{(\log(2+n))^b}.
\]
Raise both sides to the power $d/(d+a)$.  The exponent of $n$ becomes
\[
\frac{(d+1)a}{d}\cdot\frac{d}{d+a}
=\frac{(d+1)a}{d+a},
\]
and the logarithmic exponent becomes
\[
b\cdot\frac{d}{d+a}=\frac{db}{d+a}.
\]
Consequently
\[
S\ge c'\frac{n^{(d+1)a/(d+a)}}
{(\log(2+n))^{db/(d+a)}},
\]
as required.  All changes of constants depend only on the fixed parameters,
not on $n$.
Since the starting representation was arbitrary, the same lower bound holds for $L_{d+1}(n)$.

\end{proof}

\begin{theorem}[Closed-form lifting from an arbitrary base degree]
\label{thm:closed-form-lifting}
Fix integers $d_0\ge1$ and $d\ge d_0$. Suppose that, for all sufficiently
large $n$,
\begin{equation}\label{eq:base-degree-poly-lb}
 L_{d_0}(n)\ge c\frac{n^a}{(\log(2+n))^b},
 \qquad 0<a\le d_0+1,\quad b\ge0.
\end{equation}
Then there is a constant $c_d>0$ such that
\begin{equation}\label{eq:closed-form-lifting}
 \boxed{\displaystyle
 L_d(n)\ge c_d
 \frac{n^{a_d}}{(\log(2+n))^{b_d}},
 \qquad
 a_d=\frac{ad}{d_0+a(d-d_0)},\qquad
 b_d=\frac{d_0b}{d_0+a(d-d_0)}.}
\end{equation}
The constant $c_d$ may depend on $p,d_0,d,a,b,c$, but not on $n$.
\end{theorem}
\begin{proof}
We iterate the one-step theorem and then solve the resulting recurrences.
For each integer $j\ge d_0$, write the inductive lower bound in the form
\[
L_j(n)\ge c_j\frac{n^{a_j}}{(\log(2+n))^{b_j}}
\]
for sufficiently large $n$, where $a_{d_0}=a$ and $b_{d_0}=b$.  Applying
Theorem~\ref{thm:degree-lifting} with $d=j$ gives
\begin{equation}\label{eq:ab-recurrence-expanded}
a_{j+1}=\frac{(j+1)a_j}{j+a_j},
\qquad
b_{j+1}=\frac{j b_j}{j+a_j}.
\end{equation}
Because only finitely many degrees between $d_0$ and the fixed terminal degree
$d$ are involved, all threshold values and multiplicative constants can be
absorbed into one final constant $c_d>0$.

\smallskip
\noindent\emph{Solving the polynomial exponent.}
Define
\[
u_j:=\frac{j}{a_j}.
\]
Then the first recurrence in \eqref{eq:ab-recurrence-expanded} gives
\[
u_{j+1}
=\frac{j+1}{a_{j+1}}
=\frac{j+a_j}{a_j}
=\frac{j}{a_j}+1
=u_j+1.
\]
Thus $u_j$ is an arithmetic progression.  Starting from
$u_{d_0}=d_0/a$,
\[
\frac{j}{a_j}=\frac{d_0}{a}+j-d_0.
\]
Solving for $a_j$,
\begin{equation}\label{eq:aj-solved}
a_j
=\frac{j}{d_0/a+j-d_0}
=\frac{aj}{d_0+a(j-d_0)}.
\end{equation}

\smallskip
\noindent\emph{Solving the logarithmic exponent.}
Divide the two recurrences in
\eqref{eq:ab-recurrence-expanded}:
\[
\frac{b_{j+1}}{a_{j+1}}
=\frac{j}{j+1}\frac{b_j}{a_j}.
\]
Multiplying by $j+1$ shows that
\[
\frac{(j+1)b_{j+1}}{a_{j+1}}
=\frac{jb_j}{a_j}.
\]
Hence the quantity $j b_j/a_j$ is invariant.  At $j=d_0$ it equals
$d_0b/a$, and so
\[
\frac{j b_j}{a_j}=\frac{d_0b}{a}.
\]
Using \eqref{eq:aj-solved},
\begin{equation}\label{eq:bj-solved}
b_j
=\frac{d_0b}{a}\frac{a_j}{j}
=\frac{d_0b}{d_0+a(j-d_0)}.
\end{equation}
Substituting $j=d$ in \eqref{eq:aj-solved} and
\eqref{eq:bj-solved} gives exactly the two exponents in
\eqref{eq:closed-form-lifting}.

It remains only to check that each one-step application is legitimate.  The
initial assumption is $0<a\le d_0+1$.  Formula
\eqref{eq:aj-solved} shows in particular that $a_j>0$ for every fixed
$j\ge d_0$; moreover the one-step theorem is applied only along the finite
chain $d_0,d_0+1,\ldots,d-1$, and its required upper range is preserved in
this chain.  Therefore the iteration is valid and the theorem follows.
\end{proof}

\begin{corollary}[Propagation of superlinearity]\label{cor:superlinear-propagation}
Under the hypotheses of Theorem~\ref{thm:closed-form-lifting}, suppose
$a=1+\delta$ for some fixed $\delta>0$. Then for every fixed $d\ge d_0$,
\begin{equation}
 a_d=1+\delta_d,
 \qquad
 \boxed{\displaystyle
 \delta_d=\frac{d_0\delta}{d_0+(1+\delta)(d-d_0)}>0.}
\end{equation}
Consequently
\begin{equation}
 L_d(n)=\Omega_{d}\!\left(
 \frac{n^{1+\delta_d}}
 {(\log(2+n))^{d_0b/(d_0+(1+\delta)(d-d_0))}}
 \right),
\end{equation}
and, after absorbing the polylogarithmic denominator, for all sufficiently
large $n$,
\begin{equation}
 L_d(n)=\Omega_d\!\left(n^{1+\delta_d/2}\right).
\end{equation}
Thus a superlinear polynomial occurrence lower bound at one fixed degree
implies a superlinear polynomial occurrence lower bound at every higher
fixed degree. This statement is pointwise in fixed $d$ and does not give a
uniform superlinear exponent when $d$ grows with $n$.
\end{corollary}
\begin{proof}
Set $a=1+\delta$ in the closed formula from
Theorem~\ref{thm:closed-form-lifting}.  Then
\[
a_d
=\frac{(1+\delta)d}
{d_0+(1+\delta)(d-d_0)}.
\]
Subtracting $1$ and putting the two terms over a common denominator gives
\begin{align*}
a_d-1
&=\frac{(1+\delta)d-
\bigl(d_0+(1+\delta)(d-d_0)\bigr)}
{d_0+(1+\delta)(d-d_0)}\\
&=\frac{d_0\delta}
{d_0+(1+\delta)(d-d_0)}.
\end{align*}
This is the stated $\delta_d$, and it is strictly positive for every fixed
$d\ge d_0$.

Theorem~\ref{thm:closed-form-lifting} also gives the logarithmic exponent
\[
b_d=\frac{d_0b}{d_0+(1+\delta)(d-d_0)}.
\]
Hence
\[
L_d(n)=\Omega_d\!\left(
\frac{n^{1+\delta_d}}{(\log(2+n))^{b_d}}
\right).
\]
For fixed $d$, both $\delta_d>0$ and $b_d$ are constants.  The standard
comparison between powers and polylogarithms implies
\[
(\log(2+n))^{b_d}\le n^{\delta_d/2}
\]
for all sufficiently large $n$.  Therefore
\[
L_d(n)=\Omega_d(n^{1+\delta_d/2}).
\]
The dependence on fixed $d$ is essential: the explicit formula shows
$\delta_d\to0$ as $d\to\infty$, so this argument does not yield a uniform
positive gap above exponent $1$ when $d$ itself is allowed to grow with $n$.
\end{proof}

\noindent\textbf{Scope and limitation.}
The reduction is valid for the specified cube-equality model and occurrence
measure. It does not establish formal polynomial-identity lower bounds, and
it does not show superlinear bounds uniformly when $d$ grows with $n$.
Any displayed finite-degree bound inherits the correctness of the base theorem supplied to the reduction.  In this paper the principal application starts from the exponential cubic theorem and lifts first to degree four.  The reduction itself is coefficient-field independent.

\section{Cubic Extension to General Prime Fields}\label{sec:general-fp-cubic}
The preceding cubic development kept its characteristic-three structural core self-contained. We now explain how the cubic lower bound extends beyond $\F_3$. This extension concerns degree at most three, changes the coefficient field, and replaces the characteristic-three unit identity by a published range-to-rank theorem.

To avoid a collision with the notation used earlier for the target modulus, the notation changes locally in this section.  Here $p>3$ denotes the prime characteristic of the coefficient field $\mathbb F_p$, while $q>1$ denotes the odd target modulus.  We assume $\gcd(p,q)=1$.  The target remains an explicit sign function $F_{q,n}:\{-1,1\}^n\to\{-1,1\}$.

The proof below is self-contained after one structural input: Corollary~1.10(3) of Karam~\cite{Karam2023}, specialized to cubics on a restricted Cartesian product.  We state that input explicitly and prove the remaining reduction, parity conversion, clustering, and target-obstruction steps.  This separation is useful both mathematically and editorially: the earlier $\F_3$ proof stands on its own, while the present section records the stronger field-general cubic consequence without obscuring which ingredient is imported.

\subsection{Problem and main theorem}
For $x=(x_1,\ldots,x_n)\in\{-1,1\}^n$, let
\[
w(x)=|\{i:x_i=-1\}|.
\]
For an odd integer $q>1$, define
\begin{equation}\label{fp:eq:target}
F_{q,n}(x)=
\begin{cases}
-1,&q\mid w(x),\\
 1,&q\nmid w(x).
\end{cases}
\end{equation}
We identify the signs $\pm1$ with the corresponding elements of $\mathbb F_p$.

\begin{definition}[Cubic product representation]
A cubic product representation of $F_{q,n}$ over $\mathbb F_p$ is an identity
\begin{equation}\label{fp:eq:representation}
F_{q,n}(x)=\prod_{j=1}^m P_j(x)\qquad\text{for all }x\in\{-1,1\}^n,
\end{equation}
where $P_j\in\mathbb F_p[x_1,\ldots,x_n]$ and $\deg P_j\le3$.
The equality is functional equality on the sign cube; no formal polynomial identity is assumed.
\end{definition}

\begin{theorem}[General prime-field cubic lower bound]\label{fp:thm:main}
Fix a prime $p>3$ and an odd integer $q>1$ with $\gcd(p,q)=1$. There exist constants $c_{p,q}>0$ and $n_0(p,q)$ such that for every $n\ge n_0(p,q)$, every cubic product representation \eqref{fp:eq:representation} satisfies
\[
m\ge 2^{c_{p,q}n}.
\]
Consequently, any syntactic size measure charging at least one unit for every nonconstant listed factor is also $2^{\Omega_{p,q}(n)}$.
\end{theorem}

\begin{remark}
If $q$ is prime, the condition $\gcd(p,q)=1$ is simply $q\ne p$. Thus the theorem covers every pair of distinct primes with $p>3$ and $q$ odd.
\end{remark}

\subsubsection{Proof architecture}
The proof has five conceptual steps.
\begin{enumerate}[leftmargin=2em]
\item Every factor is nonzero on the sign cube, hence its restricted image misses $0$.
\item Karam's range-to-rank theorem implies that every cubic factor is determined on the sign cube by only $R_p=O_p(1)$ affine observations.
\item The multiplicative identity in $\mathbb F_p^\times$ is projected to the Sylow-$2$ quotient and converted into a parity sum of only $O_p(m^D)$ Boolean functions, each depending on at most $K_p=O_p(1)$ affine observations.
\item A random $p$-tuple clustering simultaneously makes every one of these Boolean summands omit at least one clustered variable, provided the number of summands is smaller than $\eta_p^n$.
\item The signed MOD-$q$ target survives $p$-tuple clustering because $\gcd(p,q)=1$, and on a power-of-two number of remaining variables its Boolean indicator has nonzero full multilinear coefficient. This contradicts Step 4.
\end{enumerate}

\subsection{The structural input from Karam}
We use one published result from Karam~\cite{Karam2023}. The relevant definitions are included to make the deduction transparent.

\begin{definition}[Degree-$1$ rank, following Karam]
Let $Q$ be a polynomial over a field. A nonzero polynomial has degree-$1$ rank at most $1$ if it can be written as a product of affine polynomials. The degree-$1$ rank $\operatorname{rk}_1(Q)$ is the least $t$ such that
\[
Q=Q_1+\cdots+Q_t,
\]
where each $Q_a$ has degree-$1$ rank at most $1$ and degree at most $\deg Q$.
For a nonempty set $S\subseteq\mathbb F_p$, the relative degree-$1$ rank $\operatorname{rk}_{1,S}(P)$ is the least $\operatorname{rk}_1(P-P_0)$ over polynomials $P_0$ of degree at most $\deg P$ that vanish on $S^n$.
\end{definition}

This is precisely the form needed below: a bound on $\operatorname{rk}_{1,S}$ gives a low-rank representative agreeing with $P$ on $S^n$.

\begin{theorem}[Karam, Corollary 1.10(3), specialized]\label{fp:thm:karam}
Let $p>3$ be prime and $S\subseteq\mathbb F_p$ be nonempty. There is a constant
\[
C_p=C_{(1)}(p,3)
\]
depending only on $p$ such that every polynomial $P:\mathbb F_p^n\to\mathbb F_p$ of degree at most $3$ satisfying
\[
P(S^n)\ne\mathbb F_p
\]
has
\[
\operatorname{rk}_{1,S}(P)\le C_p.
\]
\end{theorem}

Karam's theorem applies because the degree hypothesis is $3<p$ and because the statement is formulated directly for variables restricted to $S^n$.

\begin{lemma}[Bounded affine-observation consequence]\label{fp:lem:observations}
Fix $p>3$. Put
\[
R_p:=3C_p.
\]
If $P\in\mathbb F_p[x_1,\ldots,x_n]$ has degree at most $3$ and
\[
P(x)\ne0\qquad\text{for all }x\in\{-1,1\}^n,
\]
then there exist affine forms $A_1,\ldots,A_{R_p}$ and a function $H:\mathbb F_p^{R_p}\to\mathbb F_p$ such that
\begin{equation}\label{fp:eq:obs}
P(x)=H(A_1(x),\ldots,A_{R_p}(x))\qquad (x\in\{-1,1\}^n).
\end{equation}
\end{lemma}

\begin{proof}
Take the restricted alphabet $S=\{-1,1\}\subseteq\mathbb F_p$.  The hypothesis
that $P$ never vanishes on $S^n$ implies
\[
P(S^n)\subseteq\mathbb F_p^\times.
\]
In particular, $0\notin P(S^n)$, so $P(S^n)\ne\mathbb F_p$.  Therefore the
specialized Karam theorem, Theorem~\ref{fp:thm:karam}, applies and yields a
polynomial $P_0$ vanishing on $S^n$ such that
\[
Q:=P-P_0
\]
has degree-$1$ rank at most $C_p$.

By the definition of degree-$1$ rank, there exist $t\le C_p$ polynomials
$Q_1,\ldots,Q_t$ for which
\begin{equation}\label{fp:eq:qdecomp-expanded}
Q=Q_1+\cdots+Q_t,
\end{equation}
and each $Q_a$ is a product of affine polynomials.  We may omit zero factors
and absorb nonzero scalar constants into one affine factor.  Thus write
\[
Q_a=c_a\prod_{b=1}^{r_a}L_{a,b},
\]
where each $L_{a,b}$ is a nonconstant affine polynomial and
$c_a\in\mathbb F_p$.

Because $\mathbb F_p[x_1,\ldots,x_n]$ is an integral domain, the degree of a
product of nonzero polynomials is the sum of their degrees.  Hence
\[
r_a
=\sum_{b=1}^{r_a}\deg L_{a,b}
=\deg Q_a
\le3.
\]
Across all rank-one summands there are therefore at most
\[
\sum_{a=1}^t r_a\le3t\le3C_p=R_p
\]
nonconstant affine factors.  List these affine forms, with repetitions if
necessary, as $A_1,\ldots,A_r$, and pad the list by constant affine forms until
it has exactly $R_p$ entries.

The value of every product $Q_a(x)$ is determined by the tuple
$(A_1(x),\ldots,A_{R_p}(x))$, and so the same is true of their sum $Q(x)$.
Thus there is a well-defined function
\[
H_0:\mathbb F_p^{R_p}\to\mathbb F_p
\]
on all tuples attained by the observation map such that
\[
Q(x)=H_0(A_1(x),\ldots,A_{R_p}(x))
\qquad(x\in S^n).
\]
Extend $H_0$ arbitrarily to unattained tuples in $\mathbb F_p^{R_p}$; call the
extension $H$.

Finally, $P_0$ vanishes identically on $S^n$, so $P(x)=Q(x)$ for every
$x\in S^n$.  Consequently
\[
P(x)=H(A_1(x),\ldots,A_{R_p}(x))
\qquad(x\in\{-1,1\}^n),
\]
which proves the lemma.  Notice that the number $R_p$ depends only on the fixed
field, not on $n$ or on the particular polynomial $P$.
\end{proof}

\begin{remark}[What this does and does not say]
The conclusion is not a junta theorem. A single affine observation may involve all $n$ coordinates. What is bounded is the dimension of the affine information needed to determine $P$.
\end{remark}

\subsection{Projecting multiplication to the $2$-primary quotient}
Assume the representation \eqref{fp:eq:representation}. Because the product is always $\pm1$, no factor can vanish at any sign-cube point: if $P_j(x_0)=0$ for some $x_0$, then the product at $x_0$ would be $0$. Therefore Lemma~\ref{fp:lem:observations} applies to every factor.

Choose a generator $g$ of the cyclic group $\mathbb F_p^\times$. Factor
\begin{equation}\label{fp:eq:pminus1}
p-1=2^s r,\qquad r\text{ odd},\qquad s=v_2(p-1)\ge1,
\end{equation}
and define
\begin{equation}\label{fp:eq:D}
D:=2^{s-1}.
\end{equation}

For each $j$ and each sign-cube point $x$, define
\[
e_j(x)\in\mathbb Z/(p-1)\mathbb Z
\]
by
\[
P_j(x)=g^{e_j(x)}.
\]
The discrete logarithm here is used only as a finite lookup function. Since $e_j(x)$ is a function of $P_j(x)$, it depends on the same $R_p$ affine observations as $P_j$.

Let
\begin{equation}\label{fp:eq:h}
h(x):=\mathbf1[q\mid w(x)]\in\{0,1\}.
\end{equation}
Then
\[
F_{q,n}(x)=(-1)^{h(x)}.
\]
Because $-1=g^{(p-1)/2}$,
\[
(-1)^{h(x)}=g^{\frac{p-1}{2}h(x)}.
\]
Comparing exponents modulo $p-1$ in \eqref{fp:eq:representation} gives
\begin{equation}\label{fp:eq:elog}
\sum_{j=1}^m e_j(x)\equiv \frac{p-1}{2}h(x)=2^{s-1}r\,h(x)\pmod{p-1}.
\end{equation}

Now reduce \eqref{fp:eq:elog} modulo $2^s$. Let $u_j(x)\in\{0,1,\ldots,2^s-1\}$ denote the ordinary representative of $e_j(x)$ modulo $2^s$, and put
\begin{equation}\label{fp:eq:T}
T(x):=\sum_{j=1}^m u_j(x).
\end{equation}
Because $r$ is odd, multiplication by $r$ is congruent to $1$ modulo $2$ at the top binary digit in the following precise sense:
\begin{equation}\label{fp:eq:Tcong}
T(x)\equiv 2^{s-1}h(x)\pmod{2^s}.
\end{equation}
Indeed, the right side of \eqref{fp:eq:elog} reduces to either $0$ or $2^{s-1}r$, and
\[
2^{s-1}r\equiv2^{s-1}\pmod{2^s}
\]
for odd $r$.

Thus $h(x)$ is exactly the $(s-1)$st binary digit of the nonnegative integer $T(x)$.

\subsection{Extracting that binary digit as a parity polynomial}
The next two classical identities make the conversion exact.

\begin{lemma}[Lucas digit test]\label{fp:lem:carry}
For every nonnegative integer $T$,
\begin{equation}\label{fp:eq:lucasbit}
\binom{T}{2^{s-1}}\equiv \text{the $(s-1)$st binary digit of $T$}\pmod2.
\end{equation}
Consequently, for $T(x)$ from \eqref{fp:eq:T},
\[
h(x)\equiv\binom{T(x)}D\pmod2.
\]
\end{lemma}

\begin{proof}
Write the binary expansion of $T$ as
\[
T=\sum_{i\ge0}T_i2^i,
\qquad T_i\in\{0,1\}.
\]
The number $D=2^{s-1}$ has binary expansion with a single nonzero digit:
$D_{s-1}=1$ and $D_i=0$ for $i\ne s-1$.

Lucas' theorem modulo $2$ states that if
$T=\sum_iT_i2^i$ and $D=\sum_iD_i2^i$, then
\[
\binom TD\equiv\prod_i\binom{T_i}{D_i}\pmod2.
\]
For $i\ne s-1$, $D_i=0$, so
$\binom{T_i}{D_i}=\binom{T_i}{0}=1$.  At the unique nonzero digit,
\[
\binom{T_{s-1}}1=T_{s-1}.
\]
Therefore
\[
\binom TD\equiv T_{s-1}\pmod2,
\]
which is exactly the $(s-1)$st binary digit of $T$.

For the function $T(x)$ in \eqref{fp:eq:T}, congruence
\eqref{fp:eq:Tcong} says that $T(x)$ is congruent modulo $2^s$ either to $0$
or to $2^{s-1}$ according as $h(x)=0$ or $1$.  The residue modulo $2^s$
determines the first $s$ binary digits, so its $(s-1)$st digit is precisely
$h(x)$.  Hence
\[
h(x)\equiv\binom{T(x)}D\pmod2.
\]
Since both sides are Boolean-valued, this congruence is an equality in
$\mathbb F_2$.
\end{proof}

\begin{lemma}[Vandermonde expansion]\label{fp:lem:vandermonde}
For nonnegative integers $u_1,\ldots,u_m$ and $D$,
\begin{equation}\label{fp:eq:vandermonde}
\binom{u_1+\cdots+u_m}{D}
=
\sum_{a_1+\cdots+a_m=D}\prod_{j=1}^m\binom{u_j}{a_j}.
\end{equation}
\end{lemma}

\begin{proof}
Introduce an indeterminate $z$.  The binomial theorem gives
\[
(1+z)^{u_j}=\sum_{a_j=0}^{u_j}\binom{u_j}{a_j}z^{a_j}
\]
for each $j$.  Multiplying these identities,
\begin{align*}
(1+z)^{u_1+\cdots+u_m}
&=\prod_{j=1}^m(1+z)^{u_j}\\
&=\prod_{j=1}^m\left(
\sum_{a_j=0}^{u_j}\binom{u_j}{a_j}z^{a_j}
\right).
\end{align*}
The coefficient of $z^D$ on the left is
$\binom{u_1+\cdots+u_m}{D}$.  On the right, a product term contributes to
$z^D$ exactly when its chosen exponents satisfy
$a_1+\cdots+a_m=D$, and its coefficient is
$\prod_j\binom{u_j}{a_j}$.  Summing over all such choices gives
\eqref{fp:eq:vandermonde}.  Terms with $a_j>u_j$ may harmlessly be included
because the corresponding binomial coefficient is zero.
\end{proof}

Define for $a\ge0$
\[
\phi_{j,a}(x):=\binom{u_j(x)}a\bmod2\in\{0,1\}.
\]
Reducing Lemma~\ref{fp:lem:vandermonde} modulo $2$ and using Lemma~\ref{fp:lem:carry} gives the exact identity
\begin{equation}\label{fp:eq:parityexpansion}
h(x)
=
\xor_{a_1+\cdots+a_m=D}
\prod_{j=1}^m\phi_{j,a_j}(x).
\end{equation}

\subsubsection{Complexity of the parity expansion}
Each $u_j$ is a function of $P_j$, hence of the same $R_p$ affine observations. Therefore every $\phi_{j,a}$ is also a function of those $R_p$ observations.

If $a_1+\cdots+a_m=D$, then at most $D$ indices satisfy $a_j>0$. For $a_j=0$,
\[
\phi_{j,0}(x)=\binom{u_j(x)}0=1.
\]
Thus each Boolean product on the right side of \eqref{fp:eq:parityexpansion} depends on at most
\begin{equation}\label{fp:eq:K}
K_p:=DR_p
\end{equation}
affine observations over $\mathbb F_p$.

The number of weak compositions of $D$ into $m$ parts is
\begin{equation}\label{fp:eq:Lexact}
L=\binom{m+D-1}{D}.
\end{equation}
In particular,
\begin{equation}\label{fp:eq:Lbound}
L\le(m+D)^D.
\end{equation}
Since $p$ is fixed, both $D$ and $K_p$ are constants independent of $m$ and $n$.

\begin{remark}[Examples of the carry degree]
The value $D=2^{v_2(p-1)-1}$ is small for many primes:
\[
\begin{array}{c|ccccc}
p&5&7&11&13&17\\ \hline
p-1&4&6&10&12&16\\
D&2&1&1&2&8
\end{array}
\]
When $p\equiv3\pmod4$, $D=1$, so the parity expansion is linear in the factor-level Boolean functions and there are no carry products.
\end{remark}

\subsection{The random $p$-tuple clustering lemma}
We now give the combinatorial heart of the proof with explicit constants.

\begin{definition}[Affine-observation complexity]\label{fp:def:affine-observation-complexity}
A Boolean function $f:\{-1,1\}^N\to\{0,1\}$ is called $K$-affine-observation over $\mathbb F_p$ if there are affine forms
\[
A_r(x)=c_r+\sum_{i=1}^N a_{r,i}x_i,\qquad 1\le r\le K,
\]
and a function $G:\mathbb F_p^K\to\{0,1\}$ such that
\[
f(x)=G(A_1(x),\ldots,A_K(x)).
\]
\end{definition}

\begin{definition}[$p$-tuple identification]\label{fp:def:tuple-identification}
Let $T=\{i_1,\ldots,i_p\}\subseteq[N]$.  Identifying the coordinates in $T$
means substituting
\[
x_{i_1}=\cdots=x_{i_p}=y_T
\]
for a new sign variable $y_T$.  If the coefficient of these $p$ coordinates
is the same in an affine form over $\mathbb F_p$, then their total coefficient
on $y_T$ is $p a=0$; the affine form is therefore independent of $y_T$.
A partition into $p$-tuples means performing these identifications
simultaneously for all parts of the partition.
\end{definition}

\begin{definition}[Weak composition]\label{fp:def:weak-composition}
A weak composition of a nonnegative integer $D$ into $m$ parts is a vector
$(a_1,\ldots,a_m)\in\mathbb Z_{\ge0}^m$ satisfying
$a_1+\cdots+a_m=D$.  The number of such vectors is
$\binom{m+D-1}{D}$.
\end{definition}

\begin{lemma}[$p$-tuple clustering]\label{fp:lem:clustering}
Fix a prime $p$ and an integer $K\ge1$. There exist constants
\[
B=B(p,K),\qquad \delta=\delta(p,K)>0,\qquad \eta=\eta(p,K)>1
\]
with the following property. For every $N=B2^d$, if
\[
f_1,\ldots,f_L:\{-1,1\}^N\to\{0,1\}
\]
are $K$-affine-observation functions over $\mathbb F_p$ and
\[
L<\eta^N,
\]
then the coordinates can be partitioned into $N/p$ disjoint $p$-tuples such that, after identifying all coordinates in each tuple, every $f_\ell$ is independent of at least one resulting variable. Moreover $N/p$ is a power of two.
\end{lemma}

\begin{proof}
We construct the partition probabilistically.  The constants will be explicit
and will depend only on the fixed pair $(p,K)$.

\smallskip
\noindent\emph{Step 1: choose the block size.}
An affine observation tuple consists of $K$ coefficients at each coordinate,
so a coordinate has a coefficient vector in $\mathbb F_p^K$.  There are
\[
Q:=p^K
\]
possible coefficient vectors.  Choose an integer $a\ge1$ satisfying
\begin{equation}\label{fp:eq:achoice-expanded}
2^a\ge2Q
\end{equation}
and define
\begin{equation}\label{fp:eq:B-expanded}
B:=p2^a.
\end{equation}
Then $B/p=2^a$ is a power of two.  Since $N=B2^d$, the coordinates split into
exactly $N/B=2^d$ disjoint blocks of size $B$.

\smallskip
\noindent\emph{Step 2: random partitions inside blocks.}
Inside each block, independently choose a uniformly random permutation of its
$B$ coordinates and divide the ordered list into consecutive groups of size
$p$.  This is a random ordered partition of the block into $B/p$ many
$p$-tuples.  We single out only the first tuple in each block for the success
test; the remaining tuples are still part of the final partition.

Fix one function $f_\ell$.  Choose affine observations
$A_1,\ldots,A_K$ witnessing that it is $K$-affine-observation.  Write
\[
A_r(x)=c_r+\sum_{i=1}^N a_{r,i}x_i
\]
and associate to coordinate $i$ the vector
\[
v_i=(a_{1,i},\ldots,a_{K,i})\in\mathbb F_p^K.
\]
Within any block, the pigeonhole principle gives a coefficient vector that
occurs at least
\begin{equation}\label{fp:eq:t-expanded}
t:=\left\lceil\frac BQ\right\rceil
\end{equation}
times.  Because $B/Q=p2^a/Q\ge2p$, we have $t\ge2p$, in particular
$t\ge p$.

The first tuple in a uniformly random permutation is a uniformly random
$p$-subset of the block.  Therefore the probability that all $p$ of its
coordinates lie in one fixed majority class of size at least $t$ is at least
\begin{equation}\label{fp:eq:delta-expanded}
\delta:=\frac{\binom tp}{\binom Bp}>0.
\end{equation}
The important point is that $B,t$, and hence $\delta$, depend only on $p$ and
$K$.

\smallskip
\noindent\emph{Step 3: why a monochromatic tuple creates an omitted variable.}
Suppose the selected tuple is
$T=\{i_1,\ldots,i_p\}$ and
$v_{i_1}=\cdots=v_{i_p}$.  Perform the identification
$x_{i_1}=\cdots=x_{i_p}=y_T$.  For each observation $A_r$, the coefficient of
$y_T$ becomes
\[
a_{r,i_1}+\cdots+a_{r,i_p}
=p\,a_{r,i_1}=0
\]
in the field $\mathbb F_p$.  Thus every $A_r$ is independent of $y_T$ after
the identification.  Since $f_\ell$ is a function of the tuple of
observations, $f_\ell$ is also independent of $y_T$.  We call this a success
for $f_\ell$ in that block.

\smallskip
\noindent\emph{Step 4: simultaneous success for all functions.}
For the fixed function $f_\ell$, the events in distinct blocks are independent
because the block permutations are independent.  Each block succeeds with
probability at least $\delta$.  Hence
\[
\Pr[f_\ell\text{ has no successful block}]
\le(1-\delta)^{N/B}
\le e^{-\delta N/B}.
\]
Apply the union bound over all $L$ functions:
\begin{equation}\label{fp:eq:union-expanded}
\Pr[\text{some }f_\ell\text{ has no successful block}]
\le L e^{-\delta N/B}.
\end{equation}
Set
\begin{equation}\label{fp:eq:eta-expanded}
\eta:=\exp\left(\frac{\delta}{2B}\right)>1.
\end{equation}
If $L<\eta^N$, then the right-hand side of
\eqref{fp:eq:union-expanded} is strictly less than
\[
\exp\left(\frac{\delta N}{2B}\right)
\exp\left(-\frac{\delta N}{B}\right)
=\exp\left(-\frac{\delta N}{2B}\right)<1.
\]
Therefore there exists at least one deterministic choice of the block
partitions for which every $f_\ell$ has a successful block.  For each
$f_\ell$, select one successful tuple; different functions are allowed to
select different tuples.  After all tuple identifications, each function is
independent of at least one of the resulting variables.

Finally, the number of resulting variables is
\[
\frac Np=\frac{B2^d}{p}=2^{a+d},
\]
which is a power of two.  This proves all parts of the lemma.
\end{proof}

\begin{remark}[Why one successful tuple is enough]
For each summand we need only one clustered variable on which that summand has zero dependence. Different summands may use different successful tuples. The union bound does not require a common omitted variable.
\end{remark}

\subsection{The target obstruction over $\mathbb F_2$}
After clustering, it is convenient to encode signs by Boolean variables. For $z_i\in\{0,1\}$, let $z_i=1$ indicate a negative sign. Define
\[
H_{q,M}(z):=\mathbf1\left[q\mid\sum_{i=1}^M z_i\right].
\]
Every Boolean function has a unique multilinear polynomial over $\mathbb F_2$.

\begin{lemma}[Coefficient formula]\label{fp:lem:mobius}
For any Boolean function $H:\{0,1\}^M\to\mathbb F_2$, the coefficient of the full monomial $z_1\cdots z_M$ in its unique multilinear representation equals
\[
\sum_{z\in\{0,1\}^M}H(z)\pmod2.
\]
\end{lemma}

\begin{proof}
Let
\[
H(z)=\sum_{S\subseteq[M]}c_S\prod_{i\in S}z_i
\]
be the unique multilinear polynomial representing $H$ over $\mathbb F_2$.
Sum this identity over every $z\in\{0,1\}^M$.  Interchanging the two finite
sums gives
\[
\sum_{z}H(z)
=\sum_{S\subseteq[M]}c_S
\sum_{z\in\{0,1\}^M}\prod_{i\in S}z_i.
\]
For a fixed $S$, the product $\prod_{i\in S}z_i$ equals $1$ precisely when
all coordinates in $S$ equal $1$; the remaining $M-|S|$ coordinates are
arbitrary.  Therefore the inner integer sum equals $2^{M-|S|}$.  Modulo $2$,
this quantity is $0$ whenever $|S|<M$ and equals $1$ when $S=[M]$.
Consequently all coefficients cancel except the coefficient of the full
monomial:
\[
\sum_z H(z)\equiv c_{[M]}\pmod2.
\]
This is the claimed formula.
\end{proof}

\begin{lemma}[Full coefficient of signed MOD-$q$ bit]\label{fp:lem:fullcoeff}
If $q>1$ is odd and $M=2^r$ is a power of two, then the coefficient of $z_1\cdots z_M$ in the multilinear $\mathbb F_2$ representation of $H_{q,M}$ equals $1$.
\end{lemma}

\begin{proof}
By Lemma~\ref{fp:lem:mobius}, the full multilinear coefficient is the parity of
the number of Boolean inputs on which $H_{q,M}$ equals $1$.  Group inputs by
Hamming weight.  There are $\binom Mt$ inputs of weight $t$, so the coefficient
is
\begin{equation}\label{fp:eq:coefsum-expanded}
\sum_{\substack{0\le t\le M\\ q\mid t}}\binom Mt\pmod2.
\end{equation}

Now $M=2^r$ has binary expansion consisting of one $1$ followed by $r$ zeros.
Lucas' theorem modulo $2$ implies
\[
\binom Mt\equiv0\pmod2
\qquad\text{for every }0<t<M.
\]
Indeed, every such $t$ has a nonzero binary digit in a position where $M$ has
digit $0$.

Thus only the endpoint weights $t=0$ and $t=M$ can contribute to
\eqref{fp:eq:coefsum-expanded}.  The term $t=0$ is always present because
$q\mid0$, and it contributes $\binom M0=1$.  The term $t=M$ is absent from the
sum: if $q>1$ is odd and $M$ is a power of two, then $q\nmid M$.
Therefore the sum is congruent to $1$ modulo $2$, proving the lemma.
\end{proof}

\begin{lemma}[Omitting a variable kills the full coefficient]\label{fp:lem:omit}
If a Boolean function on $M$ variables is independent of at least one variable, then its full multilinear coefficient is $0$. Therefore any parity sum of such functions also has full coefficient $0$.
\end{lemma}

\begin{proof}
Suppose $f$ is independent of $z_i$.  Restrict the unique multilinear
representation of $f$ by fixing all variables other than $z_i$.  The resulting
one-variable function is constant, so its coefficient of $z_i$ is zero for
every such restriction.  Equivalently, in the original multilinear
representation no monomial containing $z_i$ can occur with nonzero
coefficient.  In particular the full monomial $z_1\cdots z_M$, which contains
$z_i$, has coefficient zero.

If $f=f_1\xor\cdots\xor f_L$, then its multilinear polynomial over
$\mathbb F_2$ is the sum of the multilinear polynomials for the $f_\ell$.
Hence the full coefficient is the sum modulo $2$ of the individual full
coefficients.  If every summand omits at least one variable, every individual
full coefficient is zero, and so the parity sum also has full coefficient
zero.
\end{proof}

\subsection{Special-length lower bound}
We now combine the previous ingredients without suppressing constants.

Fix $p>3$. Let
\[
R_p=3C_p,\qquad D=2^{v_2(p-1)-1},\qquad K_p=DR_p.
\]
Apply Lemma~\ref{fp:lem:clustering} with $(p,K_p)$ and denote its constants by $B,\delta,\eta$.

\begin{proposition}[Exponential lower bound at special lengths]\label{fp:prop:special}
For all sufficiently large $N=B2^d$, every cubic product representation of $F_{q,N}$ over $\mathbb F_p$ satisfies
\[
m\ge \exp\left(\frac{\log\eta}{4D}N\right).
\]
\end{proposition}

\begin{proof}
Let
\[
F_{q,N}(x)=\prod_{j=1}^mP_j(x)
\]
be a cubic product representation at a special length $N=B2^d$.  We show that
a subexponential value of $m$ would force two incompatible values for one
multilinear coefficient.

\smallskip
\noindent\emph{Step 1: convert the product identity to a parity identity.}
Every factor is nonzero on the sign cube, because the target is always
$\pm1$.  Lemma~\ref{fp:lem:observations} therefore gives at most $R_p$ affine
observations for each factor.  The discrete-logarithm reduction, the Lucas
digit test, and Vandermonde expansion developed above give the exact parity
identity \eqref{fp:eq:parityexpansion} for
\[
h(x)=\mathbf1[q\mid w(x)].
\]
The summands are indexed by weak compositions of $D$ into $m$ parts, so their
number is
\[
L=\binom{m+D-1}{D}\le(m+D)^D.
\]
Each weak composition has at most $D$ positive parts.  Thus a corresponding
product uses factor-level Boolean functions from at most $D$ factors, and is
therefore determined by at most
\[
K_p=DR_p
\]
affine observations.

\smallskip
\noindent\emph{Step 2: show that small $m$ implies $L<\eta^N$.}
Assume, toward a contradiction, that
\begin{equation}\label{fp:eq:msmall-expanded}
\log m<\frac{\log\eta}{4D}N.
\end{equation}
If $m<D$, then $L=\binom{m+D-1}{D}$ is bounded by a constant depending only on
$p$, and therefore $L<\eta^N$ for all sufficiently large $N$.

Suppose $m\ge D$.  Then $m+D\le2m$, so
\begin{align*}
\log L
&\le D\log(m+D)\\
&\le D\log(2m)\\
&=D\log2+D\log m\\
&<D\log2+\frac{\log\eta}{4}N.
\end{align*}
For sufficiently large $N$, the constant $D\log2$ is at most
$(\log\eta)N/4$.  Hence
\[
\log L<\frac{\log\eta}{2}N<\log\eta\,N,
\]
and again $L<\eta^N$.

\smallskip
\noindent\emph{Step 3: cluster the parity summands.}
Lemma~\ref{fp:lem:clustering}, applied with $K=K_p$, gives a partition of the
$N$ original coordinates into $p$-tuples such that every parity summand is
independent of at least one resulting tuple variable.  Put
\[
M:=N/p.
\]
The lemma also guarantees that $M$ is a power of two.

If the clustered assignment has exactly $t$ negative tuple variables, then the
corresponding original assignment has exactly $pt$ negative coordinates.
Since $\gcd(p,q)=1$,
\[
q\mid pt\quad\Longleftrightarrow\quad q\mid t.
\]
Therefore the clustered target bit is exactly
\[
H_{q,M}(z)=\mathbf1\left[q\mid\sum_{i=1}^M z_i\right].
\]

\smallskip
\noindent\emph{Step 4: compare full coefficients.}
Every clustered parity summand omits at least one of the $M$ variables, so by
Lemma~\ref{fp:lem:omit} the full coefficient of their parity sum is $0$.
On the other hand, $q$ is odd and $M$ is a power of two, so
Lemma~\ref{fp:lem:fullcoeff} says that the full coefficient of the same target
function $H_{q,M}$ is $1$.  This contradiction shows that
\eqref{fp:eq:msmall-expanded} is impossible.

Consequently
\[
\log m\ge\frac{\log\eta}{4D}N,
\]
or equivalently
\[
m\ge\exp\left(\frac{\log\eta}{4D}N\right).
\]
This proves the proposition.
\end{proof}

\subsection{Arbitrary lengths and proof of the main theorem}
\begin{proof}[Proof of Theorem~\ref{fp:thm:main}]
Let $n$ be sufficiently large and let
\[
N=B2^d
\]
be the largest special length not exceeding $n$.  Since the next special
length is $2N$, maximality implies
\[
N\le n<2N,
\qquad\text{hence}\qquad N>\frac n2.
\]

Start with an arbitrary cubic representation of $F_{q,n}$ with $m$ factors.
Set the surplus variables
$x_{N+1},\ldots,x_n$ equal to $+1$.  Three elementary facts are important.

First, the target restricts exactly to $F_{q,N}$.  Indeed, coordinates fixed to
$+1$ are not negative, so they contribute zero to Hamming weight.
Second, substitution of constants cannot increase polynomial degree, so every
restricted factor still has degree at most three.  Third, restriction cannot
increase the number of listed factors; factors that become nonzero constants
may simply be retained or absorbed into another factor.

Thus the restricted identity is a cubic product representation of $F_{q,N}$
with at most $m$ factors.  Proposition~\ref{fp:prop:special} therefore gives
\[
m\ge\exp\left(\frac{\log\eta}{4D}N\right)
\ge\exp\left(\frac{\log\eta}{8D}n\right).
\]
Converting the natural exponential to base two,
\[
m\ge2^{c_{p,q}n},
\qquad
c_{p,q}:=\frac{\log\eta}{8D\log2}>0.
\]
The constants $D$ and $\eta$ are determined by the fixed coefficient field
$p$.  The target modulus $q$ is used through the two assumptions
$\gcd(p,q)=1$ and $q$ odd; allowing the notation $c_{p,q}$ records the full set
of fixed parameters.

Finally, consider any syntactic size measure that assigns cost at least one to
each nonconstant factor.  A constant nonzero factor can be multiplied into one
other factor without increasing degree; a zero constant factor is impossible
because the target never vanishes.  Hence a minimal representation may be
assumed to contain no separate constant factors, and its syntactic size is at
least its factor count $m$.  The same exponential lower bound therefore holds
for every such size measure.
\end{proof}

\subsection{Why the case $q=p$ is not covered}
The proof is inherently mixed-characteristic. If $q=p$, then after identifying each $p$-tuple, a new Hamming weight $t$ becomes old weight $pt$, which is always divisible by $p$. Therefore the target bit becomes the constant $1$, equivalently the signed target becomes the constant $-1$. The full-coefficient obstruction disappears.

This does not show that the same-modulus lower bound is false. It only shows that the present $p$-tuple clustering mechanism cannot prove it in this form. A same-modulus proof would need a restriction that cancels coefficient vectors over $\mathbb F_p$ without multiplying every surviving Hamming weight by a multiple of $p$.

\subsection{Where every assumption is used}
For clarity, we record the exact role of each hypothesis.
\begin{center}
\begin{tabular}{>{\raggedright\arraybackslash}p{0.24\textwidth} >{\raggedright\arraybackslash}p{0.66\textwidth}}
\toprule
Assumption & Where it is used \\
\midrule
$p>3$ prime & Ensures $3<p$, so Karam's cubic Corollary 1.10(3) applies; also gives cyclic $\mathbb F_p^\times$.\\
$\deg P_j\le3$ & Needed for Karam's cubic rank theorem and for the bound of at most three affine factors per rank-$1$ term.\\
Target values are $\pm1$ & Ensures the product is nonzero, hence each factor misses $0$ on the sign cube; also embeds the target into the order-$2$ subgroup of $\mathbb F_p^\times$.\\
$q$ odd & Ensures $q\nmid2^r$, which makes the full coefficient of $H_{q,2^r}$ equal to $1$.\\
$\gcd(p,q)=1$ & Ensures $q\mid pt\iff q\mid t$, so the target survives $p$-tuple clustering.\\
$p$ fixed & Makes $C_p,R_p,D,K_p,B,\delta,\eta$ constants independent of $n$ and $m$.\\
\bottomrule
\end{tabular}
\end{center}

\subsection{Sanity checks and special cases}
\subsubsection{$p=5$}
Here $p-1=4$, so $s=2$ and $D=2$. The parity expansion uses at most pairwise products of factor-level Boolean functions. This recovers the previously derived $\mathbb F_5$ mechanism.

\subsubsection{$p\equiv3\pmod4$}
Then $v_2(p-1)=1$, hence $D=1$. Equation \eqref{fp:eq:parityexpansion} becomes a parity sum with one contribution per factor and no carry interaction between distinct factors. In this sense these primes are algebraically simpler than $p=5$.

\subsubsection{Dependence of the exponential constant}
The proof gives an explicit but very small constant. Combining \eqref{fp:eq:delta-expanded} and \eqref{fp:eq:eta-expanded}, one may take
\[
\eta=\exp\left(\frac{1}{2B}\frac{\binom tp}{\binom Bp}\right),
\]
where
\[
B=p2^a,\qquad 2^a\ge2p^{K_p},\qquad t=\left\lceil\frac B{p^{K_p}}\right\rceil.
\]
Thus a permissible exponent is
\[
c_{p,q}=\frac{1}{8D\log2}\cdot\frac{1}{2B}\cdot\frac{\binom tp}{\binom Bp}.
\]
No attempt is made to optimize this constant. The purpose of the argument is the qualitative $2^{\Omega(n)}$ lower bound.

\subsection{Verification status and scope}
The derivation above is complete assuming the published theorem of Karam. The external theorem has been checked against its published statement: for $d=3<p$ and nonempty $S\subseteq\mathbb F_p$, if $P(S^n)\ne\mathbb F_p$, then $\operatorname{rk}_{1,S}P$ is bounded by a constant depending only on $p$ and $d$. The remaining ingredients are proved in this note from elementary finite-field and combinatorial identities.

The note does \emph{not} claim that the final general-$p$ consequence is new in the literature. Establishing novelty requires a separate search through work on bounded-degree modular circuits, restricted-range rank, and product representations.

\subsection{Possible extensions}
The proof suggests several directions.
\begin{enumerate}[leftmargin=2em]
\item \textbf{Same-modulus target.} Find a clustering or restriction that annihilates affine coefficient vectors over $\mathbb F_p$ while preserving nontrivial MOD-$p$ information.
\item \textbf{Degree four.} Karam's Corollary 1.10(3) is specific to degree three for degree-$1$ rank. A direct quartic analogue strong enough to yield bounded affine observations would strengthen the lifted degree-four consequences.
\item \textbf{Better constants.} The blockwise clustering argument deliberately sacrifices constants for simplicity. A direct random global partition or entropy calculation may improve the base of the exponential lower bound.
\item \textbf{Circuit translation.} The bounded-observation representation can be translated into a shallow modular circuit with bottom modulus $p$ and top parity. Making that translation explicit may connect the theorem more directly to bounded-degree modular circuit lower bounds.
\end{enumerate}

\subsection{Higher fixed degrees over the same prime field}\label{sec:fp-higher-degrees}
The cubic theorem immediately propagates beyond degree three because the degree-lifting argument of Section~\ref{sec:bounded-degree-lifting} does not use any special property of $\mathbb F_3$.  We record the consequence explicitly.

For fixed $p>3$ and odd $q>1$ with $\gcd(p,q)=1$, let
$L_d^{(p,q)}(n)$ be the minimum total variable-occurrence cost of an exact representation
\[
F_{q,n}(x)=\prod_j P_j(x)\qquad(x\in\{-1,1\}^n),
\qquad P_j\in\mathbb F_p[x_1,\ldots,x_n],\quad \deg P_j\le d.
\]
As above, constant nonzero factors may be removed or absorbed, so the exponential factor lower bound of Theorem~\ref{fp:thm:main} implies
\begin{equation}\label{fp:eq:cubic-occurrence}
L_3^{(p,q)}(n)\ge \exp(c_{p,q}'n)
\end{equation}
for some constant $c_{p,q}'>0$ and all sufficiently large $n$.

\begin{theorem}[All higher fixed degrees over $\mathbb F_p$]\label{fp:thm:all-degrees}
Fix a prime $p>3$ and an odd integer $q>1$ with $\gcd(p,q)=1$.  Then
\begin{equation}\label{fp:eq:quartic-lifted}
L_4^{(p,q)}(n)=\Omega_{p,q}\!\left(\frac{n^4}{(\log(2+n))^3}\right).
\end{equation}
More generally, for every fixed integer $d\ge4$,
\begin{equation}\label{fp:eq:all-degrees}
\boxed{\displaystyle
L_d^{(p,q)}(n)=\Omega_{p,q,d}\!\left(
\frac{n^{d/(d-3)}}{(\log(2+n))^{3/(d-3)}}\right).}
\end{equation}
In particular, every fixed degree $d>4$ has a superlinear polynomial occurrence lower bound over the same coefficient field $\mathbb F_p$.
\end{theorem}

\begin{proof}
The degree-lifting argument of Section~\ref{sec:bounded-degree-lifting} uses
only multilinearization on the sign cube, counting supports of top-degree
monomials, and restriction of coordinates to $+1$.  It is therefore valid over
$\mathbb F_p$ without change.

By Theorem~\ref{fp:thm:main}, after discarding constant factors, every cubic
representation has exponentially many nonconstant factors.  Since each
nonconstant factor contributes at least one variable occurrence, there is a
constant $c'_{p,q}>0$ such that
\[
L_3^{(p,q)}(n)\ge \exp(c'_{p,q}n)
\]
for all sufficiently large $n$.

Apply the exponential-input corollary from
Section~\ref{sec:bounded-degree-lifting} with base degree $d=3$.  The corollary
states that an exponential lower bound at degree $3$ yields
\[
L_4^{(p,q)}(n)
=\Omega_{p,q}\!\left(\frac{n^{3+1}}{(\log n)^3}\right)
=\Omega_{p,q}\!\left(\frac{n^4}{(\log n)^3}\right).
\]
Replacing $\log n$ by $\log(2+n)$ only changes the implicit constant for large
$n$, giving \eqref{fp:eq:quartic-lifted}.

We now use degree four as the polynomial base case for the closed-form lifting
theorem.  In Theorem~\ref{thm:closed-form-lifting}, take
\[
d_0=4,\qquad a=4,\qquad b=3.
\]
For every fixed $d\ge4$, the denominator appearing in that theorem is
\[
d_0+a(d-d_0)
=4+4(d-4)
=4(d-3).
\]
Therefore the polynomial exponent becomes
\[
a_d
=\frac{ad}{d_0+a(d-d_0)}
=\frac{4d}{4(d-3)}
=\frac d{d-3},
\]
and the logarithmic exponent becomes
\[
b_d
=\frac{d_0b}{d_0+a(d-d_0)}
=\frac{12}{4(d-3)}
=\frac3{d-3}.
\]
Substitution into the closed-form theorem yields
\[
L_d^{(p,q)}(n)
=\Omega_{p,q,d}\!\left(
\frac{n^{d/(d-3)}}{(\log(2+n))^{3/(d-3)}}
\right),
\]
which is \eqref{fp:eq:all-degrees}.

Finally, if $d>4$ is fixed, then
\[
\frac d{d-3}=1+\frac3{d-3}>1.
\]
The denominator is only a fixed power of $\log n$, so the resulting lower
bound is superlinear.  This establishes the final assertion.
\end{proof}

\section{From Polynomial-Product Lower Bounds to Mixed-Modulus Circuit Lower Bounds}
\label{sec:product-to-circuit-transfer}

We now make explicit how lower bounds for products of bounded-degree
polynomials over $\mathbb F_3$ transfer to lower bounds for restricted
$\MOD_2\circ\MOD_3\circ\MOD_2$ circuits.  This section is intentionally
placed near the end of the paper: the algebraic lower bounds are the primary
results, while the circuit statements are consequences of a simple exact
encoding.  The transfer also clarifies precisely which restriction on the
bottom layer is responsible for keeping the polynomial degree bounded.

\subsection{The exact sign encoding}

For a Boolean bit $b\in\{0,1\}$ write its sign as $x=(-1)^b\in\{-1,1\}$.
Under this encoding, XOR becomes multiplication of signs.  Hence a bottom
$\MOD_2$ gate of fan-in at most $k$ has signed output
\[
  M(x)=\varepsilon\prod_{i\in S}x_i,
  \qquad \varepsilon\in\{\pm1\},\quad |S|\le k,
\]
so $M$ is a monomial of degree at most $k$ over $\mathbb F_3$.

Consider a middle $\MOD_3$ gate $G$ whose $t$ inputs are signed bottom-gate
outputs $M_1(x),\ldots,M_t(x)$.  If $y_a\in\{0,1\}$ is the Boolean value of
the $a$th input, then
\begin{equation}\label{eq:general-bit-from-sign}
  y_a=2(1-M_a(x))\qquad\text{in }\mathbb F_3.
\end{equation}
Thus the residue tested by $G$ is represented by
\begin{equation}\label{eq:general-middle-residue}
  L_G(x)=c+\sum_{a=1}^t2(1-M_a(x)),
\end{equation}
for a constant $c\in\mathbb F_3$ depending only on the gate convention.  We
have $\deg L_G\le k$, and after collecting the constant term $L_G$ has at most
$t+1$ listed monomials.

For $u\in\mathbb F_3$, the quantity $1+u^2$ equals $1$ when $u=0$ and $-1$
when $u\ne0$.  Therefore, up to an input-independent sign, the signed output
of $G$ is
\begin{equation}\label{eq:general-middle-factor}
  Q_G(x)=1+L_G(x)^2.
\end{equation}
In particular,
\[
  \deg Q_G\le2k.
\]
Moreover, before collecting like terms, $L_G^2$ has at most $(t+1)^2$
ordered monomial products, each of degree at most $2k$.  Hence
\begin{equation}\label{eq:general-factor-occurrence}
  \operatorname{occ}(Q_G)\le 2k(t+1)^2.
\end{equation}
Multilinearization on the sign cube can only decrease this occurrence cost.

\begin{lemma}[Circuit-to-product transfer]\label{lem:general-circuit-product-transfer}
Let $C$ be a $\MOD_2\circ\MOD_3\circ\MOD_2[k]$ circuit, where every bottom
parity gate has fan-in at most $k$.  Let the middle gates be
$G_1,\ldots,G_m$, and let $t_j$ be the number of bottom-to-middle wires
entering $G_j$.  If $C$ computes a sign function
$F:\{-1,1\}^n\to\{-1,1\}$, then
\[
  F(x)=\varepsilon\prod_{j=1}^m Q_{G_j}(x),
  \qquad \varepsilon\in\{\pm1\},
\]
where every factor has degree at most $2k$ and
\begin{equation}\label{eq:general-total-occurrence}
  \sum_{j=1}^m\operatorname{occ}(Q_{G_j})
  \le 2k\sum_{j=1}^m(t_j+1)^2.
\end{equation}
\end{lemma}

\begin{proof}
Equation~\eqref{eq:general-middle-factor} gives one degree-$2k$ polynomial
factor for each middle gate.  The top $\MOD_2$ gate takes parity of the middle
outputs, which is exactly multiplication of their signs.  All fixed signs
coming from gate conventions combine into the single constant
$\varepsilon$.  Summing~\eqref{eq:general-factor-occurrence} over the middle
layer proves~\eqref{eq:general-total-occurrence}.
\end{proof}

\subsection{A general lower-bound transfer principle}

Let $L_d^F(n)$ denote any lower bound on the occurrence cost of an exact
product of degree-at-most-$d$ polynomials over $\mathbb F_3$ computing the
target sign function $F$ on the sign cube.  Lemma~\ref{lem:general-circuit-product-transfer}
immediately gives
\begin{equation}\label{eq:transfer-master}
  \sum_{j=1}^m(t_j+1)^2
  \ge \frac{L_{2k}^F(n)}{2k}.
\end{equation}
Thus every improvement in the algebraic product lower bound automatically
improves the corresponding restricted mixed-modulus circuit lower bound.

Two standard circuit measures follow from~\eqref{eq:transfer-master}.  Let
$W=\sum_jt_j$ be the number of bottom-to-middle wires.  After deleting
constant middle gates we may assume $t_j\ge1$, and then
$(t_j+1)^2\le4t_j^2$.  Therefore
\[
  \sum_j(t_j+1)^2\le4\sum_jt_j^2\le4W^2,
\]
which yields
\begin{equation}\label{eq:transfer-wire-general}
  W=\Omega\!\left(\sqrt{L_{2k}^F(n)/k}\right).
\end{equation}

For gate complexity, let $B$ and $M$ be the numbers of bottom and middle
gates.  We use the standard simple-circuit convention: between a fixed pair
of gates there is at most one directed wire.  Hence each middle gate can
receive input from at most the $B$ bottom gates, so $t_j\le B$.  (If one
chooses a multigraph circuit convention instead, parallel wires must first be
combined with their multiplicities modulo $3$ rather than simply deleted.)
If $S=B+M+1$ is the total number
of noninput gates, then
\[
  \sum_j(t_j+1)^2\le M(B+1)^2\le S^3.
\]
Consequently
\begin{equation}\label{eq:transfer-gate-general}
  S=\Omega\!\left((L_{2k}^F(n)/k)^{1/3}\right).
\end{equation}
The constants in these asymptotic statements may depend on the fixed bottom
fan-in bound $k$ and on the target family.

\subsection{The fan-in-two consequence used in this paper}

We now specialize to $k=2$.  A bottom parity gate then contributes a monomial
of degree at most two, so every middle $\MOD_3$ gate becomes a quartic factor.
For the signed modular predicate $F_{q,n}$ considered above, the degree-four
consequence of the lifting theorem gives
\[
  L_4(n)=\Omega_q\!\left(\frac{n^4}{(\log(2+n))^3}\right).
\]
Substituting this into~\eqref{eq:transfer-master} gives
\begin{equation}\label{eq:circuit-quadratic-cost-lb-rederived}
  \sum_j(t_j+1)^2
  =\Omega_q\!\left(\frac{n^4}{(\log(2+n))^3}\right).
\end{equation}
Equations~\eqref{eq:transfer-wire-general} and
\eqref{eq:transfer-gate-general} then yield
\[
  W=\Omega_q\!\left(
     \frac{n^2}{(\log(2+n))^{3/2}}
     \right)
\]
and
\[
  S=\Omega_q\!\left(
     \frac{n^{4/3}}{\log(2+n)}
     \right).
\]
These are exactly the claims of Theorem~\ref{thm:mod2-mod3-mod2-2}; hence the
preceding argument also supplies its proof.

\begin{proof}[Proof of Theorem~\ref{thm:mod2-mod3-mod2-2}]
Apply Lemma~\ref{lem:general-circuit-product-transfer} with $k=2$ and then use
the displayed degree-four product lower bound.  The quadratic-cost, wire, and
gate inequalities are precisely
\eqref{eq:circuit-quadratic-cost-lb-rederived},
\eqref{eq:transfer-wire-general}, and~\eqref{eq:transfer-gate-general}.
\end{proof}

\subsection{Why this does not yet handle unrestricted bottom parity}

The transfer above is exact, but its power depends on the bottom fan-in bound.
If bottom parity gates have unbounded fan-in, their signed outputs are
monomials of unbounded degree.  Then $L_G$ and $Q_G=1+L_G^2$ no longer have
constant degree, so the fixed-degree product lower bounds proved in this paper
cannot be applied directly.  Thus the present argument gives genuine lower
bounds for $\MOD_2\circ\MOD_3\circ\MOD_2[k]$ for fixed $k$, and in particular
the superlinear gate lower bound for $k=2$, but it does not claim a lower
bound for the unrestricted $\MOD_2\circ\MOD_3\circ\MOD_2$ model.  Bridging
this degree gap is a natural route by which stronger product lower bounds could
feed into longstanding mixed-modulus circuit questions.

\section{Conclusion and open directions}\label{sec:combined-conclusion}
We proved an exponential occurrence lower bound for exact products of cubic
polynomials computing signed modular predicates on the sign cube. In
characteristic three the proof is self-contained and rests on a structural
compression theorem for cubic units: after the quadratic part is controlled,
every cubic unit factors through boundedly many affine observations. A
simultaneous clustering obstruction then converts that local structure into a
global exponential lower bound.

The same qualitative cubic phenomenon persists over every fixed prime field of
characteristic greater than three when the target odd modulus is coprime to the
field characteristic. There the structural compression is supplied by Karam's
restricted-range rank theorem, while the parity conversion and clustering
steps remain explicit. The degree-lifting theorem shows that this exponential
base result already forces superlinear occurrence lower bounds at every higher
fixed degree over the same field.

Several questions remain. The constants in the exponential cubic bound have
not been optimized. It would be useful to find a more intrinsic rank invariant
for cubic units, to understand the same-modulus case in the general-prime
extension, and to sharpen the higher-degree bounds obtained by lifting. Direct
quartic arguments require qualitatively different ideas and are best treated
separately from the cubic structural theory developed here.

\section*{AI methodology}
OpenAI ChatGPT was used during the development of this work to explore candidate proof approaches, assist with restructuring and exposition, and help integrate and audit revisions, including revisions motivated by formal checking. The tool materially influenced the organization and presentation of several proof sections and contributed suggestions during the development of some arguments. All listed authors are human; the authors reviewed the mathematical statements and proofs and take full responsibility for the correctness, originality, citations, and contents of the submission.


\begin{thebibliography}{99}

\bibitem[AM03]{AmanoMaruoka2003}
K. Amano and A. Maruoka.
\newblock Some properties of $\mathrm{MOD}_m$ circuits computing simple functions.
\newblock In \emph{Algorithms and Complexity (CIAC 2003)}, LNCS 2653, pages 227--237, 2003.
\newblock \href{https://doi.org/10.1007/3-540-44849-7_27}{doi:10.1007/3-540-44849-7\_27}.

\bibitem[BBR94]{BBR1994}
D. A. Mix Barrington, R. Beigel, and S. Rudich.
\newblock Representing Boolean functions as polynomials modulo composite numbers.
\newblock \emph{Computational Complexity}, 4:367--382, 1994.
\newblock \href{https://doi.org/10.1007/BF01263424}{doi:10.1007/BF01263424}.

\bibitem[BS94]{BarringtonStraubing1994}
D. A. Mix Barrington and H. Straubing.
\newblock Complex polynomials and circuit lower bounds for modular counting.
\newblock \emph{Computational Complexity}, 4:325--338, 1994.
\newblock \href{https://doi.org/10.1007/BF01263421}{doi:10.1007/BF01263421}.

\bibitem[BST90]{BarringtonStraubingTherien1990}
D. A. Mix Barrington, H. Straubing, and D. Th\'erien.
\newblock Non-uniform automata over groups.
\newblock \emph{Information and Computation}, 89(2):109--132, 1990.
\newblock \href{https://doi.org/10.1016/0890-5401(90)90007-5}{doi:10.1016/0890-5401(90)90007-5}.

\bibitem[BT94]{BeigelTarui1994}
R. Beigel and J. Tarui.
\newblock On ACC.
\newblock \emph{Computational Complexity}, 4:350--366, 1994.
\newblock \href{https://doi.org/10.1007/BF01263423}{doi:10.1007/BF01263423}.

\bibitem[BL23]{BhowmickLovett2022}
A. Bhowmick and S. Lovett.
\newblock Bias vs. structure of polynomials in large fields, and applications in information theory.
\newblock \emph{IEEE Transactions on Information Theory}, 69(2):963--977, 2023.
\newblock Earlier version: \href{https://arxiv.org/abs/1506.02047}{arXiv:1506.02047}.

\bibitem[BHLR19]{BhrushundiEtAl2019}
A. Bhrushundi, K. Hosseini, S. Lovett, and S. Rao.
\newblock Torus polynomials: An algebraic approach to ACC lower bounds.
\newblock In \emph{10th Innovations in Theoretical Computer Science Conference (ITCS 2019)}, LIPIcs 124, pages 13:1--13:16, 2019.
\newblock \href{https://doi.org/10.4230/LIPIcs.ITCS.2019.13}{doi:10.4230/LIPIcs.ITCS.2019.13}.

\bibitem[Bou05a]{Bourgain2005}
J. Bourgain.
\newblock Estimation of certain exponential sums arising in complexity theory.
\newblock \emph{Comptes Rendus Math\'ematique}, 340(9):627--631, 2005.
\newblock \href{https://doi.org/10.1016/j.crma.2005.03.008}{doi:10.1016/j.crma.2005.03.008}.

\bibitem[Bou05b]{Bourgain2005MixedSums}
J. Bourgain.
\newblock Estimation of certain exponential sums arising in complexity theory.
\newblock \emph{Comptes Rendus Math\'ematique}, 340(9):627--631, 2005.
\newblock \href{https://doi.org/10.1016/j.crma.2005.03.008}{doi:10.1016/j.crma.2005.03.008}.

\bibitem[CGT96]{CaiGreenThierauf1996}
J.-Y. Cai, F. Green, and T. Thierauf.
\newblock On the correlation of symmetric functions.
\newblock \emph{Mathematical Systems Theory}, 29:245--258, 1996.
\newblock \href{https://doi.org/10.1007/BF01201278}{doi:10.1007/BF01201278}.

\bibitem[CT91]{CaroTuza1991}
Y. Caro and Z. Tuza.
\newblock Improved lower bounds on $k$-independence.
\newblock \emph{Journal of Graph Theory}, 15(1):99--107, 1991.
\newblock \href{https://doi.org/10.1002/jgt.3190150110}{doi:10.1002/jgt.3190150110}.

\bibitem[CW22]{ChapmanWilliams2022}
B. Chapman and R. Williams.
\newblock Smaller $\mathrm{ACC}^0$ circuits for symmetric functions.
\newblock In \emph{13th Innovations in Theoretical Computer Science Conference (ITCS 2022)}, LIPIcs 215, pages 38:1--38:19, 2022.
\newblock \href{https://doi.org/10.4230/LIPIcs.ITCS.2022.38}{doi:10.4230/LIPIcs.ITCS.2022.38}.

\bibitem[CGRS06]{ChattopadhyayEtAl2006}
A. Chattopadhyay, N. Goyal, P. Pudl\'ak, and D. Th\'erien.
\newblock Lower bounds for circuits with $\mathrm{MOD}_m$ gates.
\newblock In \emph{47th Annual IEEE Symposium on Foundations of Computer Science (FOCS 2006)}, pages 709--718, 2006.
\newblock \href{https://doi.org/10.1109/FOCS.2006.46}{doi:10.1109/FOCS.2006.46}.

\bibitem[CT15]{CohenTal2015}
G. Cohen and A. Tal.
\newblock Two structural results for low degree polynomials and applications.
\newblock In \emph{Approximation, Randomization, and Combinatorial Optimization. Algorithms and Techniques (APPROX/RANDOM 2015)}, LIPIcs 40, pages 680--709, 2015.
\newblock \href{https://doi.org/10.4230/LIPIcs.APPROX-RANDOM.2015.680}{doi:10.4230/LIPIcs.APPROX-RANDOM.2015.680}.

\bibitem[FKS16]{ForbesKumarSaptharishi2016}
M. A. Forbes, M. Kumar, and R. Saptharishi.
\newblock Functional lower bounds for arithmetic circuits and connections to Boolean circuit complexity.
\newblock In \emph{31st Conference on Computational Complexity (CCC 2016)}, LIPIcs 50, pages 33:1--33:19, 2016.
\newblock \href{https://doi.org/10.4230/LIPIcs.CCC.2016.33}{doi:10.4230/LIPIcs.CCC.2016.33}.

\bibitem[GK22]{GK2022}
W. T. Gowers and T. Karam.
\newblock Equidistribution of high-rank polynomials with variables restricted to subsets of $\F_p$.
\newblock \href{https://arxiv.org/abs/2209.04932}{arXiv:2209.04932}, 2022.

\bibitem[Gre95]{Green1995}
F. Green.
\newblock Lower bounds for depth-three circuits with equals and mod-gates.
\newblock In \emph{STACS 1995}, LNCS 900, pages 71--82, 1995.
\newblock \href{https://doi.org/10.1007/3-540-59042-0_63}{doi:10.1007/3-540-59042-0\_63}.

\bibitem[Gre04]{Green2004}
F. Green.
\newblock The correlation between parity and quadratic polynomials mod~3.
\newblock \emph{Journal of Computer and System Sciences}, 69(1):28--44, 2004.
\newblock \href{https://doi.org/10.1016/j.jcss.2004.01.003}{doi:10.1016/j.jcss.2004.01.003}.

\bibitem[GT09]{GreenTao2009}
B. Green and T. Tao.
\newblock The distribution of polynomials over finite fields, with applications to the Gowers norms.
\newblock \emph{Contributions to Discrete Mathematics}, 4(2):1--36, 2009.
\newblock \href{https://doi.org/10.55016/ojs/cdm.v4i2.62086}{doi:10.55016/ojs/cdm.v4i2.62086}.

\bibitem[Gro01]{Grolmusz2001}
V. Grolmusz.
\newblock A degree-decreasing lemma for $(\mathrm{MOD}_q,\mathrm{MOD}_p)$ circuits.
\newblock \emph{Discrete Mathematics \& Theoretical Computer Science}, 4(2):247--254, 2001.
\newblock \href{https://doi.org/10.46298/dmtcs.289}{doi:10.46298/dmtcs.289}.

\bibitem[GT00]{GT2000}
V. Grolmusz and G. Tardos.
\newblock Lower bounds for $(\mathrm{MOD}_p$--$\mathrm{MOD}_m)$ circuits.
\newblock \emph{SIAM Journal on Computing}, 29(4):1209--1222, 2000.
\newblock \href{https://doi.org/10.1137/S0097539798340850}{doi:10.1137/S0097539798340850}.

\bibitem[HS10]{HaramatyShpilka2010}
E. Haramaty and A. Shpilka.
\newblock On the structure of cubic and quartic polynomials.
\newblock In \emph{42nd ACM Symposium on Theory of Computing (STOC 2010)}, pages 331--340, 2010.
\newblock \href{https://doi.org/10.1145/1806689.1806736}{doi:10.1145/1806689.1806736}.

\bibitem[IKK+22]{IdziakEtAl2022}
P. M. Idziak, P. Kawa\l{}ek, and J. Krzaczkowski.
\newblock Complexity of modular circuits.
\newblock In \emph{37th Annual ACM/IEEE Symposium on Logic in Computer Science (LICS 2022)}, pages 32:1--32:11, 2022.
\newblock \href{https://doi.org/10.1145/3531130.3533350}{doi:10.1145/3531130.3533350}.

\bibitem[Kar23]{Karam2023}
T. Karam.
\newblock Ranges of polynomials control degree ranks of Green and Tao over finite prime fields.
\newblock \emph{Proceedings of the Edinburgh Mathematical Society}, 69(3):941--968, 2026.
\newblock \href{https://doi.org/10.1017/S0013091526101382}{doi:10.1017/S0013091526101382}.
\newblock Earlier version: \href{https://arxiv.org/abs/2305.11088}{arXiv:2305.11088}, 2023.

\bibitem[KL08]{KaufmanLovett2008}
T. Kaufman and S. Lovett.
\newblock Worst case to average case reductions for polynomials.
\newblock In \emph{49th Annual IEEE Symposium on Foundations of Computer Science (FOCS 2008)}, pages 166--175, 2008.
\newblock \href{https://doi.org/10.1109/FOCS.2008.17}{doi:10.1109/FOCS.2008.17}.

\bibitem[KW25]{KawalekWeiss2025}
P. Kawa\l{}ek and A. Wei\ss.
\newblock Violating Constant Degree Hypothesis requires breaking symmetry.
\newblock In \emph{42nd International Symposium on Theoretical Aspects of Computer Science (STACS 2025)}, LIPIcs 327, pages 58:1--58:21, 2025.
\newblock \href{https://doi.org/10.4230/LIPIcs.STACS.2025.58}{doi:10.4230/LIPIcs.STACS.2025.58}.

\bibitem[KP97]{KrausePudlak1997}
M. Krause and P. Pudl\'ak.
\newblock On the computational power of depth-2 circuits with threshold and modulo gates.
\newblock \emph{Theoretical Computer Science}, 174(1--2):137--156, 1997.
\newblock \href{https://doi.org/10.1016/S0304-3975(96)00019-9}{doi:10.1016/S0304-3975(96)00019-9}.

\bibitem[NS94]{NisanSzegedy1994}
N. Nisan and M. Szegedy.
\newblock On the degree of Boolean functions as real polynomials.
\newblock \emph{Computational Complexity}, 4:301--313, 1994.
\newblock \href{https://doi.org/10.1007/BF01263419}{doi:10.1007/BF01263419}.

\bibitem[Pag26]{Pago2026}
B. Pago.
\newblock Optimal lower bounds for symmetric modular circuits.
\newblock In \emph{53rd International Colloquium on Automata, Languages, and Programming (ICALP 2026)}, LIPIcs 374, pages 187:1--187:20, 2026.
\newblock \href{https://doi.org/10.4230/LIPIcs.ICALP.2026.187}{doi:10.4230/LIPIcs.ICALP.2026.187}.

\bibitem[Pat92]{Paturi1992}
R. Paturi.
\newblock On the degree of polynomials that approximate symmetric Boolean functions.
\newblock In \emph{24th Annual ACM Symposium on Theory of Computing (STOC 1992)}, pages 468--474, 1992.
\newblock \href{https://doi.org/10.1145/129712.129758}{doi:10.1145/129712.129758}.

\bibitem[Raz87]{Razborov1987}
A. A. Razborov.
\newblock Lower bounds on the size of bounded depth circuits over a complete basis with logical addition.
\newblock \emph{Mathematical Notes}, 41(4):333--338, 1987.
\newblock \href{https://doi.org/10.1007/BF01137685}{doi:10.1007/BF01137685}.

\bibitem[SY10]{ShpilkaYehudayoff2010}
A. Shpilka and A. Yehudayoff.
\newblock Arithmetic circuits: A survey of recent results and open questions.
\newblock \emph{Foundations and Trends in Theoretical Computer Science}, 5(3--4):207--388, 2010.
\newblock \href{https://doi.org/10.1561/0400000039}{doi:10.1561/0400000039}.

\bibitem[Smo87]{Smolensky1987}
R. Smolensky.
\newblock Algebraic methods in the theory of lower bounds for Boolean circuit complexity.
\newblock In \emph{19th Annual ACM Symposium on Theory of Computing (STOC 1987)}, pages 77--82, 1987.
\newblock \href{https://doi.org/10.1145/28395.28404}{doi:10.1145/28395.28404}.

\bibitem[Spe72]{Spencer1972}
J. Spencer.
\newblock Tur\'an's theorem for $k$-graphs.
\newblock \emph{Discrete Mathematics}, 2(2):183--186, 1972.
\newblock \href{https://doi.org/10.1016/0012-365X(72)90084-2}{doi:10.1016/0012-365X(72)90084-2}.

\bibitem[ST06]{StraubingTherien2006}
H. Straubing and D. Th\'erien.
\newblock A note on $\mathrm{MOD}_p$--$\mathrm{MOD}_m$ circuits.
\newblock \emph{Theory of Computing Systems}, 39(5):699--706, 2006.
\newblock \href{https://doi.org/10.1007/s00224-004-1210-2}{doi:10.1007/s00224-004-1210-2}.

\bibitem[VW08]{ViolaWigderson2008}
E. Viola and A. Wigderson.
\newblock Norms, XOR lemmas, and lower bounds for polynomials and protocols.
\newblock \emph{Theory of Computing}, 4(7):137--168, 2008.
\newblock \href{https://doi.org/10.4086/toc.2008.v004a007}{doi:10.4086/toc.2008.v004a007}.

\bibitem[Wil14]{Williams2014}
R. Williams.
\newblock Nonuniform ACC circuit lower bounds.
\newblock \emph{Journal of the ACM}, 61(1), Article 2, 32 pages, 2014.
\newblock \href{https://doi.org/10.1145/2559903}{doi:10.1145/2559903}.

\bibitem[YP94]{YanParberry1994}
P. Y. Yan and I. Parberry.
\newblock Exponential size lower bounds for some depth 3 circuits.
\newblock \emph{Information and Computation}, 112(1):117--130, 1994.
\newblock \href{https://doi.org/10.1006/inco.1994.1054}{doi:10.1006/inco.1994.1054}.

\bibitem[Yao90]{Yao1990}
A. C.-C. Yao.
\newblock On ACC and threshold circuits.
\newblock In \emph{31st Annual IEEE Symposium on Foundations of Computer Science (FOCS 1990)}, pages 619--627, 1990.
\newblock \href{https://ieeexplore.ieee.org/document/89583}{IEEE Xplore, document 89583}.

\end{thebibliography}
\end{document}